\documentclass[12pt,preprint]{aastex}

\slugcomment{ApJ, accepted}

\shorttitle{Far--Infrared Star Formation Rate Indicators}
\shortauthors{Calzetti et al.}

\begin{document}

\title{The Calibration of Monochromatic Far--Infrared Star Formation Rate
Indicators.\altaffilmark{1}}

\author{D. Calzetti\altaffilmark{2}, S.-Y. Wu\altaffilmark{2}, S. Hong\altaffilmark{2}, 
R. C. Kennicutt\altaffilmark{3}, J. C. Lee\altaffilmark{4}, D.A. Dale\altaffilmark{5}, 
C. W. Engelbracht\altaffilmark{6}, L.  van Zee\altaffilmark{7},  B. T. Draine\altaffilmark{8}, 
C.-N. Hao\altaffilmark{9,3},  K. D. Gordon\altaffilmark{10},  J. Moustakas\altaffilmark{11}, E.J. Murphy\altaffilmark{12}, M. Regan\altaffilmark{10}, A. Begum\altaffilmark{3}, 
M. Block\altaffilmark{6}, J.  Dalcanton\altaffilmark{13}, J. Funes\altaffilmark{14}, 
A. Gil de Paz\altaffilmark{15}, B. Johnson\altaffilmark{3}, S. Sakai\altaffilmark{16}, E. Skillman\altaffilmark{17},  F. Walter\altaffilmark{18}, D. Weisz\altaffilmark{17}, B. Williams\altaffilmark{13}, Y. Wu\altaffilmark{19} }

\altaffiltext{1}{Based on observations obtained with the Spitzer Space
Telescope, which is operated by JPL, CalTech, under NASA Contract
1407. } 
\altaffiltext{2}{Dept. of Astronomy, University of Massachusetts, Amherst, MA 01003; calzetti@astro.umass.edu}
\altaffiltext{3}{Institute of Astronomy, Cambridge University, Cambridge, U.K.}
\altaffiltext{4}{Carnegie Observatories of Washington, Pasadena, California}
\altaffiltext{5}{Dept. of Physics and Astronomy, University of Wyoming, Wyoming}
\altaffiltext{6}{Steward Observatory, University of Arizona, Arizona}
\altaffiltext{7}{Dept. of Astronomy, University of Indiana, Indiana}
\altaffiltext{8}{Princeton University Observatory, Peyton Hall, Princeton, New Jersey}
\altaffiltext{9}{Tianjin Astrophysical Center, Tianjin Normal University, 300387 Tianjin, China}
\altaffiltext{10}{Space Telescope Science Institute, Baltimore, Maryland}
\altaffiltext{11}{Center for Astrophysics and Space Sciences, University of California, San Diego, California}
\altaffiltext{12}{Spitzer Science Center, Caltech, California}
\altaffiltext{13}{Dept. of Astronomy, University of Washington, Washington}
\altaffiltext{14}{Vatican Observatory, University of Arizona, Arizona}
\altaffiltext{15}{Departamento de Astrofisica, Universidad Computense de Madrid, Spain}
\altaffiltext{16}{Dept. of Astronomy, UCLA, California}
\altaffiltext{17}{Dept. of Astronomy, University of Minnesota, Minnesota}
\altaffiltext{18}{Max Planck Institut f\"ur Astronomie, Heidelberg, Germany}
\altaffiltext{19}{IPAC, CalTech, Pasadena, California}

\begin{abstract}
Spitzer data at 24, 70, and 160~$\mu$m and ground--based H$\alpha$ images are analyzed 
for a sample of 189 nearby star--forming and starburst galaxies to investigate whether reliable 
star formation rate (SFR) indicators can be defined using the monochromatic infrared dust emission centered at 70 and 160~$\mu$m. We compare recently published recipes 
for SFR measures using combinations of the 24~$\mu$m and 
observed H$\alpha$ luminosities with those using 24~$\mu$m luminosity alone. 
From these comparisons, we derive a reference SFR indicator for use in our analysis. 
Linear correlations between 
SFR and the 70~$\mu$m and 160~$\mu$m luminosity are found for 
L(70)$\gtrsim$1.4$\times$10$^{42}$~erg~s$^{-1}$ and L(160)$\gtrsim$2$\times$10$^{42}$~erg~s$^{-1}$, corresponding to SFR$\gtrsim$0.1--0.3~M$_{\odot}$~yr$^{-1}$, and calibrations of SFRs based on 
L(70) and L(160) are proposed. Below those two 
luminosity limits, the relation between SFR and 70~$\mu$m (160~$\mu$m) luminosity is non--linear 
and SFR calibrations become problematic.  
A more important limitation is the dispersion of the data around the mean trend, which 
increases for increasing wavelength. The scatter of the 70~$\mu$m  (160~$\mu$m) 
data around the mean is about 25\% (factor $\sim$2) larger than the scatter of the 24~$\mu$m data.  
We interpret this increasing dispersion 
as an effect of the increasing contribution to the infrared emission of dust heated by stellar populations 
not associated with the current star formation. Thus, the 70 (160) $\mu$m luminosity can be reliably used to trace SFRs  in large galaxy samples, but will be of limited utility for 
individual objects, with the exception of infrared--dominated galaxies. The non--linear relation 
between SFR and the 70 and 160~$\mu$m emission at faint galaxy luminosities  suggests a variety  of mechanisms affecting the infrared emission for decreasing luminosity, such as increasing transparency of the interstellar medium, decreasing 
effective dust temperature, and decreasing filling factor of star forming regions across the galaxy. 
In all cases, the calibrations 
hold for galaxies with oxygen abundance higher than roughly 12$+Log$(O/H)$\sim$8.1. At lower metallicity the infrared luminosity no longer reliably  
traces the SFR because galaxies are less dusty and more transparent. 
\end{abstract}


\section{Introduction}
The star formation rate (SFR) is one of the principal parameters that needs to be measured 
in star--forming regions and galaxies, in order to characterize their evolution. Extensive 
efforts have been made over the past couple of decades to derive SFR indicators from 
luminosities at a variety of wavelengths, spanning from the UV, where the recently formed 
massive stars emit the bulk of their energy, to the infrared, where the dust-reprocessed light 
from those stars emerges, to the radio, which is mostly a tracer of supernova activity \citep[e.g.,][for earlier papers, see \citet{Kennicutt1998}]{Kennicutt1998,Yun2001, Kewley2002, Ranalli2003, Hirashita2003, Bell2003, Kewley2004, Calzetti2005, 
Schmitt2006, Moustakasal2006, AlonsoHerrero2006, Calzetti2007, Salim2007, Persic2007, Rosa2007, Kennicutt2007, Bigiel2008, Rieke2009, Calzetti2009}.

In recent years, investigations of monochromatic (i.e., based on a single band 
measurement) SFR indicators based on the infrared emission from 
galaxies have experienced a new resurgence, thanks to the high--sensitivity and high--angular resolution data 
provided by the Spitzer Space Telescope, which have yielded both deeper distant galaxy surveys and 
more accurate information on the relation between dust and stellar emission in nearby galaxies.
Deep surveys are often characterized by 
limited information across the infrared wavelength range and monochromatic SFRs are an important  tool 
for these projects. 

The rest--frame mid--infrared emission from dust in galaxies, in particular the emission detected 
in the 8~$\mu$m and 24~$\mu$m Spitzer bands, has been analyzed by a number of authors  
\citep[][]{Roussel2001,Forster2004,Boselli2004,Calzetti2005, Wu2005, AlonsoHerrero2006, PerezGonzalez2006, 
Rellano2007,  Calzetti2007, Zhu2008, Rieke2009, Salim2009}, and a general correlation (but also a number of 
caveats) between mid--IR infrared emission and SFR has been found.   

Over the next few years, new facilities, both from space (e.g., the Herschel Space Telescope) and 
from the ground (ALMA and the Large Millimeter Telescope, to mention just two) will open new 
windows of sensitivity 
at even longer wavelengths than those explored by Spitzer, and will, in turn, provide even more powerful tools 
for probing the evolution of the rate at which 
galaxies have assembled their 
%luminous 
gas and dust components. Deep surveys will  be able to probe the 
dust emission from galaxies at rest--frame infrared wavelengths that are close to the peak emission, 
and to the bulk of the infrared energy budget. Herschel will, for instance, enable us to probe the 
peak dust emission from galaxies ($\sim$60--150~$\mu$m) up to redshift z$\sim$2, while the Rayleigh--Jeans 
tail of the dust emission will be the dominion of sub--millimeter and millimeter  facilities. 

Exploring the viability, and limitations,  of using the monochromatic emission close to the infrared 
peak and at longer wavelengths as SFR indicators is thus timely for providing a reference for those future 
surveys. In this paper we investigate the use of the 70~$\mu$m and 160~$\mu$m emission,   the 
two longest wavelength Spitzer bands, from nearby galaxies as SFR diagnostics. 

The present paper is organized as follows: Section~2 introduces the
sample of local star--forming galaxies and the data used for the present analysis; Section~3 
describes the quantities used in this work; Section~4 
compares a variety of existing SFR indicators in the optical/infrared, to derive  `reference' 
SFRs for our galaxies, which are then compared with the Spitzer 70~$\mu$m and 160~$\mu$m emission from 
the same galaxies and with expectations from models in Sections~5 and 6, respectively.  A discussion of the results and the conclusions are given in Section~7.  Throughout the paper, we adopt a value of the Hubble constant 
H$_0$=70~km~s$^{-1}$~Mpc$^{-1}$. 

\section{Sample Description}

We combine data from two Spitzer Legacy surveys, LVL \citep[Local Volume Legacy,][]{Dale2009,Lee2009} and 
SINGS  \citep[Spitzer Infrared Nearby Galaxies Survey,][]{Kennicutt2003}, with the local starburst 
galaxy sample of the Spitzer MIPS GTO programs \citep{Engelbracht2008} and the Luminous 
Infrared Galaxies (LIRGs) sample of  \citet{AlonsoHerrero2006}, in order to span close to 5 
orders of magnitude in SFR surface density (SFSD=SFR/area), and thus cover as wide as possible a range 
that may characterize star--forming galaxies at all redshifts. 

Galaxies with MIPS detections at 24, 70, and 160~$\mu$m and H$\alpha$ measurements are preferentially selected, though, in order to broaden the sample   
to low-luminosity and low-metallicity objects, galaxies with MIPS upper limits  are included if H$\alpha$($\lambda$0.6563~$\mu$m) or P$\alpha$($\lambda$1.876~$\mu$m) 
data exist for them. In general, when both MIPS and H$\alpha$ measurements are present, extinction 
corrected SFRs will be derived by combining the observed 24~$\mu$m and H$\alpha$ luminosities (see 
section~4.2); for the few MIPS 24~$\mu$m upper limits in our sample, we will rely on the extinction 
corrected (via the H$\alpha$/P$\alpha$ ratio) P$\alpha$ luminosity, or will assume, for low 
metallicity objects, that the SFR is practically unobscured by dust and use the observed H$\alpha$ 
luminosity.  We also require that measurements of oxygen abundance be available for each galaxy, 
either from the spectroscopy of \citet{Moustakas2009} or from the literature 
\citep{Marble2010,Engelbracht2008}, as the infrared luminosity is sensitive to the metal 
abundance in galaxies \citep[][]{Cannon2005,Engelbracht2005,Cannon2006a,Cannon2006b,Walter2007,Calzetti2007,Rellano2007}, 
and this is one of the parameters investigated in the present paper. Finally, images in the emission lines of 
H$\alpha$ or H$\alpha+$[NII] need to be available, to measure the extent of the emitting region 
in the warm ionized gas; we use the line emitting area to normalize all luminosities, in order to 
remove any dependence on distance from our results, and to ensure that galaxies are not 
scaled by mass (or by global luminosity). In this paper, we use the definition of 
luminosity surface density (LSD=luminosity/area), and infrared luminosity surface density (IRSD), to remove any dependence of luminosity on the galaxy distance or size.

The selection criteria and observation strategy for the 258 galaxies in the LVL sample are described 
in \citet{Dale2009} and \citet{Lee2009}; IRAC and MIPS total fluxes for each galaxy are reported 
in \citet{Dale2009}, while the foreground--galactic--extinction--corrected H$\alpha+$[NII] total fluxes 
and [NII]/H$\alpha$ ratios and other general information on the galaxies, including distances, are reported in \citet[][]{Kennicutt2008}.  The main criterion for the LVL sample is that galaxies need to be 
within the local 11~Mpc volume. %\citep[]{Kennicutt2008,Dale2009,Lee2009}. 
Oxygen abundances from the literature are currently available for 108 of the LVL galaxies  \citep{Marble2010}; the additional 
requirement to have narrow--band optical images available for measuring emission 
region sizes further reduces the sample to 80 galaxies\footnote{The final LVL H$\alpha$ sample will include 174 galaxies.}. Of these, 
11 are starbursts already included in 
the MIPS--GTO sample (see below); we thus consider our final LVL sample as consisting of 69 galaxies. 

The SINGS sample contains 75 galaxies closer than about 30~Mpc, representative of a wide 
range of properties in terms of morphology, luminosity, dust temperature, etc. \citep{Kennicutt2003}. 
Of those 75 galaxies, 33 are in common with the LVL sample. The number of SINGS galaxies 
included in our sample is further reduced by excluding non--star--forming, 
early type galaxies, mostly Sy2--dominated ellipticals or S0's (NGC584, NGC855, NGC1266, 
NGC1316, NGC1404, NGC4552, NGC5195), and one galaxy, NGC5033, for which optical narrow--band 
data are not available.  We thus have a final sample of 67 SINGS galaxies, 31 of which in common with 
the LVL sample. Total MIPS fluxes for the SINGS galaxies are listed in \citet{Dale2007}, and the 
H$\alpha+$[NII] fluxes plus the [NII]/H$\alpha$ ratios are listed in \citet{Kennicutt2009}. Oxygen 
abundances are from \citet{Moustakas2009}. 

In the LVL and SINGS samples we retain four galaxies,  M81DwA, HolmbergIX, UGC6900, and 
UGC9128, which are undetected or marginally detected in the  MIPS bands, but have H$\alpha$ measurements; these galaxies are metal poor, and consequently, dust--poor, thus likely to have 
most of their star formation light emerge directly at UV/optical  wavelengths, unobscured by dust. 

The areas used to normalize luminosities and calculate LSDs are 
derived, for each galaxy, from the area occupied by the ionized gas emission as traced by the H$\alpha$. For the LVL and SINGS samples, ionized gas emission sizes are measured directly from our images, and are defined as the semi--major axis of the ellipse that includes 2/3 of the total H$\alpha$  emission flux. Given the uniform nature of the LVL and 
SINGS imaging strategies, this implies that the sizes obtained for those galaxies are internally consistent. The `2/3 ionized gas emission' radius typically includes all or most of the 
high surface brightness emitting regions in a galaxy. 
This definition for the emitting area is, therefore,  
especially useful for our analysis: it ensures some level of uniformity even when including galaxies for which the H$\alpha$  image depths vary from object to object. 
This is the case for  some of the 
galaxies described in the next samples (see below), where the only H$\alpha$ images available 
are from the literature.  

The MIPS--GTO sample of starbursts is described in \citet{Engelbracht2008}: the galaxies cover a wide range in metal content, from very metal--poor starbursts, like IZw18 and SBS0335$-$052, to metal--rich examples like  IC342. Of the 65 (we consider SBS0335$-$052E and SBS0335$-$052W a single galaxy in this paper) galaxies listed in \citet{Engelbracht2008}, we only keep the 55 that satisfy one of the following two criteria: 1. they are detected in all three MIPS bands (we exclude 
NGC1614 and NGC3256 which are part of the LIRGs sample, see below); or 2. they are detected and measured in either H$\alpha$ or P$\alpha$ \citep{Calzetti2007}. 
The latter criterion ensures inclusion in the sample of metal--poor galaxies, like IZw18, SBS0335$-$052, 
UGCA292, and HS0822$+$3542, which are marginally or un--detected in MIPS because their dust content is low. Ten of the starbursts in the MIPS--GTO sample have  images in the P$\alpha$ ($\lambda$1.876~$\mu$m) line \citep{Calzetti2007}, which we use to derive SFRs,  
after extinction correction using the H$\alpha$ emission, and to measure the size of the line emitting region when H$\alpha$ images are not available. 
For much of our analysis, we give emphasis to starbursts with {\em both} infrared broad--band and optical/nearIR narrow--band detections, 
since our `reference' SFR indicators generally require information on both (see next section). The  distances of the 55 starbursts are typically 
within 100~Mpc, except for one case (Tol2138$-$405, located at 246~Mpc). 
Total MIPS flux values, and literature values for distances, oxygen abundances, and H$\alpha$ 
fluxes are listed in  \citet{Engelbracht2008}. We supplement, where possible, H$\alpha$ measurements 
from the literature, when not present in  \citet{Engelbracht2008}, and we end up with  line emission measurements for a total of 45 galaxies. 

In order to obtain the emitting areas for the starbursts, we have searched the literature for images of the ionized gas emission (either H$\alpha$ or P$\alpha$); we were able to retrieve such information for a total of 44 starburst galaxies. For the 34 galaxies with  
H$\alpha$ images available through NED\footnote{ NED is the NASA/IPAC Extragalactic Database which is operated by the Jet Propulsion Laboratory, California Institute of Technology, under contract with the National Aeronautics and Space Administration.} in FITS format and/or with P$\alpha$ images, we derive the ionized gas emission size in the same fashion as for the LVL and SINGS galaxies, with 
the proviso that non--uniform depths from image to image will reflect 
into larger uncertainties for the sizes (next section). Finally, sizes derived from published hard-copy images only (10 galaxies) have 
larger uncertainties still, as discussed in section~3. Overall, the values for the sizes of the ionized gas 
emitting region of the starbursts are the least homogeneous in our sample; variations in sizes 
of up to 30\% may be expected from differences in the depths and characteristics of the published images. 

Ionized gas emission sizes for the 160 total LVL (67),  SINGS (69, 31 of which are in common with LVL), and starburst galaxies (55) are listed in Table~\ref{tbl-1}, together with other derived quantities relevant to the present analysis. 

Our sample of 160 star--forming and starburst galaxies is augmented with 29 LIRGs  from \citet{AlonsoHerrero2000, AlonsoHerrero2001,AlonsoHerrero2002,AlonsoHerrero2006}, to expand our 
SFR surface density range at the high end, covering in total almost 5 orders of magnitude with the 
combined sample. Selection criteria for the sample are presented in 
\citet{AlonsoHerrero2006}, together with measurements of the P$\alpha$ flux, extinction corrections, as well as information on the physical extent of the star forming area for 24 LIRGs. For a few of the LIRGs 
with extended ionized gas emission, size information is taken from \citet{Hattori2004}. To the 24 LIRGs, 
we add five additional LIRGs (NGC1614, 
NGC3256, NGC3690, NGC5653, and Zw049.057) with data from \citet{AlonsoHerrero2000, AlonsoHerrero2001,AlonsoHerrero2002}.  Infrared measurements 
at 25, 60, and 100~$\mu$m from IRAS and distances (which we rescale to our adopted value of the Hubble constant) for each galaxy are from \citet{Sanders2003} and \citet{Surace2004}.  For two 
of the LIRGs, NGC1614 and NGC3256, MIPS measurements are available from \citet{Engelbracht2008}. 
Although individual oxygen abundance measurements are not available for the majority of the 
LIRGs, their metallicities are characteristic of high--metallicity galaxies \citep{AlonsoHerrero2006}. Basic 
parameters for the 29 LIRGs are listed in Table~\ref{tbl-2}. 

The final sample consists of 189 nearby star--forming galaxies, which we divide into a `high--metallicity' sample, consisting of 142 objects with oxygen abundances 12$+Log$(O/H)$\gtrsim$8.1 (113 normal star--forming and starburst and 29 LIRGS), and a `low--metallicity' sample of 47 objects with oxygen abundance 12$+Log$(O/H)$<$8.1 (Table~\ref{tbl-1}).  We do not consider the sample large enough to be reasonably divided into more 
than two metallicity bins. The separating value in metallicity is chosen after \citet{Engelbracht2005}, \citet{Jackson2006}, 
\citet{Draine2007}, and 
\citet{Engelbracht2008}, where it is found that galaxies roughly below this metallicity value tend to be underluminous in their PAH emission (as measured in the Spitzer--IRAC 8~$\mu$m band). 
We do not expect the striking trend of the PAH bands to be also present in the 
thermal--dust--emission--dominated longer wavelength bands. However, we do expect that, as the metal and dust content of galaxies decreases, the galaxies become increasingly transparent, i.e., their dust opacity decreases, and the infrared dust emission will be a progressively less accurate tracer of the current SFR.  We thus select the value 12$+Log$(O/H)$\sim$8.1 as a convenient `transition' value for dust content in our galaxies. In section~7 (Discussion and Conclusions), we also discuss the implications of further dividing the high--metallicity sample into two subsamples, in order to investigate the role of metal abundance in driving some of the observed scatter in the data.  

\begin{figure}
\figurenum{1}
\plotone{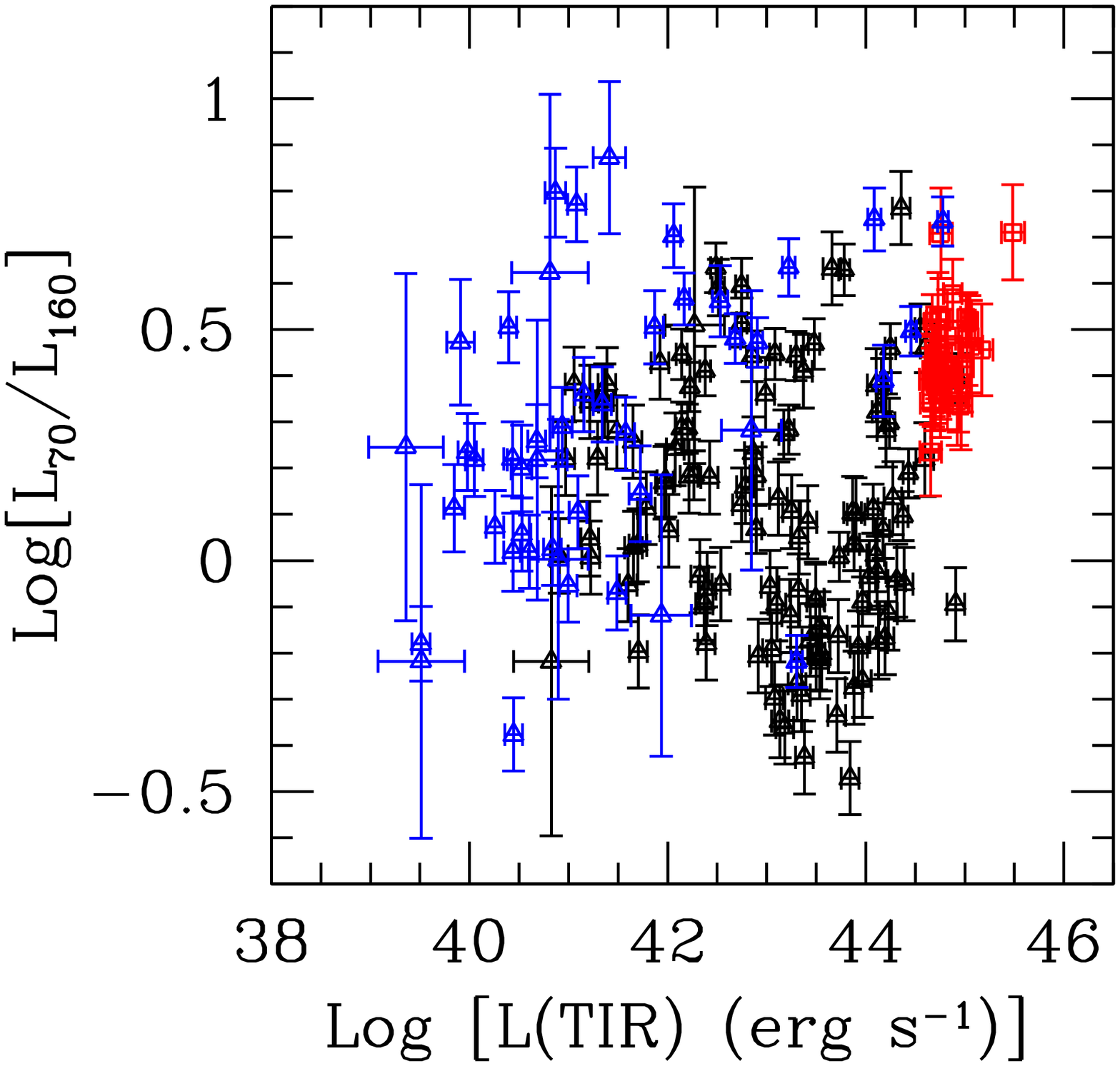}
%\plotone{figure16.eps}
\caption{The logarithm of the ratio of 70--to--160~$\mu$m luminosity for the 189 galaxies in our sample, as a function of the total infrared luminosity TIR. The sample covers almost 6 orders of magnitude 
in infrared luminosity. Symbol colors identify the star--forming and starburst high--metallicity 
(black symbols) and low--metallicity  (blue symbols) galaxies,  and the LIRGs (red).  
%(left) the infrared luminosity surface density 
%($\Sigma_{TIR}$=L(TIR)/area) and (right) the SFR surface density 
%($\Sigma_{SFR}$) for the star--forming and starburst high--metallicity (black symbols) and low--%metallicity  (blue 
%symbols) galaxies,  and the LIRGs (red). Two indicative mean trends are shown for the high 
%metallicity datapoints, in both plots; the trends are not fits to the data. 
\label{fig1}}
\end{figure}

The range of infrared luminosity covered by our sample is shown in Figure~\ref{fig1}, together with the range of 70--to--160~$\mu$m luminosity ratio, the latter being a rough proxy for effective dust temperature \citep[e.g.][]{Engelbracht2008}. The combination 
of low--metallicity and high--metallicity star forming and starburst galaxies plus the addition of the LIRGs 
enables us to explore almost six orders of magnitude in total infrared luminosity, from $\sim$3$\times$10$^{39}$~erg~s$^{-1}$ to $\sim$3$\times$10$^{45}$~erg~s$^{-1}$ (the derivation of fluxes in `MIPS--equivalent' bands for the LIRGs is discussed in the next section). We also explore one order of 
magnitude range in the 70--to--160~$\mu$m luminosity ratio, from about 1/3 to roughly 6 times 
in L(70)/L(160). The LIRGs occupy a relatively small range in luminosity ratio, with values in the 
range L(70)/L(160)$\sim$2--5, as expected for these warm (relatively to many galaxies) systems. 

\section{Luminosities, Sizes,  and Uncertainties}

%For the LVL, SINGS, and MIPS--GTO  samples, data from Spitzer MIPS are available at 
%24, 70, and 160~$\mu$m. Measured fluxes and uncertainties are reported in 
%\citet{Dale2008,Dale2007,Engelbracht2008} for the LVL, SINGS, and MIPS--GTO samples, respectively. 
 We derive total infrared luminosities, L(TIR), in the range 3--1100~$\mu$m for the galaxies in the LVL, SINGS, and MIPS--GTO samples using 
equation~4 in \citet{DaleHelou2002}. Most of the galaxies are also bright enough to have IRAS detections, at least at 25, 60, and 100~$\mu$m, but we concentrate on the 
MIPS data alone. All infrared fluxes  are used in this work as directly 
measured  \citep[different from the approach of, e.g.,][who use color--corrected fluxes]{Engelbracht2008}. 

For most of the LIRGs, only IRAS infrared data are available, typically with detections at 25, 60, and 
100~$\mu$m.  Exceptions are NGC1614 and NGC3256, which have published MIPS measurements 
\citep{Engelbracht2008}, and NGC2369, ESO320-G030, and Zw049.057, for which \citet{Rieke2009} published infrared SEDs. Since our reference luminosities are those in the MIPS bands, we need 
to interpolate the IRAS band fluxes of all the other LIRGs to derive MIPS-equivalent ones. The 
difference between the luminosity at MIPS 24~$\mu$m and at IRAS 25~$\mu$m is small
 \citep[see, also][]{Kennicutt2009}, and we adopt the best fit:
\begin{equation}
Log [L(24)] = Log [L(25)] - 0.028,
\end{equation}
where L($\lambda$) = 4 $\pi$ d$^2$ $(c/\lambda)f_\nu(\lambda)$, in units of erg~s$^{-1}$, is the monochromatic luminosity at wavelength $\lambda$, f$_{\nu}$($\lambda$) is the flux density per unit 
frequency, and $d$ is the galaxy's distance. The difference between the 60~$\mu$m IRAS band and the 70~$\mu$m MIPS band is larger than between the  25~$\mu$m  and 24~$\mu$m bands, and we use a 
%simple linear
polynomial  interpolation between 
the 60 and 100~$\mu$m IRAS bands to derive a 70~$\mu$m--equivalent flux density:
%\begin{equation}
%Log[L(70)] = Log [L(60)] + [{Log(70/60)\over Log(100/60)}] Log [L(100)/L(60)].
%\end{equation}
\begin{equation}
f_{\nu}(70) = f_{\nu}(60) (a_0 + a_1 x + a_2 x^2 + a_3 x^3 + a_4 x^4 + a_5 x^5),
\end{equation}
with f$_{\nu}$($\lambda$) the flux density in Jy, $x$=f$_{\nu}$(60)/f$_{\nu}$(100), and 
the vector\newline  
a$_n$=[1.5078, 0.9356, $-$3.8900, 4.0576, $-$1.7617, 0.2712]. 
Tests using starburst galaxies from the MIPS--GTO sample with IRAS 60/100 colors similar to the
 LIRGs \newline  (0.4$\le$f$_{\nu}$(60)/f$_{\nu}$(100)$\le$1.3) show 
that the above interpolation reproduces the observed L(70) with a typical uncertainty of less than 
10\% (i.e., less than 0.04~dex in logarithmic scale). We also compare our interpolated values with MIPS--equivalent fluxes derived from the template SEDs of \citet{Rieke2009}, by matching LIRGs to the templates according to TIR luminosity and SED shape (using the IRAS fluxes). We typically find 
that the values derived from the templates 
differ from those derived from equations~1 and 2 by $\sim$1--5\% at 24~$\mu$m and less than 
12\% at 70~$\mu$m. 

We use the SED templates  also to extrapolate fluxes at 160~$\mu$m for the LIRGs with only IRAS measurements. Nine of the 29 LIRGs have long--wavelength 
measurements at 850~$\mu$m from SCUBA on 
the JCMT \citep{Dunne2000} and one, NGC3690/IC694, at 1.2~mm from the IRAM--30~m 
telescope \citep{Braine1998}. Interpolations 
between these long wavelengths and the IRAS 100~$\mu$m band to recover the MIPS 
160~$\mu$m emission are complicated by the presence of a cool, T$\approx$20~K, dust component in the IR emission of LIRGs \citep[e.g.][]{Dunne2001}. The MIPS~160 $\mu$m band is close to the emission peak of 
this cool dust component, thus any interpolation will be subject to the uncertainty of the relative, 
and unknown, ratio of the cool and warm dust emission components. A simple linear interpolation 
between 100~$\mu$m and 450~$\mu$m \citep[using the roughly constant 450--to--850 ratio of 7.9 
found by][]{Dunne2001}  provides estimated 160~$\mu$m fluxes that are between 30\% and 75\% 
lower than those estimated from the template SEDs, in agreement with the presence of the 20~K 
emitting dust component. To avoid a significant underestimate in the 160~$\mu$m flux, we adopt the results from the extrapolation of the SED templates as our `bona fide' 160~$\mu$m fluxes for the LIRGs. 
The resulting 70--to--160~$\mu$m luminosity ratio for the LIRGs has median value L(70)/L(160)$\sim$2.7, to be compared 
with the median ratio L(70)/L(160)$\sim$3.5 of the five galaxies for which direct MIPS measurements 
are available. The difference is only $\sim$30\% (0.11~dex), implying that our procedure overestimates L(160) by that amount at most. This is well within the typical uncertainty for our  measurements.  

For the wavelength--integrated infrared luminosity L(TIR) of the LIRGs, we compare the results of 
equation~4 of  \citet{DaleHelou2002}, using the MIPS--equivalent fluxes obtained with the SED  
templates and equations~(1) and (2) with the results of their  equation~5, using the IRAS bands. The two equations are known to provide different L(TIR) values for the more quiescently star forming 
galaxies and for galaxies with cooler dust    \citep{DaleHelou2002}, because the 
MIPS 160~$\mu$m band traces cool dust more accurately than the IRAS 100~$\mu$m band 
(the two being the longest wavelength bands from Spitzer and IRAS, respectively). The offset is 
dependent on the 60/100 color  \citep[see, e.g.][]{Kennicutt2009}, and 
L(TIR) values obtained with the two methods for the starburst galaxies from the sample of 
\citet{Engelbracht2008} provide an empirical correction:
\begin{equation}
%Log [L(TIR)_{MIPS}] = Log [L(TIR)_{IRAS}] + 0.20 Log [f_{\nu}(60)/f_{\nu}(100)],
Log [L(TIR)_{MIPS}] = Log [L(TIR)_{IRAS}] - 0.35 Log [f_{\nu}(60)/f_{\nu}(100)] - 0.15,
\end{equation}
which gives consistent L(TIR) values with a scatter typically less than 25\%.  A list of 
`MIPS--equivalent' quantities for the LIRGs is given in Table~\ref{tbl-2}. 

We finally apply equations~(1)--(3) and the SED templates of \citet{Rieke2009} to  recover, from the IRAS measurements,  the 24, 70, and 160~$\mu$m fluxes, and the TIR luminosity  of NGC3034\footnote{We recover the following flux densities for NGC3034: f$_{\nu}$(24)=287.4~Jy, f$_{\nu}$(70)=1622.~Jy, f$_{\nu}$(160)=639.6~Jy;  the IRAS 25, 60, and 100~$\mu$m data used to derive the `MIPS--equivalent' fluxes are from \citet{Sanders2003}.}, a starburst  galaxy included in both  the  SINGS and LVL 
samples, but which is saturated in all three MIPS bands. 

Galaxy--integrated emission line data include H$\alpha+$[NII] for the 
LVL \citep{Kennicutt2008} and SINGS \citep{Kennicutt2009} samples; those papers also report the 
[NII]/H$\alpha$ ratios appropriate for each galaxy. Typical uncertainties on the final 
integrated H$\alpha$ fluxes are 
about 10\%--15\%. Similar data are also available for 39 of the MIPS--GTO galaxies, mostly from the 
literature \citep{Engelbracht2008}, although measurements for 11 galaxies are from the LVL data \citep{Kennicutt2008}, and, for 10 galaxies, P$\alpha$ fluxes are also available \citep{Calzetti2007}.  Larger uncertainties, of order 20\%, are expected for the H$\alpha$ fluxes of the starbursts, because of the non--uniform nature of the data. Finally, the only ionized gas data readily available for the LIRGs are 
the P$\alpha$ data presented in \citet{AlonsoHerrero2006}, together with estimates of the dust 
extinction values. 

\begin{figure}
\figurenum{2}
\plotone{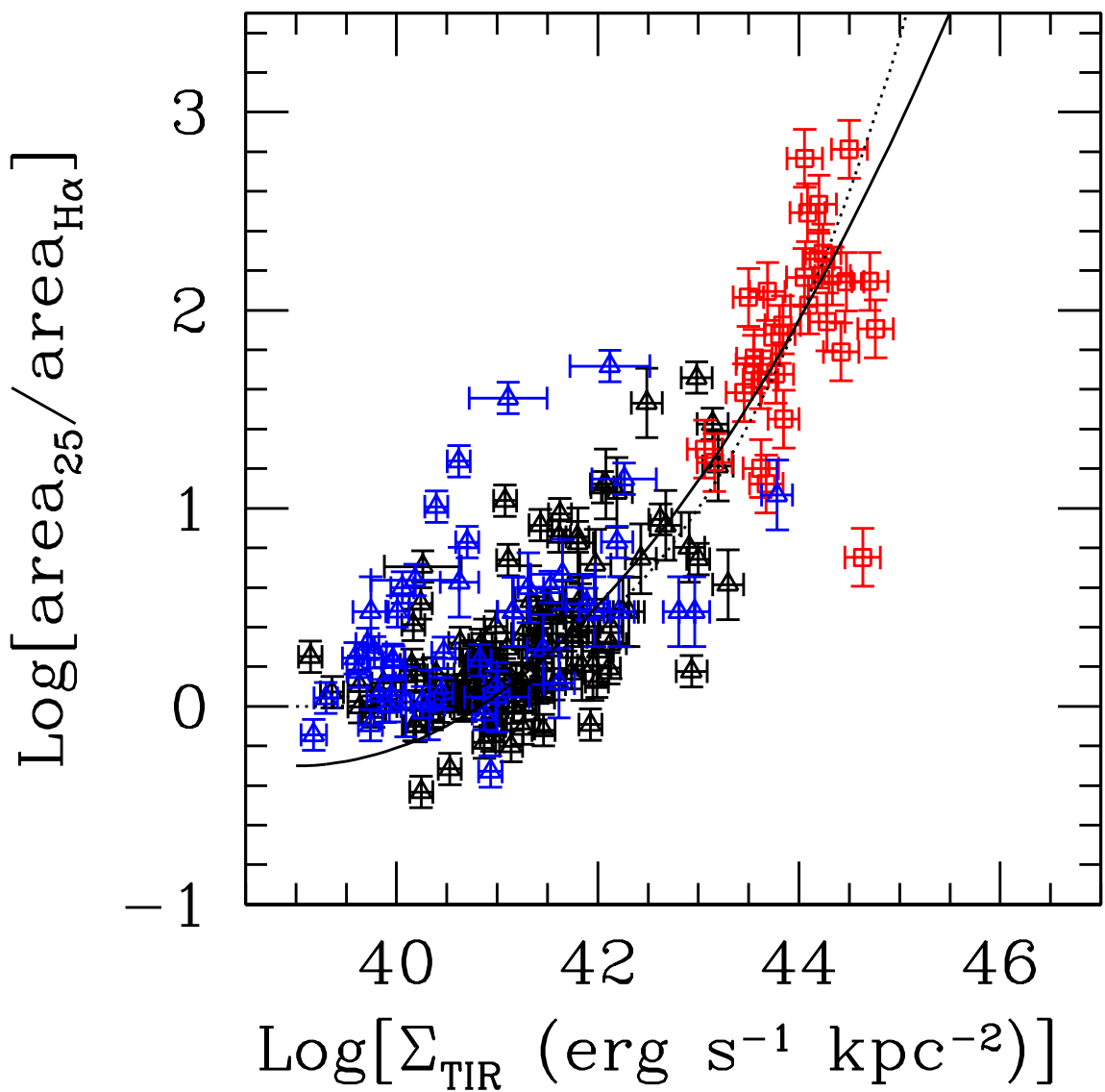}
%\plotone{figure1.eps}
\caption{The ratio of the area defined by the R$_{25}$ radius to the ionized gas (H$\alpha$) emitting area as a function of  the infrared luminosity surface density 
(IRSD: $\Sigma_{TIR}$=L(TIR)/area) for our sample galaxies, where the area used to normalize the infrared luminosity is area$_{H\alpha}$. Symbol colors are as in Figure~\ref{fig1}. 
%the star--forming and starburst high--metallicity 
%(black symbols) and low--metallicity  (blue 
%symbols) galaxies,  and the LIRGs (red). 
Two indicative mean trends are shown for the high 
metallicity datapoints; the trends are not fits to the data. 
%(left) the infrared luminosity surface density 
%($\Sigma_{TIR}$=L(TIR)/area) and (right) the SFR surface density 
%($\Sigma_{SFR}$) for the star--forming and starburst high--metallicity (black symbols) and low--%metallicity  (blue 
%symbols) galaxies,  and the LIRGs (red). Two indicative mean trends are shown for the high 
%metallicity datapoints, in both plots; the trends are not fits to the data. 
\label{fig2}}
\end{figure}

Uncertainties for the line--emitting region sizes range from about 10\% for the galaxies for which 
we perform a direct measure on the images (mostly the LVL and SINGS samples), to $\sim$15\%--20\% for the  
34 MIPS--GTO galaxies with available FITS images, to about 20\%--30\% for the starburst galaxies for which images 
were available only in hard--copy published form. For the LIRGs, the region size information is 
from \citet{Hattori2004} and  \citet{AlonsoHerrero2006}, with an estimated uncertainty of about 20\%. For 11 of the 
MIPS--GTO starbursts, line--emission images are not available, either from archives (e.g., NED) or 
from the literature. We observe, however, that the size of the star forming region decreases relative to the 
size of the stellar emission, defined by R$_{25}$, as the  infrared luminosity surface density  
(IRSD=L(TIR)/area) 
increases (Figure~\ref{fig2}). This had already been reported by 
\citet{WangHelou1992} and \citet{LehnertHeckman1996}, as an 
effect of increasing compactness of the region for increasing SFSD. Indeed, 
the infrared luminosity traces the SFR fairly accurately at high luminosities \citep{Hunter1986,Devereux1990,LehnertHeckman1996}, and only at low 
luminosities this correlation breaks, i.e., in a regime where the dust column density is low and most 
of the star formation light is unprocessed by dust; a similar trend to that of Figure~\ref{fig2} is, 
in fact, observed when the IRSD is replaced by the SFSD in the correlation with 
area$_{25}$/area$_{H\alpha}$. 
We use the median 
value measured for starbursts (area$_{25}$/area$_{H\alpha}\sim$3)  to ascribe a star--forming size 
to the 11 MIPS--GTO starbursts for which only R$_{25}$ is available (Table~\ref{tbl-1}).

In addition to normalizing luminosities by the emission area, we also normalize by total stellar mass 
in the galaxy as a way to remove distance dependencies. The 3.6~$\mu$m emission from the 
galaxies is used here as a proxy for the stellar mass, since, at these long wavelengths, the light is 
dominated by the photospheric emission from low--mass stars; hence the measurement is insensitive to the details of the galaxy's star formation history, and dust attenuation is small \citep[see, e.g.,][for a discussion on the dust extinction curve at these wavelengths]{Nishiyama2009}. Furthermore,  the contribution from dust emission is also negligible, even if  the 3.3~$\mu$m PAH emission band is 
 included in the Spitzer--IRAC 3.6~$\mu$m band \citep{Pahre2004,Calzetti2005,Dale2007,Engelbracht2008, Leroy2008}. The list of 3.6~$\mu$m fluxes for each galaxy is given in the same publications where the MIPS data are listed. For the LIRGs, we use the 
K$_S$--band measurements (retrieved from NED), which, after extinction--correction using the starburst 
curve of  \citet{Calzetti2000}, we extrapolate to 3.6~$\mu$m--equivalent fluxes using the starburst models from Starburst99 
\citep[2007 Update; ][]{Leitherer1999}, with ages in the range 20--100~Myr; we note that  
at these long wavelengths the spectral slope is basically insensitive to age, up to about 10~Gyr. The inferred `3.6~$\mu$m--equivalent' luminosities for the LIRGs 
correspond to extinction--corrected values, which are the appropriate values to use in our 
analysis. In the presence of significant extinction at 
3.6~$\mu$m, the observed luminosities are expected to be lower than those 
listed in Table~\ref{tbl-2}.

\section{Reference Star Formation Rates}

When testing the suitability of a luminosity (monochromatic and non--monochromatic)  as a SFR indicator, the 
first task is to decide what the `reference' SFR tracer is going to be for the selected sample. 
In this work, we give preference to `reference' SFR tracers which incorporate infrared luminosities, 
as the goal is to test the 70~$\mu$m luminosity as a SFR indicator, and remaining close to 
this wavelength allows us to avoid biases due to large differences in the dust optical depth 
probed.

In addition to the historical SFR tracers based on the bolometric infrared emission \citep{Hunter1986,Lonsdale1987,Rowan1989,Devereux1990,Sauvage1992,LehnertHeckman1996,Kennicutt1998}, 
a number of calibrations have appeared in the literature in recent years based on 
the Spitzer MIPS 24~$\mu$m monochromatic band emission, or, equivalently, on the 
IRAS 25~$\mu$m emission \citep[e.g., ][]{Calzetti2005,Wu2005,AlonsoHerrero2006,PerezGonzalez2006,Calzetti2007,Zhu2008,Rieke2009,Kennicutt2009}. Composite indicators based  on a combination of the H$\alpha$ optical recombination line emission or UV stellar continuum and one infrared band emission, at 8~$\mu$m or 24~$\mu$m, have also been calibrated \citep{Calzetti2007,Kennicutt2007,Zhu2008,Leroy2008,Kennicutt2009}, since the infrared bands capture the SFR from regions obscured by dust, while the optical recombination line (or UV stellar continuum) captures the SFR from regions that are relatively unobscured by dust. The composite indicators are indeed the most appropriate to use in the case of low--metallicity galaxies, where a large fraction of the star formation emerges virtually unaffected by dust re--processing.  

We intentionally avoid SFR indicators based on the total infrared emission, L(TIR), because the 70~$\mu$m wavelength region is close to the peak emission for many galaxies, and L(70) 
is thus a significant contributor to L(TIR); this causes a degeneracy between L(70) and L(TIR). Furthermore, the dust contributing to the long--wavelength infrared emission can be heated by evolved 
stellar populations, as well as by the currently star--forming ones; the `extra' contribution to the infrared 
emission will produce an overestimate of the true SFR \citep{Lonsdale1987, Sauvage1992, Buat1996}. Although the magnitude of this contribution  is uncertain and its impact on the infrared emission not 
unambiguously established 
\citep[see, e.g.,][]{Kewley2002}, its potential dependence on the relative ratio of evolved/star--forming 
stellar populations in the galaxy is sufficient reason for not using L(TIR) as a reference SFR in the 
present work. 

\subsection{The Infrared Spectral Energy Distribution of Star--forming Galaxies}

Most findings indicate that the correlation between L(24) and SFR is non--linear, in the sense that 
the infrared emission at 24~$\mu$m is overluminous in more actively star--forming systems relative to a 
linear scaling between L(24) and SFR. Two non--exclusive 
interpretations are possible for this result. The first interpretation attributes the overluminosity 
at the bright end of the 24~$\mu$m emission to the higher mean dust temperatures of more active 
systems \citep{Calzetti2007}; in this case, the infrared spectral energy 
distribution (SED) becomes `bluer' and L(24)/L(TIR) increases at higher SFRs. The second interpretation suggests that, as  the dust optical depth increases with increasing SFR,  standard dust extinction correction methods based, e.g., on the Balmer decrement or other hydrogen line ratios, 
are insufficient to recover the intrinsic values of  optical or near--infrared indicators 
(H$\alpha$, P$\alpha$, etc.), thus artificially leading to the 
interpretation that L(24) is overluminous relative to those indicators \citep{Rieke2009}. 

\begin{figure}
\figurenum{3}
\plottwo{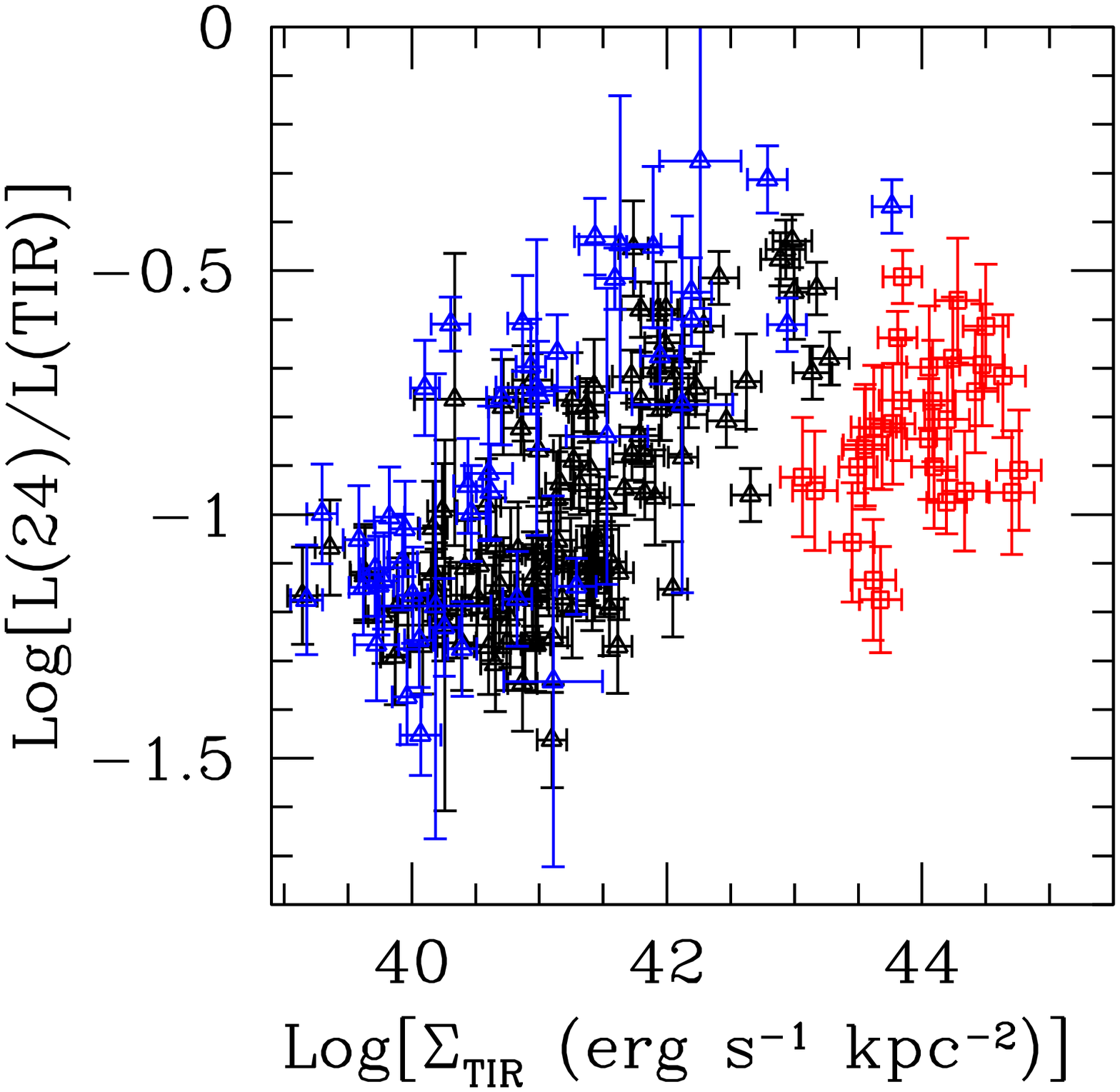}{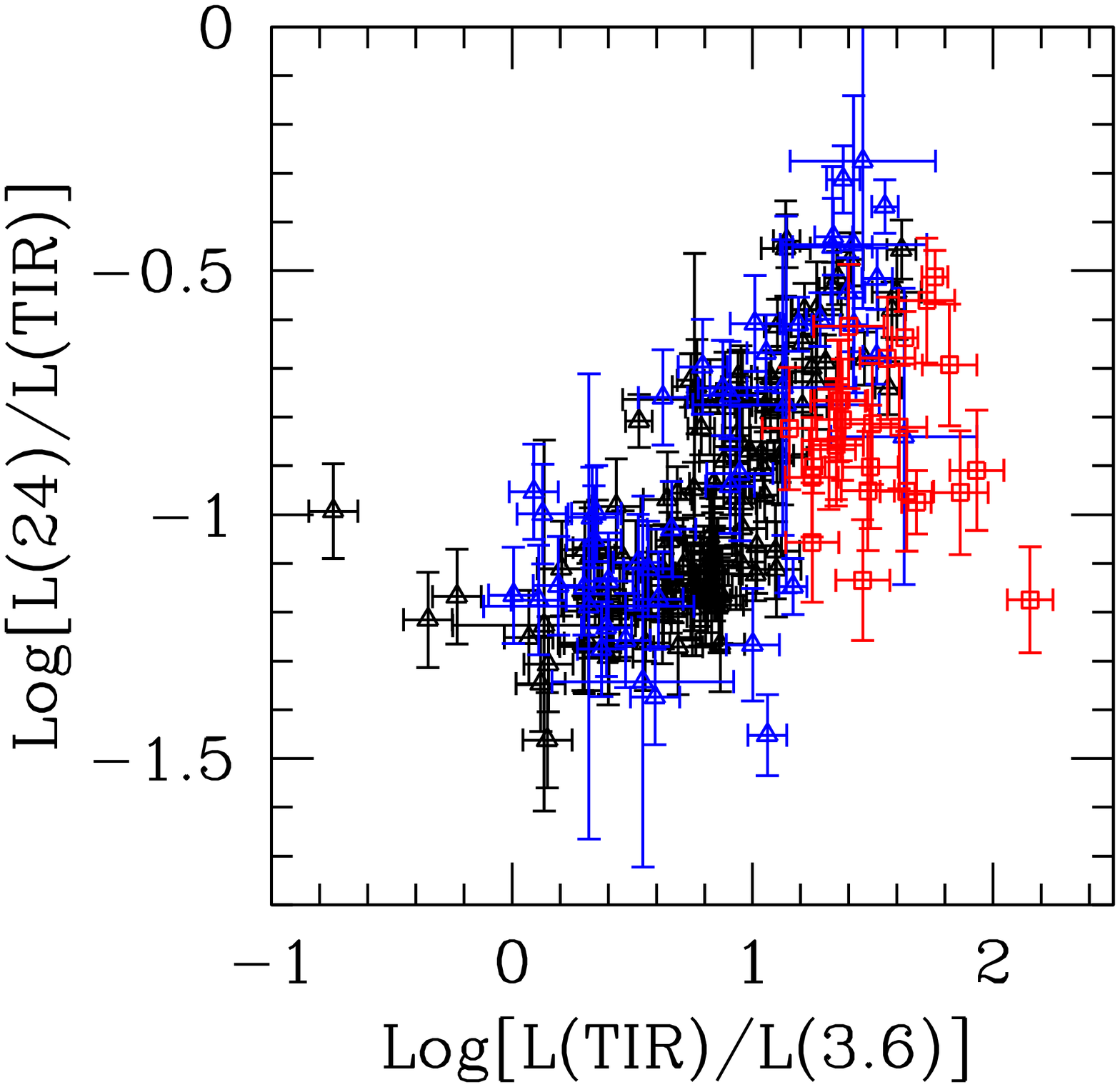}
%\plottwo{figure2a.eps}{figure2b.eps}
\caption{The L(24)/L(TIR) ratio as a function of the IRSD (left) and of the 
infrared luminosity per unit stellar luminosity at 3.6~$\mu$m (a proxy for mass in stars, right) for 
the galaxies in our sample. The color scheme for the datapoints is as in Figure~\ref{fig1}.
%Different colors separate high--metallicity star--forming and starburst 
%galaxies (black) from the low--metallicity ones (blue) from the LIRGs (in red). 
\label{fig3}}
\end{figure}

In order to 
investigate the nature of the L(24) overluminosity, we plot the ratio L(24)/L(TIR) as a function of the infrared luminosity surface density (Figure~\ref{fig3}, left). A general increase of the 24~$\mu$m luminosity as a 
function of IRSD is observed for normal and starburst galaxies, both at low and 
high metallicities; however, a significant break in the trend is observed for the LIRGs in our sample.  
These galaxies show significantly `cooler' infrared SEDs (lower L(24)/L(TIR), by roughly 0.4--0.5~dex) 
than one would infer  from their luminosity surface density if they followed the same trend as 
normal star forming and starburst galaxies. Low 
metallicity galaxies mark a trend similar to that of the high metallicity ones, but at lower average 
IRSD, as expected if the low metallicity galaxies have lower dust content and lower global 
infrared emission at fixed L(24)/L(TIR) \citep{Cannon2005,Cannon2006a,Cannon2006b,Walter2007}. Although 
some of the low metallicity galaxies are hotter than the `hottest' high metallicity galaxies, we do 
not consider this significant enough in our sample, in light of the broad scatter in properties shown 
by our low metallicity galaxies. 

The LIRGs still show a deviation, albeit less extreme than in the previous case,  from the trend of the other star--forming and starburst galaxies when the ratio L(24)/L(TIR) is plotted as a function of L(TIR)/L(3.6), i.e.,  the infrared luminosity per unit stellar mass (Figure~\ref{fig3}, right).  Thus, the observed deviation is not due to a bias in our choice of the emitting region, but is still present when  a different normalization to the infrared luminosity is used. An overestimate of L(TIR) for 
the LIRGs may potentially produce the observed deviation, but we 
would need such an overestimate to be in the range 3--100$\times$, in order  to reconcile the LIRGs with 
the other starburst galaxies in Figure~\ref{fig3}; this is unlikely to be the case,  as our typical uncertainty on L(TIR) is around 25\% (see discussion in section~3).

\begin{figure}
\figurenum{4}
\plottwo{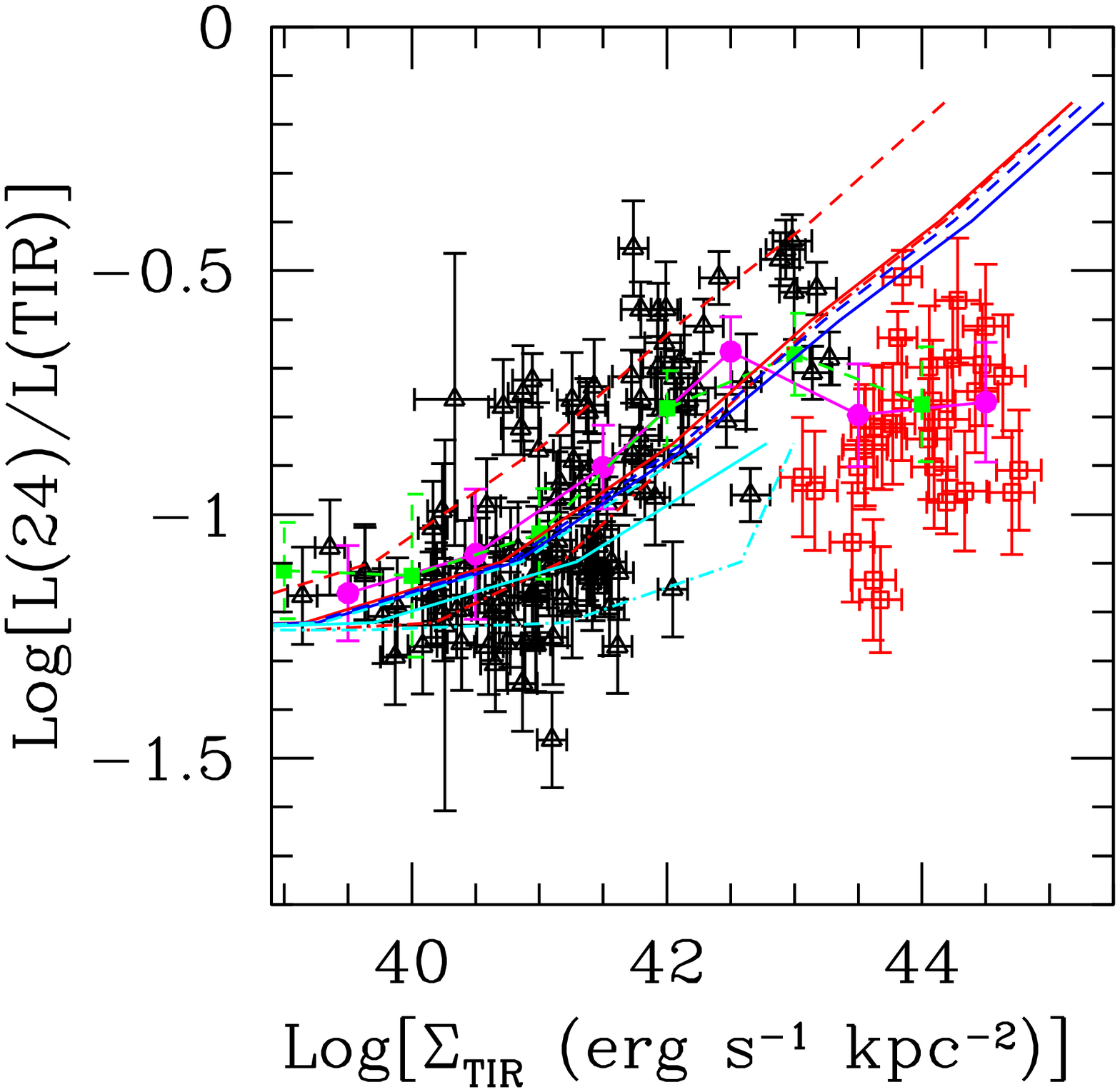}{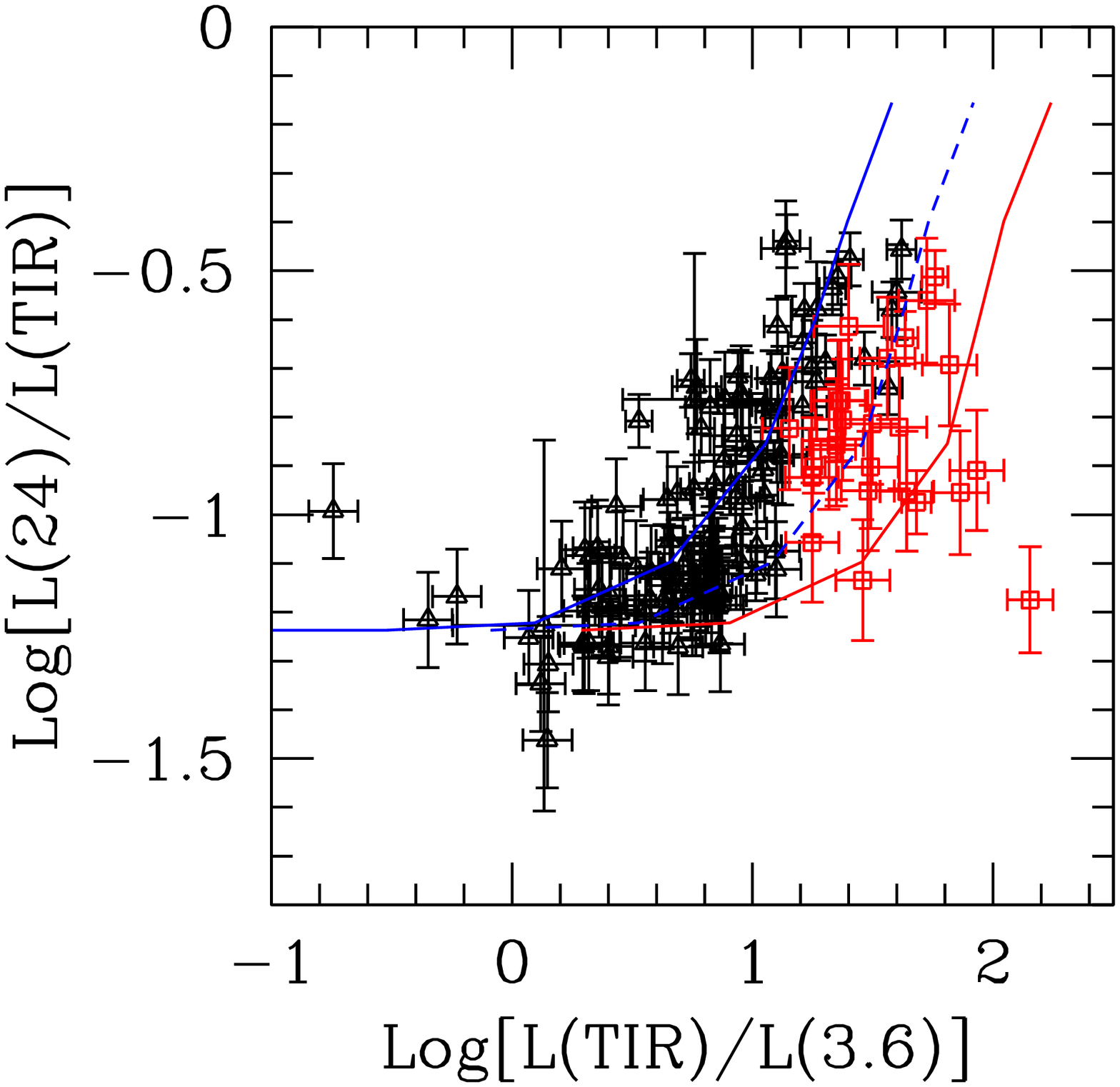}
%\plottwo{figure3.eps}{figure3b.eps}
\caption{The same as Figure~\ref{fig3} for the high--metallicity galaxies, only. (Left) Mean values of the data binned in 1~dex intervals are shown in magenta and in green (with 1~$\sigma$ error bars), corresponding to binning intervals shifted by 0.5 dex between the magenta and green 
points. The expected 
trends for a 100~Myr old, a 1~Gyr,  and a 10~Gyr  constant star formation population 
reddened by increasing amounts of foreground dust for larger SFRD, with a starburst attenuation curve 
 \citep[default dust model, see][]{Calzetti2007}, are marked by the red continuous, the  blue dash, 
 and the blue continuous lines,  respectively.  The case of exponentially decreasing star formation 
with e--folding times of $\tau=$5~Gyr (dashed cyan line) and 2~Gyr (continuous cyan line) are  
shown for the default dust model. Variations on the default dust model are also shown: 
a 100~Myr constant star formation population attenuated by constant, large dust opacity (A$_V>$10~mag) independent of TIR luminosity or SFRD (red dot-dash line); a 100~Myr constant star 
formation population attenuated by our default dust model, but with a 10\% filling factor for the IR--emitting regions within the galaxy (red dashed line); and a $\tau=$2~Gyr exponentially decreasing star formation model, with the stellar population homogeneously mixed with the dust (cyan dot--dash line). 
(Right) Model lines for a 100~Myr old  (red continuous line), a 1~Gyr old (blue dashed line) and a  
10~Gyr old (blue continuous line) constant star formation population are overplotted on the data for 
L(24)/L(TIR) versus the infrared emission per unit stellar luminosity measured at 3.6~$\mu$m.
\label{fig4}}
\end{figure}

We concentrate the rest of the discussion on the high--metallicity sample, to avoid the impact of 
low dust content on the observed IR luminosity.  Figure~\ref{fig4} (left) shows the comparison 
between data and expectations from models for L(24)/L(TIR) as a function of the  infrared 
luminosity surface density. For clarity, in the left panel of Figure~\ref{fig4} we also show the data 
after they are binned in 1~dex intervals along the IRSD axis (for two binning schemes shifted by 
0.5~dex relative to each other).  
The models are the same as described in \citet{Calzetti2007}, which we summarize briefly here. Stellar populations  SEDs are from the Starburst99 models, for a stellar initial mass function (IMF) given by a broken power law as described in  \citet{Kroupa2001} \citep[see, also,][]{Chabrier2003}. The dust extinction and geometry are modeled with the empirical prescription of \citet{Calzetti2001}, which is accurate for starburst galaxies, but will lose accuracy both at the low end and high end of the SFSD range \citep{Calzetti1994,Meurer1999,Calzetti2000,Buat2002,Goldader2002,Bell2003,Buat2005,Calzetti2005,Seibert2005}. However, we expect, to first approximation, that this simple prescription will at least guide in evaluating trends. The infrared SED is modeled according to the prescription of \citet{DraineLi2007}, according to which the fraction of infrared light emerging in each band is a function of the starlight intensity. The starlight intensity in the \citet{DraineLi2007} models is related to the SFSD of  
the stellar population models: larger SFSDs correspond to higher starlight intensity, which in turn produces warmer infrared  emission, i.e., higher L(24)/L(TIR) \citep{Calzetti2007}.  

\citet{Calzetti2007} model sub--kpc regions in galaxies, thus comparisons with 
whole galaxies need to be performed with some care.  The fiducial models from \citet{Calzetti2007},  
adopt as default a 100~Myr constant star formation population attenuated by the starburst  curve 
of \citet{Calzetti2001}. This requires differential extinction between stars and 
ionized gas, with E(B$-$V)$_{star}\sim$0.44~E(B$-$V)$_{gas}$; furthermore, the models relate 
the intensity of the SFSD to the opacity of the medium, in the sense that 
more actively star--forming systems also tend to be more dust--obscured and stronger infrared 
emitters \citep{Wang1996,Heckman1998,Hopkins2001,Calzetti2001}.  While this may be true 
on average, whole galaxies may encompass a wider range of dust emission characteristics. For 
instance, in sub--kpc regions or HII regions larger SFRs (and higher IR emission) imply higher stellar 
field intensities, which produce hotter infrared SEDs \citep{DraineLi2007}. In a galaxy, higher  IR emission  can also be accomplished with larger filling factors of the infrared emitting regions; 
conversely, lower IRSDs do not necessarily need to be accompanied by an increased transparency 
of the interstellar medium. Whole galaxies offer a wider range of physical scenarios than HII regions 
also in terms of the stellar population mix.  

\begin{figure}
\figurenum{5}
\plottwo{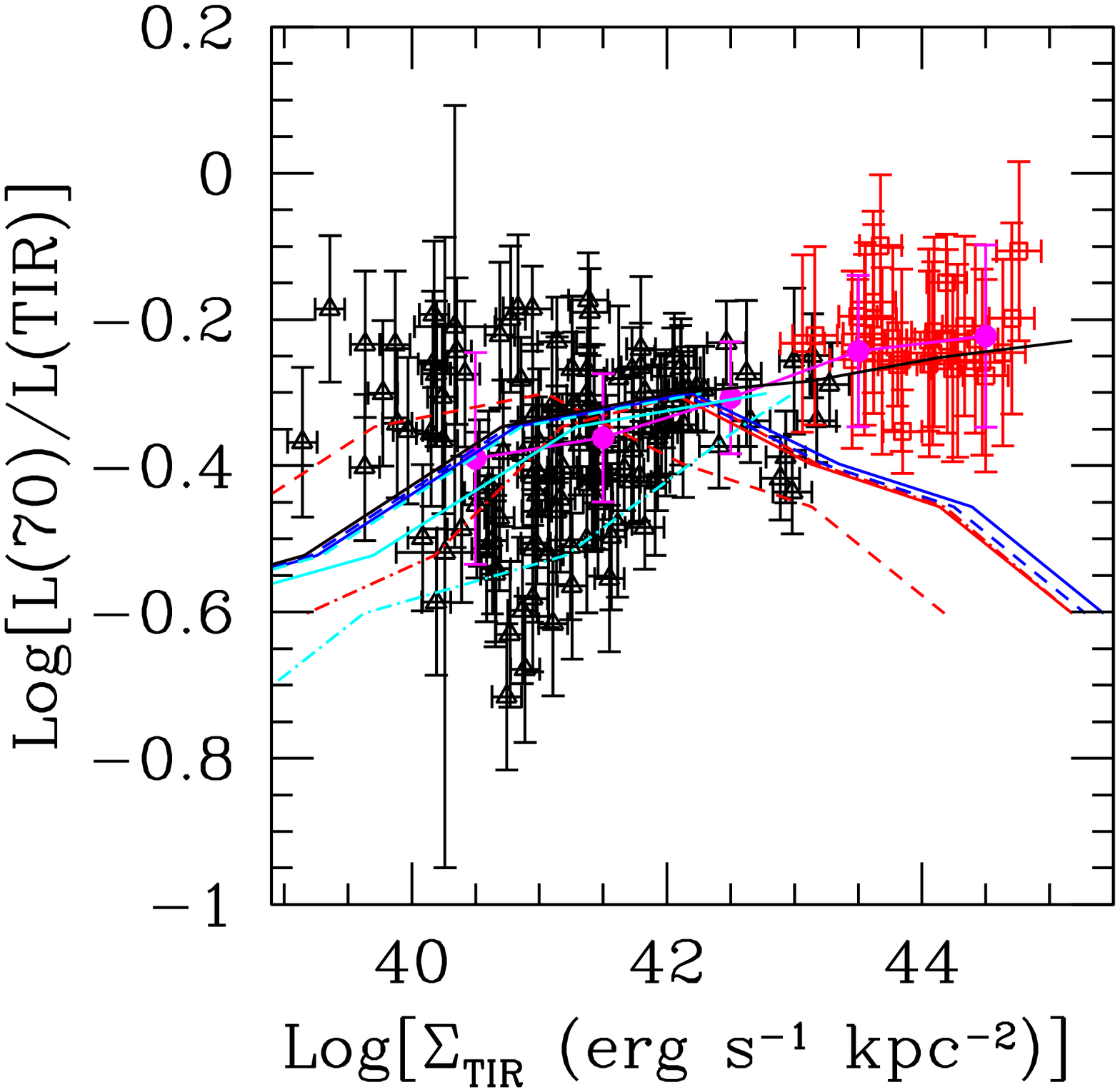}{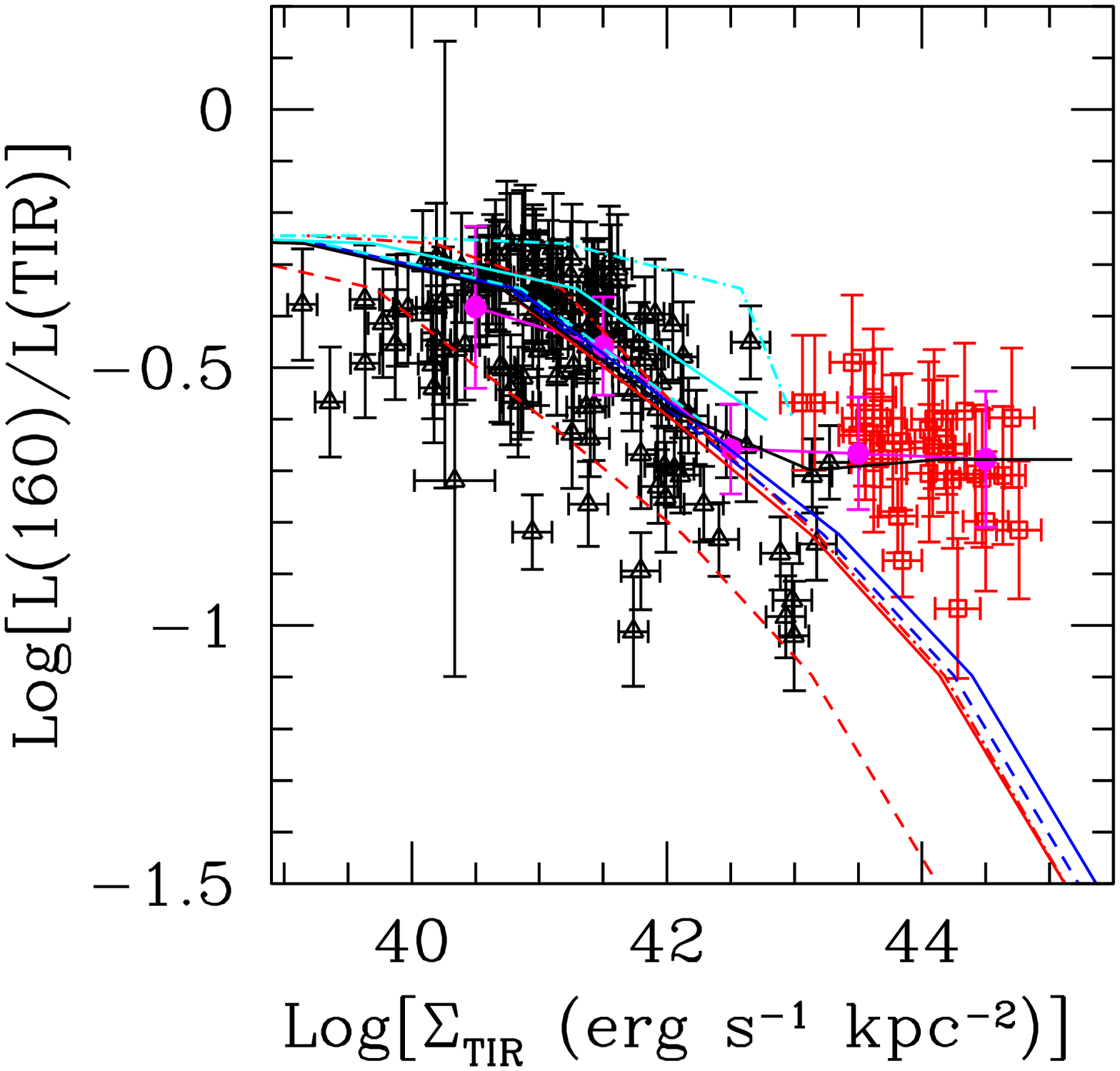}
%\plottwo{figure10.eps}{figure10b.eps}
\caption{The 70~$\mu$m--to--TIR and the 160~$\mu$m--to--TIR ratios as a function of the IRSD for the metal--rich galaxies in our sample. As in Figure~\ref{fig4}, the LIRGs are marked with red symbols. Mean values of the data binned in intervals of 1~dex each are shown in magenta. The model lines are the 
same as in Figure~\ref{fig4} (left).  According to those models, 
the peak of the infrared emission moves {\em in and out} of the 70~$\mu$m band, i.e., the IR SED becomes of sufficiently high effective temperature that the peak IR emission moves to shorter wavelengths than the MIPS 70~$\mu$m band, for the range of luminosity surface densities (LSDs)  spanned by our sample 
 \citep[][see their Figure~15]{DraineLi2007}. For the same models, the 160~$\mu$m emission becomes 
 progressively fainter, as the IRSD increases. 
Modified models for both L(70)/L(TIR) and L(160)/L(TIR) that account for the observed 
properties of the luminous galaxies, and the 
LIRGs in particular, require those luminosity ratios to remain approximately constant beyond IRSD $\Sigma_{TIR}\approx$10$^{42-42.5}$~erg~s$^{-1}$~kpc$^{-2}$ (black line). 
\label{fig5}}
\end{figure}

Indeed, we can span a wider 
range in the parameter space shown in  Figure~\ref{fig4} if we add additional 
simple scenarios to our `default' model. We include, as examples, the cases of: (1) constant  and large (A$_V>$10~mag) 
dust opacity, independent of the SFSD; (2) 10\%  filling factor by area of the SFR 
relative to the total area considered (e.g., disk galaxies, where star formation is concentrated 
along the spiral arms, and there is little star formation in the interarm regions); and (3) a variation 
in the underlying stellar population, where  the fiducial 
100~Myr old constant star formation population of \citet{Calzetti2007} is accompanied by both 
1~Gyr and 10~Gyr constant star formation models and by stellar population models with exponentially 
decreasing star formation and e--folding times of  $\tau=$2~Gyr and $\tau=$5~Gyr. We finally add a 
model in which the decreasing star formation population with 2~Gyr e--folding time is homogeneously 
mixed with dust; here the extinction curve is adopted to be the mean Milky Way one,  with 
stars and ionized gas affected by the same dust column densities \citep{Calzetti2001}, i.e., no 
differential extinction is introduced. These scenarios account for much of the 
observed dispersion of the star--forming and starburst galaxies in Figure~\ref{fig4} 
(left). For instance, galaxies that have  very low L(24)/L(TIR) ratios for IRSD$>$10$^{41}$~erg~s$^{-1}$~kpc$^{-2}$ are generally of morphological type Sab or earlier; their infrared SEDs (see, also, 
Figure~\ref{fig5}) are best described by a model of decreasing star formation with the stellar population 
mixed with the dust (cyan dashed lines in both Figures~\ref{fig4} and \ref{fig5}). In these types of galaxies, in fact, the bolometric emission tends to be dominated by evolved stellar populations; these populations are likely 
responsible for a non--negligible portion of the infrared luminosity and tend to heat the dust to colder temperatures 
than actively star--forming populations \citep[e.g.,][]{Helou1986}.  

The LIRGs  tend to have their infrared emission dominated by the current burst of star formation \citep{Scoville2000}, thus by relatively young stellar populations with ages $\lesssim$100~Myr. Yet, their 
L(24)/L(TIR) ratios  fall outside of the range spanned by the appropriate models of
 Figure~\ref{fig4} (Left). 
This ratio is lower than expected for the measured infrared emission even 
when the latter is normalized to the galaxy's stellar mass (Figure~\ref{fig4}, right). Our models 
(see above) go through the locus occupied by the LIRGs
for constant star forming populations of $\approx$100~Myr--1~Gyr age. However, 
this apparent agreement is mis--leading, since LIRGs have evolved underlying 
galaxies, and the stellar populations contributing to the 3.6~$\mu$m emission need to be at 
least a few Gyr old \citep{Scoville2000}, requiring a shift by at least a factor of 2 along the horizontal 
axis of Figure~\ref{fig4}, right--hand--side panel. 

Supporting evidence that the LIRGs in our sample have cool SEDs, comparable to those of less 
luminous star--forming/starburst galaxies is provided by the dependence of the L(70)/L(TIR) 
and the L(160)/L(TIR) ratios on the IRSD (Figure~\ref{fig5}). The models of \citet{DraineLi2007}, combined with the prescription of \citet{Calzetti2007} to relate the SFSD to the starlight intensity, predict that by the time the IRSD of LIRGs is reached, the peak of the IR SED should have moved out of the MIPS 70~$\mu$m band, towards shorter wavelengths, thus following first an increase in the L(70)/L(TIR) 
ratio from low--luminosities, to a peak around $\Sigma_{TIR}\approx$10$^{42}$erg~s$^{-1}$~kpc$^{-2}$, to be followed by a steady decrease. Similarly, the same models predict a steady decrease of the contribution of the 160~$\mu$m emission to the total far--infrared luminosity as the peak emission moves to 
shorter wavelengths at higher luminosity. The data are, instead, consistent with a roughly constant L(70)/L(TIR) over, at least, 4 orders of magnitude from $\Sigma_{TIR}\sim$10$^{41}$erg~s$^{-1}$~kpc$^{-2}$ to $\sim$10$^{45}$erg~s$^{-1}$~kpc$^{-2}$, and possibly over the full range covered by our data. The trend and dispersion of the data  at low IRSD 
($\Sigma_{TIR}<$10$^{42}$erg~s$^{-1}$~kpc$^{-2}$) can be, at least partially, accounted for by the model scenarios discussed above. The high luminosity trend, on the other hand, cannot be accounted for by any of our simple models. A similar argument can be applied to the L(160)/L(TIR) ratio (Figure~\ref{fig5}, right), which, after a decrease reasonably consistent with 
the models for increasing luminosity up to  $\Sigma_{TIR}\sim$10$^{42.5}$erg~s$^{-1}$~kpc$^{-2}$, flattens 
out for the highest values of the IRSD, indicating the presence of an extra contribution from cool dust 
to the infrared SED. When combined with the findings  for L(24)/L(TIR) (Figure~\ref{fig4}), we  
infer that, for the LIRGs, the energy absorbed by the dust itself in the mid--IR  \citep[self--absorption;][]{Rieke2009} is re--emitted
at longer infrared wavelengths, thus producing the cool IR SEDs observed by \citet{Dunne2001}. For the 
purpose of this work, we will modify the models of \citet{DraineLi2007}, by imposing that L(70)/L(TIR)=constant=0.5 and L(160)/L(TIR)$\sim$constant=0.21 for $\Sigma_{TIR}>$10$^{42.5}$erg~s$^{-1}$~kpc$^{-2}$  (indicated by the black lines in both panels of Figure~\ref{fig5}). 

The overall conclusion from the above discussion is that normal star--forming and starburst galaxies tend to have `hotter' dust, i.e., increasing L(24)/L(TIR) ratios, for increasing IRSD, in 
agreement with the predictions of \citet{DraineLi2007} coupled with a direct correlation between 
SFR and stellar field intensity \citep{Calzetti2007}. However, the trend `flattens' around an 
IRSD$\sim$0.3--1$\times$10$^{43}$~erg~s$^{-1}$~kpc$^{-2}$ (Figure~\ref{fig4}, left).  
This roughly corresponds 
to L(TIR)$\approx$1--3$\times$10$^{44}$~erg~s$^{-1}$$\sim$3--8$\times$10$^{10}$~L$_{\odot}$ or L(24)$\approx$2--6$\times$10$^{43}$~erg~s$^{-1}$$\sim$0.6--2$\times$10$^{10}$~L$_{\odot}$. The conversion between luminosity/area and luminosity should, however, be taken with caution, as it carries a large scatter, almost an order of magnitude.  
Galaxies with IRSDs or luminosities above these values, which in our sample include 
mostly LIRGs, show some evidence for flattening or decreasing L(24)/L(TIR) and
 flattening L(70)/L(TIR) and 
L(160)/L(TIR) values with increasing IRSD, corresponding to `cooler' dust emission  than 
expected from the intensity of the infrared emission \citep{Dunne2000,Dunne2001}. This trend 
is likely a manifestation 
of self-absorption by dust even at 24~$\mu$m \citep[][see the Discussion section for a quantification of this effect]{Rieke2009}.  An inflection in the L(24)/L(TIR) trend above L(TIR)$\sim$10$^{11}$~~L$_{\odot}$ has indeed been previously observed by \citet[][see their Figure~8]{Rieke2009}. As suggested by these authors, presence 
of self--absorption even at 24~$\mu$m in LIRGs will hamper the usefulness of standard SFR indicators based on optical--infrared hydrogen recombination lines at these high luminosity values (see below). 

\subsection{Comparisons Between Calibrations}

Several calibrations of SFRs using the 24~$\mu$m emission of galaxies (or star--forming 
regions in galaxies) have been recently published, using both simple proportionality between 
SFR and L(24)  and non--linear relations. The calibrations are based on a variety 
of galaxy and/or region samples, and stellar IMFs. For sake of 
comparison among the different calibrations, we report below most of the published ones, all of them 
converted to a common IMF and luminosity scale. In this section, we will use luminosity, rather than our preferred (because distance-- and size--independent) luminosity/area, since most published calibrations are derived from luminosities. We choose the Kroupa \citep{Kroupa2001} 
IMF, which has a slope $-2.3$ for stellar masses in the range 0.5--100~M$_{\sun}$ and $-1.3$ for 
stellar masses in the range 0.1--0.5~M$_{\sun}$. For this IMF, Starburst99 stellar population models with solar metallicity and constant SFR give the following relation between number of ionizing photons and SFR:
\begin{equation}
SFR (M_{\sun}~yr^{-1})= 7.41\times 10^{-54} N_{ion} (s^{-1}),
\end{equation}
which corresponds to\footnote{The calibration coefficient in equation~5 is about 3\% higher than what reported in \citet[][their equation~6]{Calzetti2007}. The present calibration is based on a $>$1~Gyr age constant star formation stellar population, while the calibration in \citet{Calzetti2007} is based on a 100~Myr age constant star formation population.  This difference is, however, smaller than the $\sim$12\% uncertainty to be expected for variations in the physical conditions of the emitting regions, i.e., variations in electron temperature between 5,000~K and 15,000~K, consistent with the variations in metallicity typical of our galaxy sample \citep{Osterbrock2006}.} : 
\begin{equation}
SFR (H\alpha) (M_{\sun}~yr^{-1}) = 5.45\times 10^{-42} L(H\alpha) (erg~s^{-1}), 
\end{equation}
where the H$\alpha$ luminosity  is extinction--corrected. For reference, adopting a \citet{Salpeter1955} IMF in the stellar mass range 0.1--100~M$_{\sun}$ would increase the calibration coefficient in equation~5 by a factor 1.51. 

Linear relations between SFR and L(24) have been published by \citet{Wu2005,Zhu2008,Rieke2009}. Reported on a common scale, the three relations\footnote{The luminosity range of applicability for the calibrations by \citet{Wu2005} and \citet{Zhu2008} is not  explicitly provided by the authors. We report ranges as derived from the luminosity range of the main samples from those authors.} are:
\begin{equation}
SFR(24)_{W05,l}= 2.75\times 10^{-43} L(24)\ \ \ \ \ \ \ \ \ \ \ \ \ \ \ \ \  \ \ \ \ \ \ \ \ \ \ 1\times10^{42}\lesssim L(24)\lesssim 1\times 10^{44}, 
\end{equation}
\begin{equation}
SFR(24)_{Z08,l}= 2.46\times 10^{-43} L(24)\ \ \ \ \ \ \ \ \ \ \ \ \ \ \ \ \  \ \ \ \ \ \ \ \ \ \ 4\times10^{41}\lesssim L(24)\lesssim 2\times 10^{44},
\end{equation}
\begin{eqnarray}
SFR(24)_{R09}&=& 2.04\times 10^{-43} L(24),\ \ \ \ \ \ \ \ \ \ \ \ \ \ \ \ \  \ \ \ \ \ \ \ \ \ \ 4\times10^{42}\le L(24)\le 5\times 10^{43},\nonumber \\
                                    &=& 2.04\times 10^{-43} L(24) \times [2.03\times 10^{-44} L(24)]^{0.048}\ \ \ \ \ \ \ L(24)>5\times10^{43},
\end{eqnarray}
where the  24~$\mu$m luminosities are 
in units of erg~s$^{-1}$. The linear \citep[with a small non--linear correction at the high 
luminosities for the derivation of][]{Rieke2009} relations are averages of the calibrations derived in 
each paper. The galaxy samples and the reference calibrators are different in the three papers: \citet{Wu2005} use both radio and the extinction--corrected H$\alpha$ luminosity of the galaxies in common between the SDSS and the Spitzer First Look Survey sample; \citet{Zhu2008} use extinction--corrected UV and 
H$\alpha$ luminosities and FIR emission of the galaxies in the SDSS and the Spitzer SWIRE galaxy sample \citep{Lonsdale2003}; \citet{Rieke2009} use both P$\alpha$ and FIR emission on templates of LIRGs and ULIRGs with data of various origins. Despite these differences, the three calibrations are remarkably close, within $\sim$10\%--35\% of each other, consistent with the calibration uncertainities of 0.1--0.15~dex quoted by the authors. 

Linear calibrations are derived under the assumption that the 24~$\mu$m luminosity increases 
proportionally to the SFR. The alternative scenario discussed in the previous section predicts the 
24~$\mu$m luminosity to increase proportionally faster than the SFR, as a result of the increasing 
mean dust temperature, and corresponding shift to bluer wavelengths of the infrared SED. This 
trend may simply imply, for whole galaxies, a progressive shift for the dominant population heating 
the dust from evolved stars to young stellar populations; this argument is expected to hold on average, rather than on a galaxy--by--galaxy basis. As seen in section~4.1, the effective dust temperature appears to reach a plateau as a function of increasing TIR luminosity around L(24)$\sim$a~few~$\times10^{43}$~erg~s$^{-1}$.
 
\begin{figure}
\figurenum{6}
\plottwo{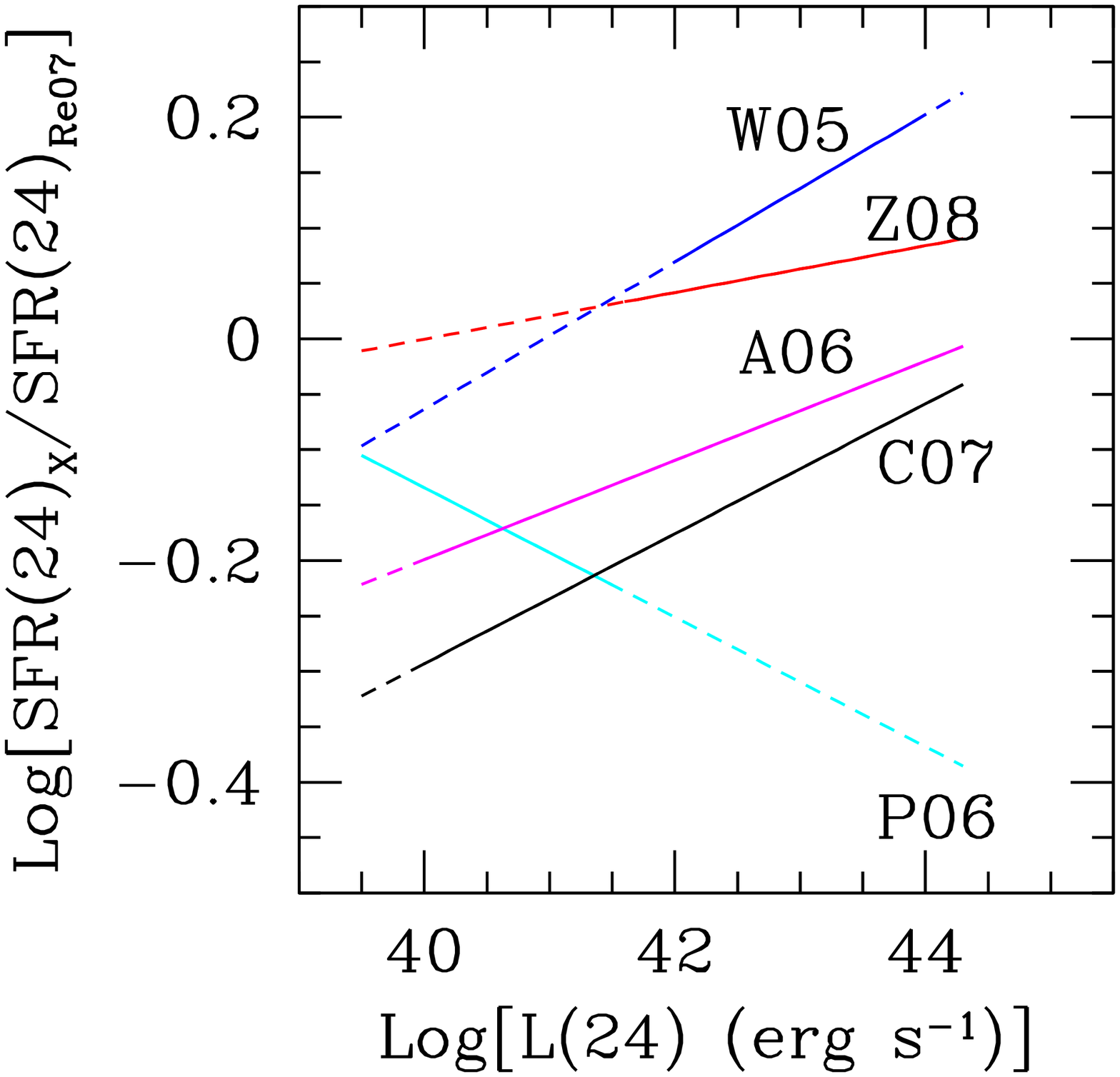}{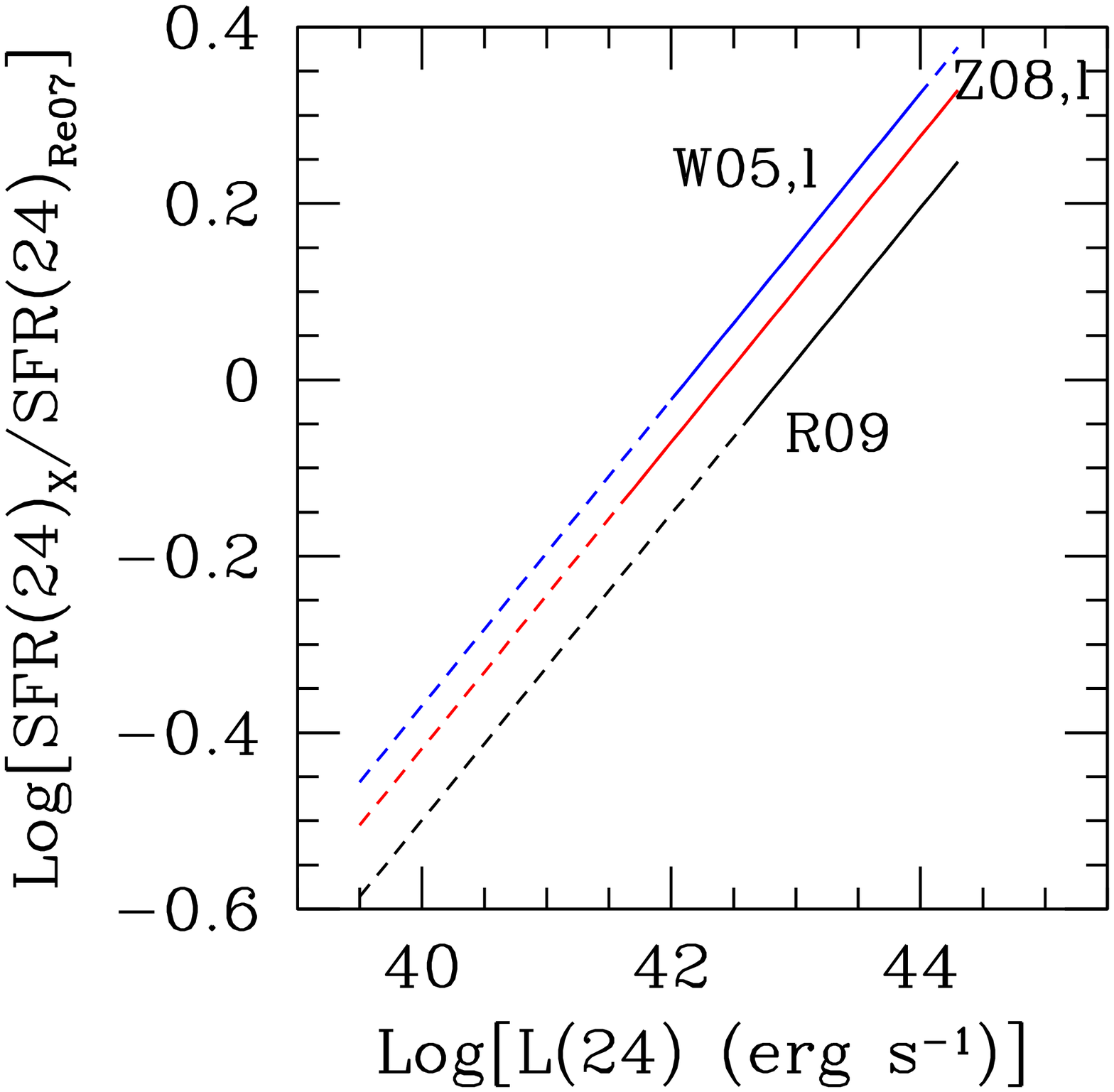}
%\plottwo{figure5.eps}{figure5b.eps}
\caption{The ratio of SFR calibrations using the 24~$\mu$m luminosity, as a function of the observed L(24) (in erg~s$^{-1}$), for the luminosity range spanned by our high--metallicity galaxies sample (excluding LIRGs), 39.5$\lesssim$Log[L(24)]$\lesssim$44.3. {\bf (Left).} The ratio of the non--linear calibrations of \citet[][W05, blue line]{Wu2005}, \citet[][A06, magenta line]{AlonsoHerrero2006}, \citet[][P06, cyan line]{PerezGonzalez2006}, \citet[][C07, black line]{Calzetti2007}, \citet[][Z08, red line]{Zhu2008} to the non--linear calibration of \citet{Rellano2007}. The continuous lines show the luminosity range used by the authors to derive their calibrations, while the dashed lines extend those calibrations to the entire luminosity range spanned by our high--metallicity sample. The non--linear calibration of \citet{Rellano2007} is used as reference because it has been derived for the entire luminosity range displayed in our Figure (see equation~14). {\bf (Right).} The ratio of the linear calibrations of \citet[][Wu05,l, blue line]{Wu2005},  \citet[][Z08,l, red line]{Zhu2008}, and \citet[][R09, black line]{Rieke2009} to the non--linear calibration of \citet{Rellano2007}. As in the left panel, continuous and dashed lines indicate range of 
derivation and extrapolation of the calibrations, respectively.
\label{fig6}}
\end{figure}

Direct fits of L(24) as a function of a variety of reference SFR indicators do show such non--linear 
behavior \citep{Wu2005,AlonsoHerrero2006,PerezGonzalez2006,Calzetti2007,Zhu2008}, and 
 the relation between 
SFR and L(24) requires the mediation of a power law with less--than--unity values. Two of the 
published calibrations involving whole galaxies are from \citet{Wu2005,Zhu2008}, which expressed 
in terms of our default IMF become:
\begin{equation}
SFR(24)_{W05}= 1.07\times 10^{-38} L(24)^{0.893}\ \ \ \ \ \ \ \ \ \ \ \ \ \ \ \ \  \ \ \ \ \ \ \ \ \ \ 1\times10^{42}\lesssim L(24)\lesssim 1\times 10^{44},
\end{equation}
\begin{equation}
SFR(24)_{Z08}= 8.10\times 10^{-37} L(24)^{0.848}\ \ \ \ \ \ \ \ \ \ \ \ \ \ \ \ \  \ \ \ \ \ \ \ \ \ \ 4\times10^{41}\lesssim L(24)\lesssim 2\times 10^{44},
\end{equation}
where the units are the same as the previous linear equations. Calibrations involving either HII--emitting regions in galaxies \citep{PerezGonzalez2006,Calzetti2007,Rellano2007} or combinations of HII 
regions and LIRGs \citep{AlonsoHerrero2006} (called `HII-regions' calibrations from now 
on) are given by\footnote{As in the case of  \citet{Wu2005} and \citet{Zhu2008}, most authors do not explicitly provide the luminosity range of validity of their SFR calibrations. As before, we report limits derived from the luminosity range of the main samples from each paper.} :
\begin{equation}
SFR(24)_{P06}= 9.01\times 10^{-34} L(24)^{0.768}\ \ \ \ \ \ \ \ \ \ \ \ \ \ \ \ \  \ \ \ \ \ \ \ \ \ \ 1\times10^{38}\lesssim L(24)\lesssim 3\times 10^{41},
\end{equation}
\begin{equation}
SFR(24)_{A06}= 5.83\times 10^{-38} L(24)^{0.871}\ \ \ \ \ \ \ \ \ \ \ \ \ \ \ \ \  \ \ \ \ \ \ \ \ \ \ 1\times10^{40}\lesssim L(24)\lesssim 3\times 10^{44},
\end{equation}
\begin{equation}
SFR(24)_{C07}= 1.31\times 10^{-38} L(24)^{0.885}\ \ \ \ \ \ \ \ \ \ \ \ \ \ \ \ \  \ \ \ \ \ \ \ \ \ \ 1\times10^{40}\lesssim L(24)\lesssim 3\times 10^{44},
\end{equation}
\begin{equation}
SFR(24)_{Re07}= 5.66\times 10^{-36} L(24)^{0.826}\ \ \ \ \ \ \ \ \ \ \ \ \ \ \ \ \  \ \ \ \ \ \ \ \ \ \ 1\times10^{38}\lesssim L(24)\lesssim 3\times 10^{44}.
\end{equation}
Typical uncertainties for the calibrations of equations 9--14 are 0.01--0.02 in the exponent \citep[$\sim$0.06 for][]{Wu2005}) and 6\%--15\% in the calibration constant. Within these uncertainties, and within the typical scatter of 0.2--0.3~dex of galaxy and HII regions samples, all calibrations mostly agree with each other,  when limited to the luminosity ranges where they were derived. Deviations from one to the other are around a factor $\sim$2 (Figure~\ref{fig6}, left panel). However, customary use does not carry the uncertainties for the SFR calibrations; this can produce larger discrepancies than a factor $\sim$2 when the calibrations are  extrapolated beyond their range of validity,  as exemplified by the left panel of Figure~\ref{fig6}. 

Galaxy--wide SFR(24) non--linear calibrations \citep{Wu2005,Zhu2008} are marginally, 
25\%--60\%, albeit systematically, higher than HII regions SFR(24) calibrations at the high 
luminosity end (Figure~\ref{fig6}, left panel). The same is observed for galaxy--wide linear calibrations (equations~6--8, and Figure~\ref{fig6}, right panel), with deviations around a factor $\sim$2 relative to HII regions non--linear calibrations, which is, again, systematic, albeit marginal within the uncertainties. In order to account for this difference, we can speculate that, in whole galaxies, the contrast between the contributions of recent star formation and old stellar populations in the heating of dust is lower than in their HII regions. This possibly leads to a lower average dust temperature (and lower, on average, L(24)/L(TIR)) and, consequently, to a higher calibration constant for SFR(24). However, this speculation cannot be easily verified at this time; it will have to await for observations with the Herschel Space Telescope, in order to obtain long infrared wavelength images of sufficient high angular resolution to isolate HII--emission--dominated regions in external galaxies. 

The comparison of the linear SFR(24) calibrations (equations~6--8, all of them for whole galaxies) with the non--linear ones \citep[Figure~\ref{fig6}, right, compared with the non--linear calibration of][]{Rellano2007} shows, again, general agreement within the existing uncertainties,  when considered in the luminosity range for which those calibrations were derived. However, extending the linear calibrations to the lowest luminosity values of our high--metallicity sample 
show significant discrepancies from the non--linear calibrations; specifically, the linear calibrations predict SFRs that are a factor $\sim$2--4 lower than predicted by the non--linear calibrations. This is expected in light of the fact that galaxies become increasingly transparent for decreasing infrared luminosity, and a larger fraction of the light from recent star formation emerges directly 
at UV and optical wavelengths. Thus, the linear SFR(24) calibrations will  underestimate   
the true SFR below L(24)$\approx$3$\times$10$^{42}$~erg~s$^{-1}$ (Figure~\ref{fig6}, right) . 

Non--linear calibrations may provide a better alternative for extending the SFR(24) to low luminosity values, because their functional form accounts for two physical effects: increasing 
transparency of the medium and lower effective dust temperature towards lower SFSDs \citep{DraineLi2007}. However, they are also potentially problematic. The first problem is that for sufficiently low dust content, even the non--linear SFR(24) will end up under--estimating the true SFR \citep{Calzetti2007}. The second problem is that  low global SFRs can be obtained both for less active, metal rich (but gas--poor) galaxies and for active, low--metallicity, low--luminosity galaxies. In the latter case, as the metal content of the system decreases, the 
 effective dust temperature increases \citep[e.g.][and Figure~\ref{fig4}]{Hunter1989,Walter2007,Engelbracht2008}, thus complicating any trend. A third problem is that 
 non--linear calibrations between SFR and luminosity are `object--dependent', in the sense that their validity relies on the presence of the 
physical conditions that lead to the non--linear relation itself. Thus, calibrations obtained for 
whole galaxies will in general not be applicable to regions within those galaxies, and viceversa. 

\begin{figure}
\figurenum{7}
\plotone{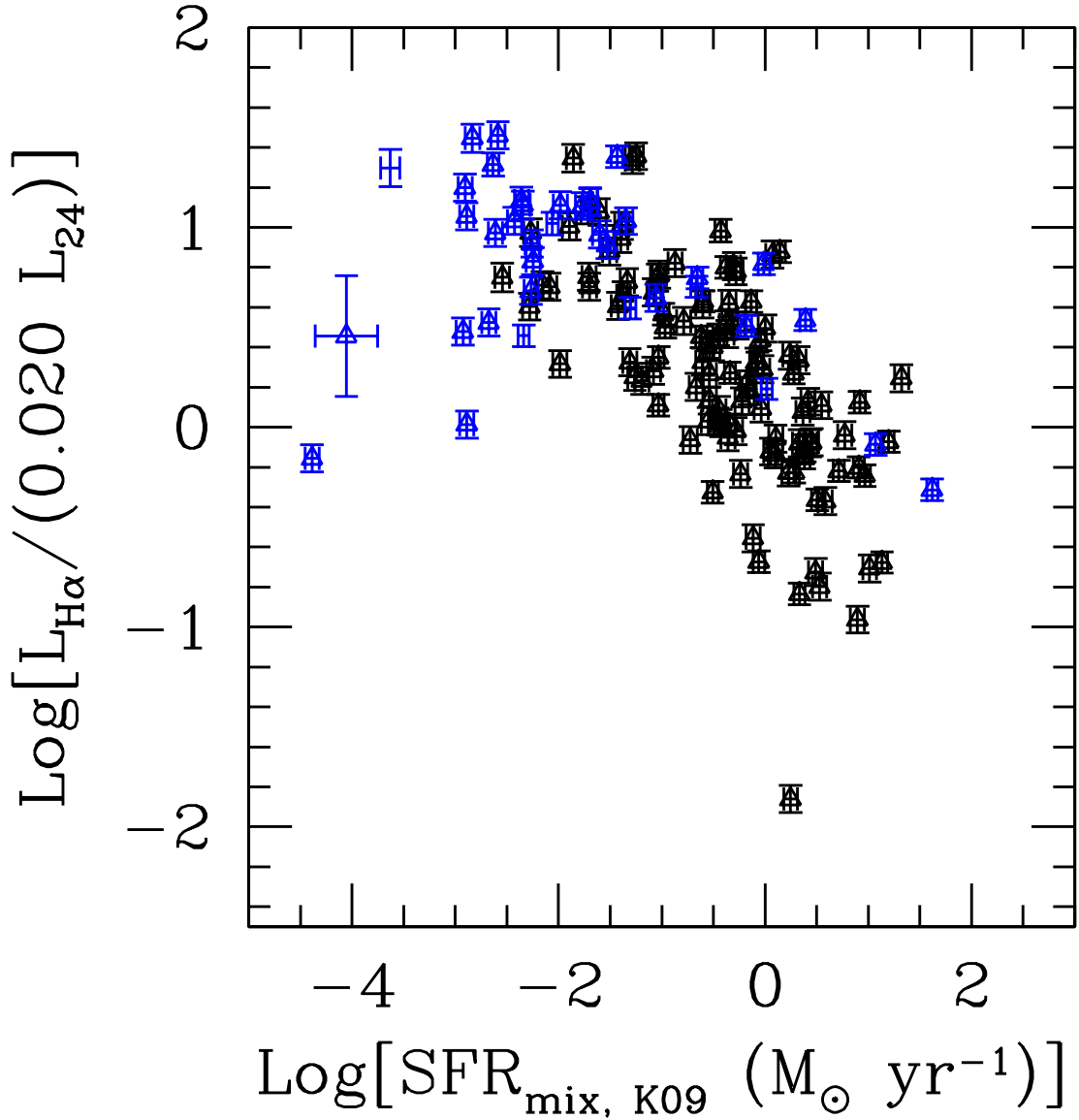}
%\plotone{figure4.eps}
\caption{The ratio between the H$\alpha$ luminosity and the scaled 24~$\mu$m luminosity 
for the galaxies in our sample (excluding LIRGs) as a function of SFR (equation~15). Blue 
points mark low--metallicity galaxies, and 
black points the high--metallicity ones. The scaling factor for the 24~$\mu$m luminosity used 
on the vertical axis of this figure is from equation~15 \citep{Kennicutt2009}. This luminosity ratio 
measures the fraction of the star formation that emerges from the galaxy unattenuated by dust. Not 
surprisingly, there is a  correlation between this luminosity ratio and SFR, in the sense that 
more actively star--forming galaxies are fainter in the optical relative to the  infrared emission \citep[see, 
also,][]{Calzetti2007}. In our normal star--forming and starburst sample, most of the 
galaxies tend  to be relatively transparent, meaning that a large fraction of the light from 
recent star formation emerges unabsorbed by dust. 
\label{fig7}}
\end{figure}

Indeed, the decreasing dust opacity of galaxies for decreasing SFR \citep{Wang1996,Calzetti2007} calls 
into question the usefulness of an infrared--based SFR indicator at the low luminosity end of 
our sample, especially in the low--metallicity regime. For many of the galaxies in our sample 
star formation appears to `shine' unimpeded by dust absorption. This is evidenced in Figure~\ref{fig7}, where the ratio of H$\alpha$ to the 24~$\mu$m luminosity \citep[rescaled according to the proportionality factor of][between the two luminosities]{Kennicutt2009} is plotted as a function of the SFR as defined in \citet{Kennicutt2009} for the normal star forming and starburst galaxies in our sample. For most of these galaxies, the amount of  star formation emerging at optical wavelengths (as traced by H$\alpha$) is larger than the star formation emerging in the infrared (as traced by the 24~$\mu$m emission). For this reason, composite SFR indicators, 
involving the combination of an optical and an infrared tracer of SFR, have been proposed \citep{Calzetti2007,Kennicutt2007,Zhu2008,Kennicutt2009}. For whole galaxies, \citet{Kennicutt2009} propose \citep[see, also][]{Zhu2008}:
\begin{equation}
SFR_{mix,K09} = 5.45\times 10^{-42} [L(H\alpha)_{obs} + 0.020 L(24)] \ \ \  \ \ \ \ \ \ \  \ \ \ \ \ \ \ \ 3\times10^{38}\lesssim L(24)\lesssim 3\times 10^{44},
\end{equation}
where the calibration is expressed in terms of our selected IMF. For HII--dominated regions and starburst galaxies, \citet{Calzetti2007} propose:
\begin{equation}
SFR_{mix,C07} = 5.45\times 10^{-42} [L(H\alpha)_{obs} + 0.031 L(24)] \ \ \ \ \ \ \ \ \  \ \ \ \ \ \ \ \ 3\times10^{38}\lesssim L(24)\lesssim 3\times 10^{44}.
\end{equation}
In both equations, L(H$\alpha$)$_{obs}$ is the observed (non--extinction--corrected) luminosity 
at H$\alpha$. 
The difference between the two equations has been suggested to be due to a difference in the underlying stellar population heating the dust: continuous star formation no older than about 100~Myr for starburst galaxies (and instantaneous bursts 
only a few Myr old for HII~regions) and much longer lived star formation, of--order 10~Gyr, for whole galaxies \citep{Kennicutt2009}. 

\begin{figure}
\figurenum{8}
\plotone{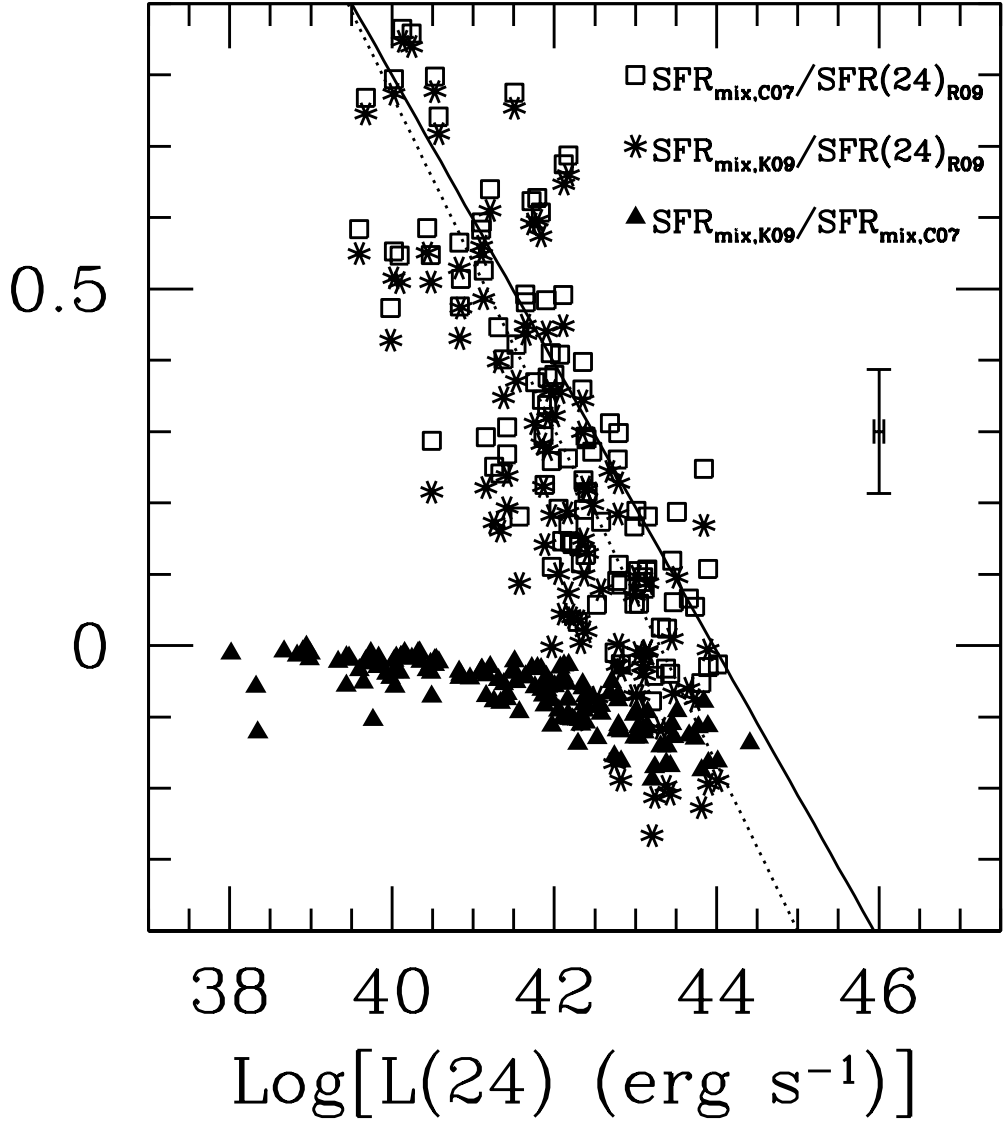}
%\plotone{figure6.eps}
\caption{The logarithm of the ratio of SFR calibrations: SFR$_{mix,K09}$/SFR$_{mix,C07}$ 
(filled triangles, 
equations~15 and 16), SFR$_{mix,K09}$/SFR(24)$_{R09}$ (asterisks, equations~15 and 8), 
and SFR$_{mix,C07}$/SFR(24)$_{R09}$ (empty squares, equations~16 and 8) as a function 
of the 24~$\mu$m luminosity (LIRGs are excluded). Fits to the data from the latter 
two ratios are shown. The calibration SFR(24)$_{R09}$ is extended below its limit of validity, 
Log[L(24)]$\sim$42.6, for illustrative purposes only. A typical 1~$\sigma$ errorbar is shown as well.
\label{fig8}}
\end{figure}

Our sample spans about 6 orders of magnitude in SFRs, from $\sim$10$^{-4}$~M$_{\sun}$~yr$^{-1}$ 
all the way up to LIRGs producing stars at a pace close to a hundred M$_{\odot}$~yr$^{-1}$. The stellar 
populations dominating the dust heating will thus transition from those characteristic of normal 
star forming spirals or irregulars (with timescales for star formation of--order 10~Gyr) to those 
characteristic of young starbursts (timescales of order 100 Myr or less). In particular, we expect 
that as we approach the luminosity of the LIRGs, and consequently high dust opacities in the 
galaxies, the inferred SFRs will need to coincide with those derived from SFR(24). In Figure~\ref{fig8}, 
we compare both equations~15 and 16 among themselves and  with the SFR(24) calibration of 
\citet{Rieke2009}, for the galaxies in our sample (excluding LIRGs). The difference between equations~15 and 16 is typically less than $\sim$15--20\% up to L(24)$\sim$4$\times$10$^{42}$~erg~s$^{-1}$, 
and becomes larger for luminosities above that value. However, this is also approaching the luminosity 
range where SFR$_{mix}$ needs to reach, as asymptotic value, SFR(24), as galaxies become more 
and more dominated by infrared emission. We elect to use equation~15 to derive SFRs for galaxies 
with luminosity L(24)$\lesssim$4$\times$10$^{42}$~erg~s$^{-1}$, and equation~16 for higher luminosity; 
this `transition' luminosity corresponds to SFR$\sim$1~M$_{\sun}$~yr$^{-1}$, which is a transition point where  
SFRs are equally contributed by the H$\alpha$ and 24~$\mu$m luminosities (Figure~\ref{fig7}).  
In our sample, that transition SFR corresponds to $\Sigma_{SFR}\approx$0.05~M$_{\sun}$~yr$^{-1}$~kpc$^{-2}$, albeit with a large uncertainty, since the relation between SFR and $\Sigma_{SFR}$ 
has over one order of magnitude scatter. 
The exact value of the transition point is not crucial, as 
the difference between SFR values (equations~15 and 16) around that point are $\sim$25\% 
($\sim$0.1~dex in logarithmic scale), and 
physical systems are expected to transition from star formation timescales of many Gyrs down to 
a few hundred Myrs in a smoother way than what we describe here. 

Equation~16 is almost identical to the linear part of Equation~8 for L(H$\alpha$)$_{obs}\ll$0.031~L(24). 
The closeness of the two equations becomes even more obvious if we recall that we measure a mean ratio L(H$\alpha$)/L(24) =0.031 for the metal--rich regions/galaxies  in the  sample of \citet{Calzetti2007}, rather than 0.038, as measured by \citet{Rieke2009} in the same luminosity range 10$^7$~L$_{\sun}\le$L(24)$\le$10$^{10}$~L$_{\sun}$. The difference is small, about 20\%. The ratio L(H$\alpha$)/L(24) =0.031, when combined with the median 
ratio L(24)/L(TIR)$\sim$0.16 \citep[see Figure~\ref{fig3}, and also][]{Rieke2009} for the LIRGs, yields a mean value for L(H$\alpha$)/L(TIR)$\sim$0.005, consistent with expectations for a 100~Myr old constant star formation population \citep[from the models of][where we assume L(TIR)$\sim$L(bol), the bolometric luminosity]{Leitherer1999}. It is worth noticing that the ratio L(H$\alpha$)/L(24) =0.038 determined by \citet{Rieke2009} yields values for L(H$\alpha$)/L(TIR)$\sim$0.006, consistent with a 30~Myr old constant star formation population. Both calibrations are, thus, consistent with infrared emission from galaxies whose emission is  dominated by the young stellar population component, as expected for starbursts and LIRGs 
\citep[e.g., ][and references therein]{Scoville2000}.  We use our own measured value  of the L(H$\alpha$)/L(24) 
mean ratio, and we replace it in equation~8, to derive SFRs in the high luminosity range of our sample.

The lowest luminosity LIRG in our sample has L(24)$\sim$4$\times$10$^{43}$~erg~s$^{-1}$, and the relative 
contributions of L(24) and L(H$\alpha$)$_{obs}$ to the SFR now weight in favor of L(24), 
with  L(H$\alpha$)$_{obs}$/(0.031 L(24))$<$0.14.  Thus, for our purposes, we can ignore the 
contribution of L(H$\alpha$)$_{obs}$ to the SFR at the high luminosity end, and we will use the re-normalized equation~8 to 
measure SFRs for the LIRGs in our sample. Of the two expressions in equation~8, we only use the 
non--linear part, in agreement with the prescription of \citet{Rieke2009} for our range of luminosities. 
The non--linear correction in equation~8 attempts to correct for the presence of self--absorption in the LIRGs and brighter galaxies  \citep[][]{Rieke2009}. Derivations of SFR(24) for whole galaxies by other authors do not include such non--linear term although their luminosity range often includes LIRGs (equations~6, 7, 9, 10, and 12); the presence of a non--linear term can remain unnoticed if optical or infrared hydrogen recombination lines (H$\alpha$, P$\alpha$) are used as reference SFR indicators, since, even when extinction corrected, the luminosity  of those  lines will be underestimated in the presence of high dust opacity (as typical of LIRGs or brighter galaxies: L(24)$>$a~few~10$^{43}$~erg~s$^{-1}$).  We note that in our sample the highest luminosity 
LIRG (L(24)$\sim$8.3$\times$10$^{44}$~erg~s$^{-1}$) shows a  difference between the SFRs derived from the re-normalized equation~8 and its linear version, equation~16, by less than 15\%; for 27 of 29 LIRGs, this 
difference is less than 10\%. 

In summary, we derive SFRs for our sample of galaxies as follows:
\begin{eqnarray}
SFR&=& 5.45\times 10^{-42} [L(H\alpha)_{obs} + 0.020 L(24)],\ \ \ \ \ \ \ \ \ \ \ \ \ \ \ \ \ \ L(24)< 4\times 10^{42},\nonumber \\
         &=&5.45\times 10^{-42} [L(H\alpha)_{obs} + 0.031 L(24)], \ \ \ \ \ \ \ \ \ \ \ \ \ \ \ \ \ \  4\times10^{42}\le L(24)< 5\times 10^{43},\nonumber \\
          &=& 1.70\times 10^{-43} L(24) \times [2.03\times 10^{-44} L(24)]^{0.048}\ \ \ \ \ \ \ \ \ \ L(24)\ge 5\times10^{43},
\end{eqnarray}
where the SFR is in units of M$_{\odot}$~yr$^{-1}$, and luminosities are in units of erg~s$^{-1}$. 
The transition regions between the three regimes are not sharp, and 
especially the one between equation~15 and 16 may begin at lower luminosity than the one 
used here. However, differences between the average SFRs at the transition points are about or 
less than 0.1~dex, which is smaller than both our typical error bar  and the spread in the data points.

Equation~17 attempts to capture a general behavior of galaxies in regard to their dust opacity. 
Below L(24)$\sim$4.0$\times10^{42}$~erg~s$^{-1}$ (L(TIR)$\approx$10$^{10}$~L$_{\odot}$), 
the young, star--forming regions in galaxies become increasingly transparent at optical  and UV wavelengths for decreasing infrared luminosity \citep[e.g.,]{Buat2007}, and optical emission lines are useful tools for measuring their dust opacity \citep{Kennicutt2009}. At higher infrared luminosities the star--forming regions become on average increasingly opaque, and, above  
L(24)$\sim$5$\times$10$^{43}$  (L(TIR)$\approx$10$^{11}$~L$_{\odot}$), their infrared 
emission becomes the most reliable 
SFR indicator \citep[e.g.]{Rieke2009}. In between L(TIR)$\approx$10$^{10}$~L$_{\odot}$ and L(TIR)$\approx$10$^{11}$~L$_{\odot}$ is a transition regime, where the SFR traced by optical and infrared line emission (e.g., P$\alpha$) is still non--negligible \citep{Calzetti2007}, although it becomes vanishingly small as the upper luminosity limit is reached. This argument should be interpreted as `mean behavior', as galaxies generally contain large numbers of star--forming regions, with different levels of dust opacity. For instance, many galaxies with L(24)$>$5$\times$10$^{43}$ still show measurable emission at blue wavelengths, specifically H$\beta$($\lambda$4861~\AA)    \citep{Kennicutt2009}. 
 
For the 10 starburst galaxies with P$\alpha$ data, we use this emission line, after correction for dust extinction, to measure the SFRs, adopting a ratio H$\alpha$/P$\alpha$=7.82 for the high--metallicity galaxies and H$\alpha$/P$\alpha$=8.73 for the low--metallicity ones \citep{Calzetti2007}.

\section{Calibration of SFR(70)}

In order to establish whether the 70~$\mu$m luminosity can be used as a SFR indicator, and to check its  limitations, we first concentrate on the 142 galaxies that constitute the high--metallicity sample. 
We will discuss the effect of metallicity on our calibrations later in this section. 

\begin{figure}
\figurenum{9}
\plottwo{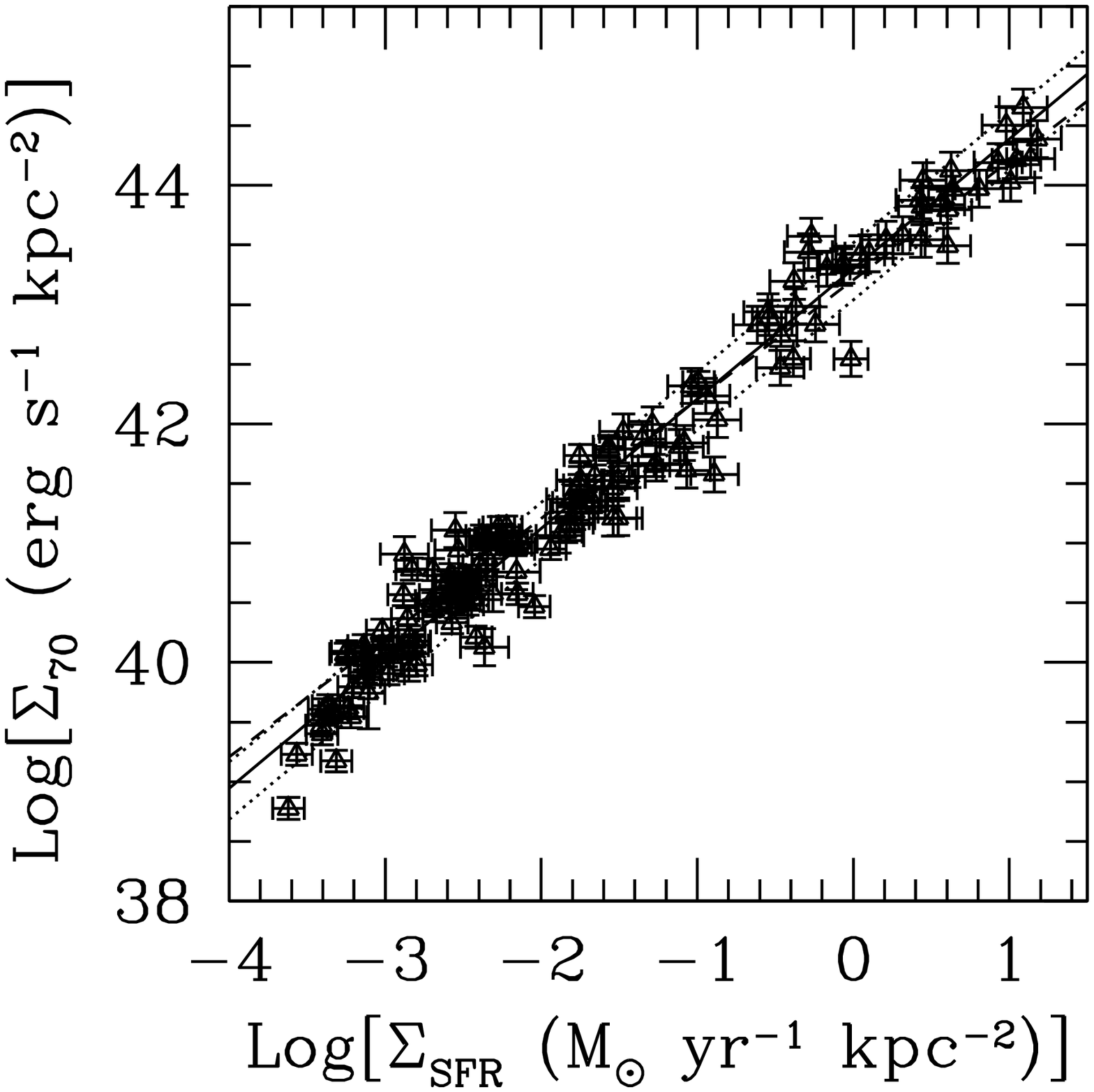}{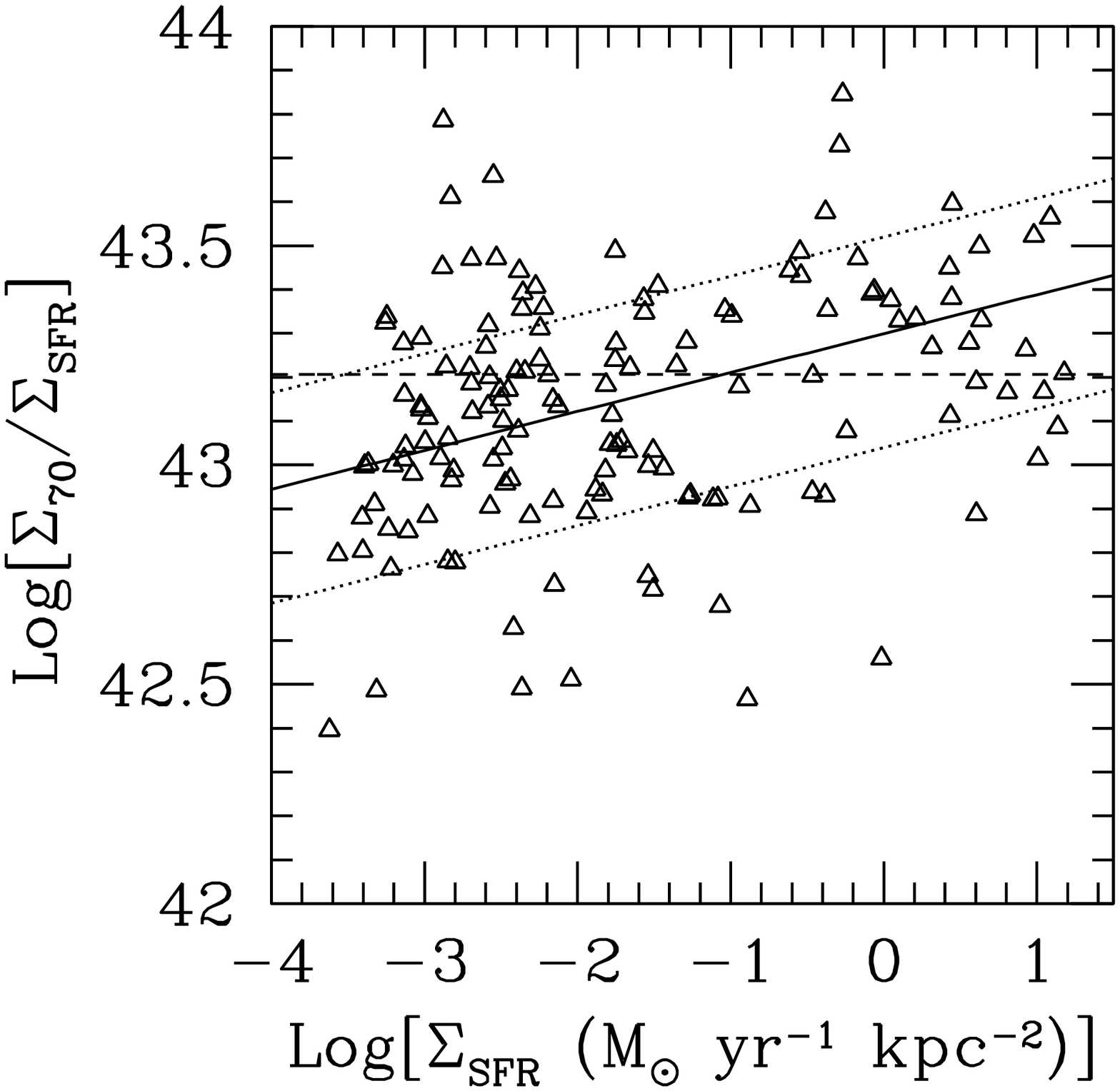}
%\plottwo{figure7a.eps}{figure7b.eps}
\caption{{\bf (Left)} The 70~$\mu$m LSD,  $\Sigma_{70}$=L(70)/area, as a function of the 
SFSD, $\Sigma_{SFR}$, in log-log scale. In both 
cases, the area used is the H$\alpha$ emitting area (Figure~\ref{fig2}).  Only the high--metallicity 
galaxies are reported (142 datapoints, with their 1~$\sigma$ errorbars), together with the best fit (continuous line) and the 1~$\sigma$ envelope to the data (dotted lines, which correspond to the range 
$-$0.26,$+$0.22~dex). The 1~$\sigma$ envelope is calculated as the standard deviation of the 
logarithm of the ratio relative to the mean trend. The best fit with unity slope is shown as a dashed  
line. {\bf (Right)} The same data and best fit lines are shown now for the ratio $\Sigma_{70}$/$\Sigma_{SFR}$ as a function of the 
SFSD. Errorbars are omitted for clarity. This plot better shows the deviation of the datapoints from 
a one--to--one relation.
\label{fig9}}
\end{figure}

The 70~$\mu$m LSD tightly correlates with the SFSD (Figure~\ref{fig9}) 
over the full 5 orders of magnitude spanned by our high--metallicity sample. The best fit through the data points is:
\begin{equation}
Log [\Sigma_{70} (erg~s^{-1}~kpc^{-2})] = (1.089\pm 0.013) Log [\Sigma_{SFR} (M_{\odot}~yr^{-1}~kpc^{-2})] + (43.303\pm0.029),
\end{equation}
with a 1~$\sigma$ dispersion of the data around the best fit line of ($-$0.26,$+$0.22)~dex. 
Although the best fit is non--linear, we need to recall that data with lower SFSD often correspond to galaxies with star formation scattered across the disk. These systems generally tend to be less dust opaque than systems with large and concentrated SFRs \citep[Figures~\ref{fig4} and \ref{fig5}, and, e.g.,][]{Wang1996,Calzetti2007}. Thus, the 70~$\mu$m emission becomes  `underluminous' because the  dust emission does not fully trace the light from star formation.  
%and the infrared SED peaks at longer wavelengths than 70~$\mu$m \citep{DraineLi2007}. 
Furthermore, the peak of the 
infrared emission shifts to longer wavelengths for decreasing IRSD values \citep[Figure~\ref{fig5}, and][]{Draine2007},  exacerbating the trend. If we constrain the fitting range to increasingly large SFSDs, the slope in equation~18 tends asymptotically to 
unity; it is about 2~$\sigma$ away from unity for $Log$($\Sigma_{SFR}$)$> -$1, and less than 1~$\sigma$ away from unity for  $Log$($\Sigma_{SFR}$)$> -$0.6. 

\begin{figure}
\figurenum{10}
\plottwo{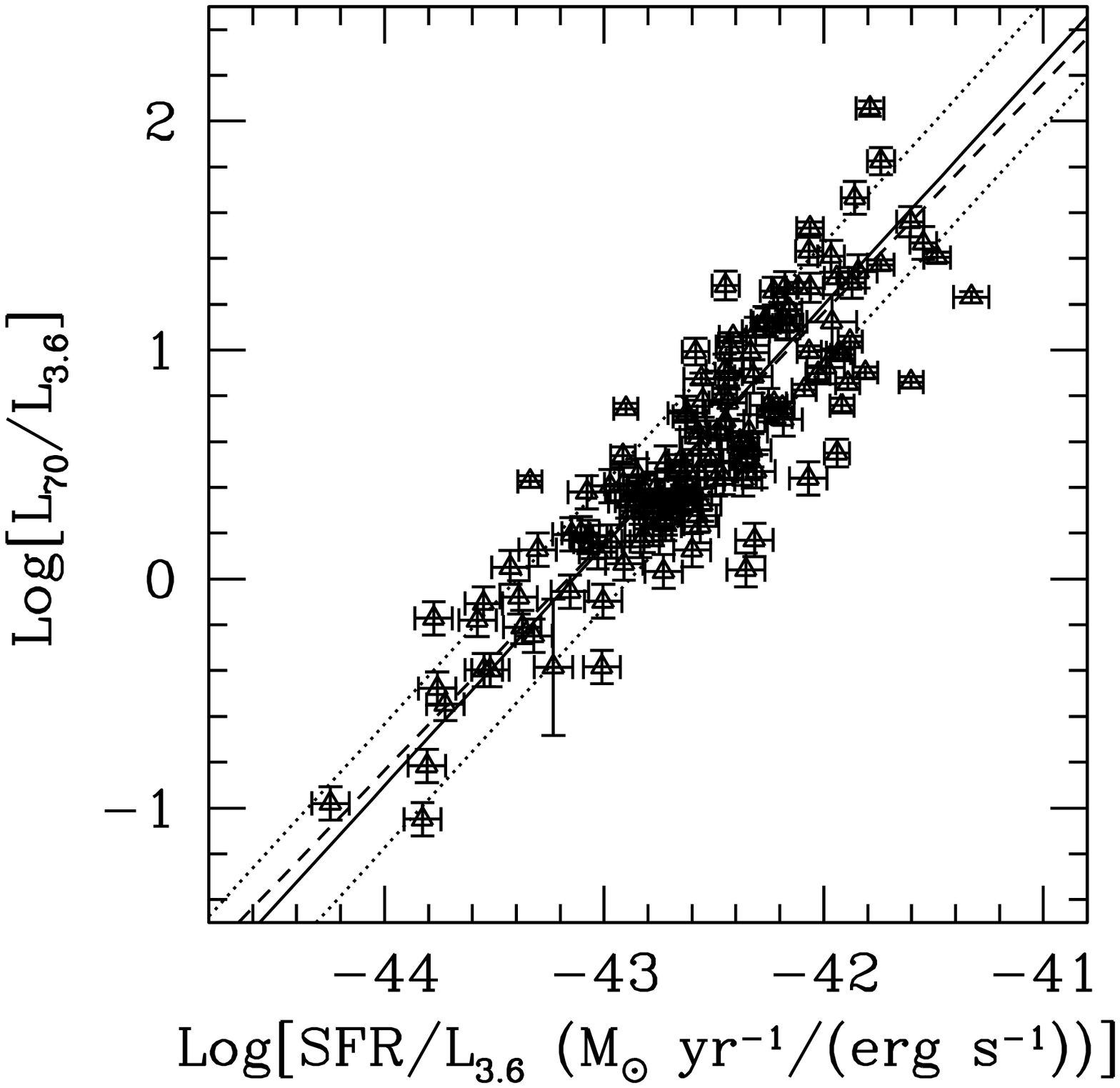}{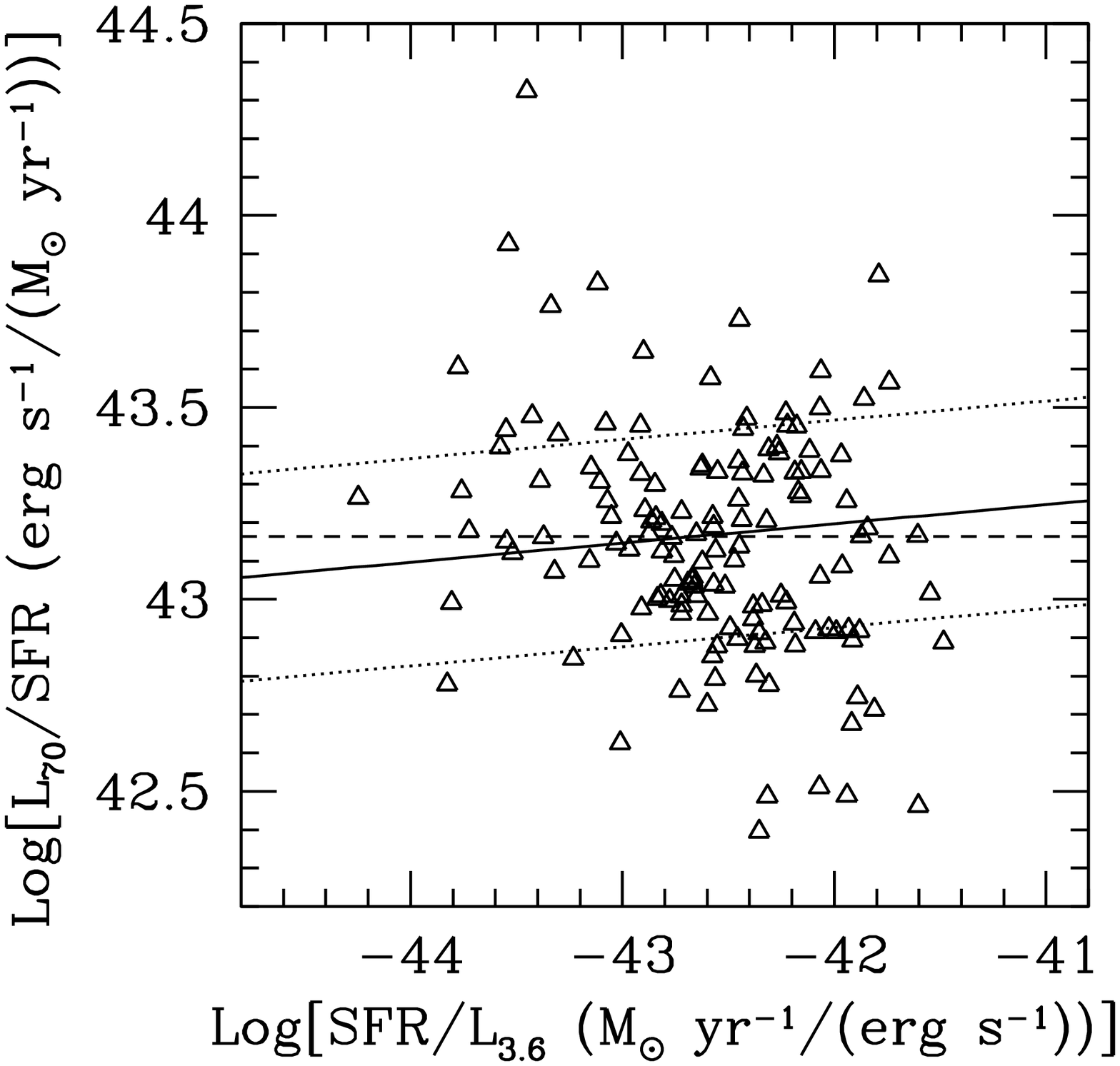}
%\plottwo{figure8a.eps}{figure8b.eps}
\caption{{\bf (left)} and  {\bf (right).} The same as Figure~\ref{fig9}, with the data normalized by the 
3.6~$\mu$m luminosity, instead of the emitting area. The same data are shown, together with the best fit (continuous line) and the 1~$\sigma$ envelope to the data (dotted lines, which correspond to the 
range $-$0.27,$+$0.27~dex). The best fit with unity slope is shown as a dashed line in both panels.
\label{fig10}}
\end{figure}

An independent test is to use the luminosities or SFRs normalized by the 3.6~$\mu$m luminosity, L(3.6) (Figure~\ref{fig10}), since this proxy for stellar mass is not dominated by the current star formation.
% thus it will tend to erase any dependency of the L(70)--SFR relation on the overall galaxy opacity. 
The best fit line through the data is:
 \begin{equation}
Log [L(70)/L(3.6)] = (1.05\pm 0.02) Log[SFR/L(3.6)] + (45.30\pm0.02),
\end{equation}
where SFR is from equation~17, and the units of SFR/L(3.6) are M$_{\odot}$~yr$^{-1}$~(erg~s$^{-1}$)$^{-1}$.
The slope in equation~19 is also slightly above unity, only 2.5~$\sigma$ away from it, and quickly converges to unity for a fitting range that excludes the faintest galaxies, in agreement with our 
conclusions above.

\begin{figure}
\figurenum{11}
\plottwo{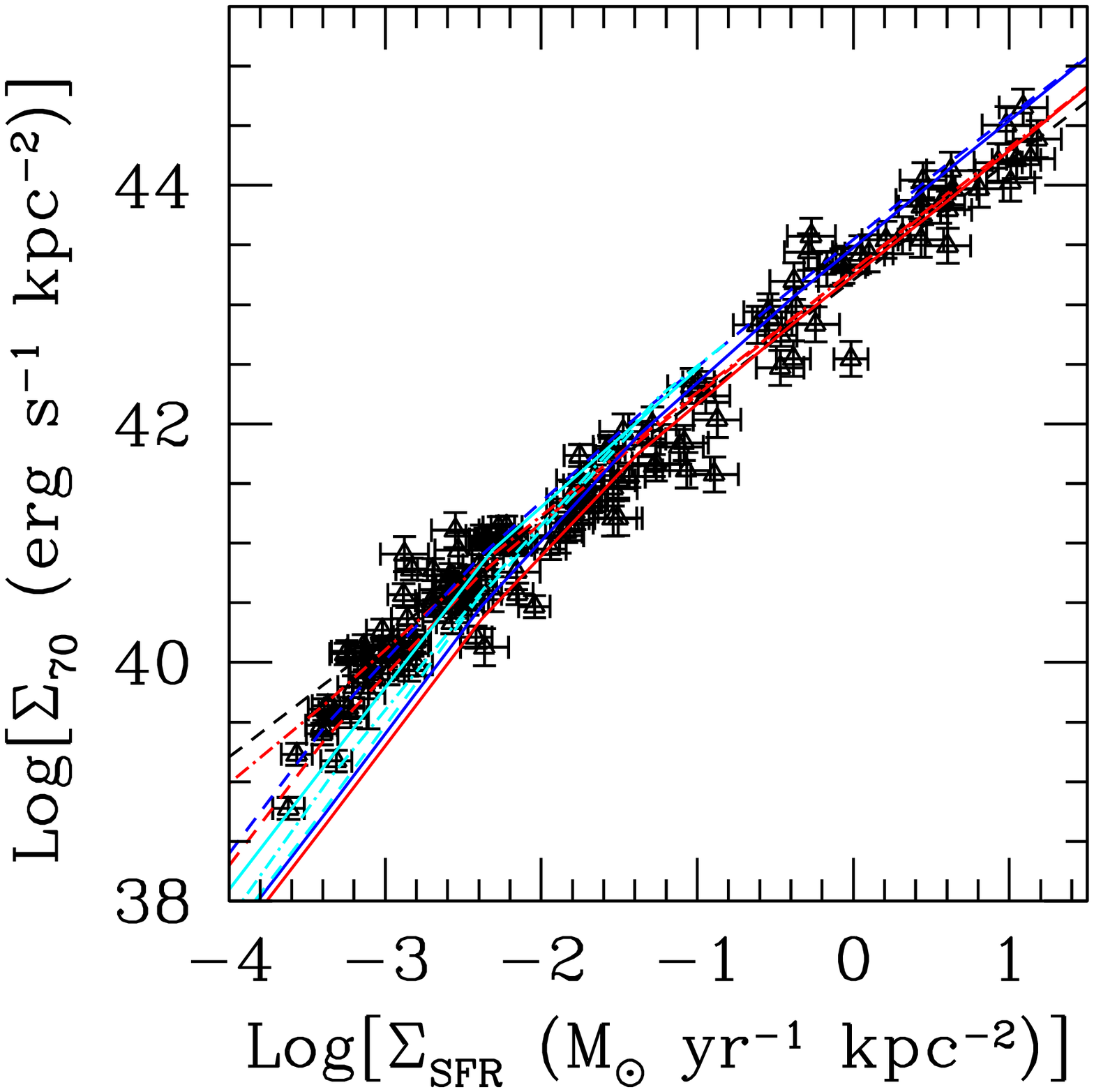}{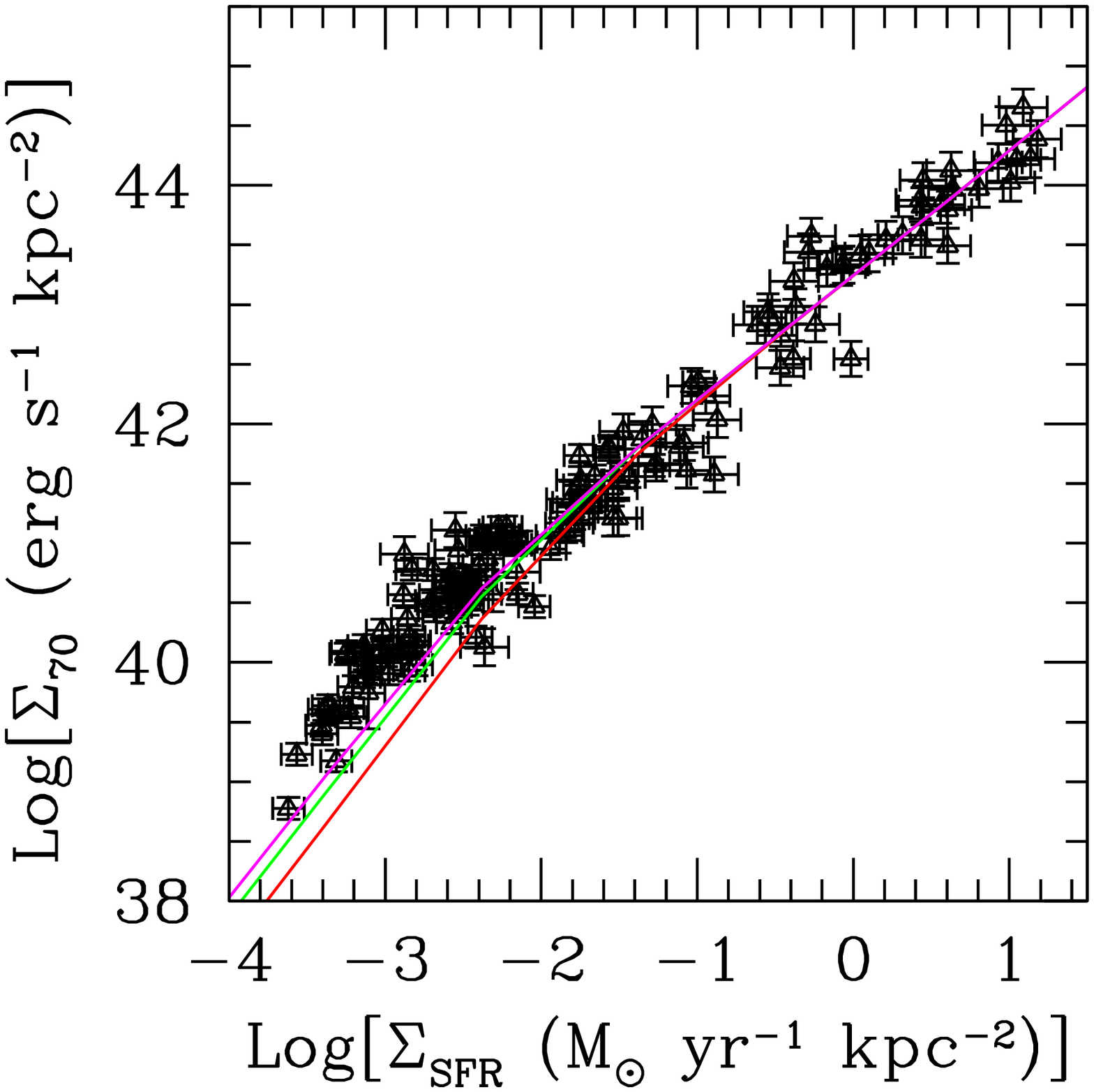}
%\plottwo{figure9.eps}{figure9b.eps}
\caption{The same data on the 70~$\mu$m LSD, $\Sigma_{70}$, versus
 the SFSD, $\Sigma_{SFR}$ as in Figure~\ref{fig9}. {\bf Left.} The data are compared with the  simple galaxy models described in 
 Figure~\ref{fig4} (see text for more details). The best linear fit with unity slope is shown as a black 
 dashed line (from Figure~\ref{fig9}). All stellar 
population models are for solar metallicity: 10~Gyr old (blue continuous) and  
100~Myr old (red continunous) constant star formation populations, and 
exponentially decreasing star formation (cyan; dash: $\tau$=5~Gyr; continuous: $\tau$=2~Gyr), 
together with our default dust model, in which the dust extinction 
is described by the starburst curve \citep{Calzetti2001}, the dust emission comes from a medium of increasing opacity, and the peak infrared emission progressively moves into the 70~$\mu$m band from longer wavelengths \citep{DraineLi2007} for increasing SFSD.   
Dashed red and blue lines  show the same default dust models, but the star--forming regions have 
 a 10\%  filling factor within the galaxies, for stellar population with 100~Myr and 10~Gyr constant star formation, respectively. The case of a constant and large dust opacity  (A$_V>$10~mag) with SFSD 
 is shown as a red dot-dash line for a 100~Myr constant star formation population. The cyan 
dot-dash line is a decreasing star formation model, with $\tau$=2~Gyr, embedded 
in a uniform distribution of dust (mixed dust--star case).  {\bf Right.} The 100~Myr constant star formation model is here reproduced for three different dust attenuation curves: the starburst curve (red line), 
used in our default model,  according to which stars and ionized gas display differential attenuation 
\citep[by a factor 0.44, see][]{Calzetti1994,Calzetti2000}; the Milky Way (green line) and SMC 
(magenta line) extinction curves, with foreground dust geometry, and stars and gas attenuated by the same dust column density. 
\label{fig11}}
\end{figure}

The simple stellar population and dust emission models briefly presented in section~4.1 can further help clarify some of the observed trends. At the high luminosity end, the models have slope roughly unity by construction (Figure~\ref{fig5}), because in this regime TIR is a good proxy for SFR; thus the trend 
 $\Sigma_{70}$--$\Sigma_{SFR}$ does not provide independent information. 
 For SFSD$\gtrsim$0.5~M$_{\odot}$~yr$^{-1}$~kpc$^{-2}$, the opacity of the 
 system is high and most of the stellar light is reprocessed by dust into the infrared. 
 Thus, the details of the extinction model are unimportant, and the normalization 
 factor between the SFSD and the total infrared emission is provided by the stellar population. 
The mean trend of the  70~$\mu$m surface
 brightness is well represented by a model where a 100~Myr constant star formation population heats 
 the dust (Figure~\ref{fig11}), which is how we describe LIRGs (equation~17).

 As the SFSD decreases, the average opacity of the galaxy also tends to decrease \citep{Wang1996,Heckman1998,Hopkins2001,Calzetti2001}. The opacity models of \citet[][constructed for HII regions]{Calzetti2007},  however, under--predict the observed $\Sigma_{70}$ by a factor 
 $\approx$2--3, for 
 $\Sigma_{SFR}\lesssim$0.01~M$_{\odot}$~yr$^{-1}$~kpc$^{-2}$ (Figure~\ref{fig11}). 
 All the other models, the extreme case of constant and large dust opacity  for the 
 entire SFSD range,  the model 
 with a filling factor of the emitting regions of about 10\% of the normalizing area, and the stellar 
 population models with exponentially decreasing star formation, mark a range that brackets 
 the spread of the datapoints. The mean trend of the data at low SFSDs is  better represented by the model  with a 10\% filling factor of the emitting regions, for constant star formation. This is perhaps not surprising, since star formation in low surface brightness galaxies tends to display a morphology of isolated knots distributed over a large, non--contiguous area, while our definition of  `ionized gas emitting area' (section~2) measures a single 
 size per galaxy, irrespective of the distribution and filling factor of the emitting regions. In the case of exponentially decreasing star formation, the mean trend of the data requires higher filling factors, 
 around 30\%--50\% for the model with e--folding time $\tau=$2~Gyr, but still around 15\% for the 5~Gyr e--folding time model. 
 
Our solutions are not unique, and different combinations of dust opacity models and filling factors can provide similar trends. For instance, the extreme case of a foreground dust geometry for the galaxy, 
with a Small Magellanic Cloud extinction curve \citep{Bouchet1985} and common dust column density to both stars and ionized 
gas  gives higher values of the infrared emission at low luminosities than the starburst attenuation curve (Figure~\ref{fig11}, right). Thus, the low luminosity data points can be accounted for by 
a larger filling factor of the emitting regions, $\sim$30\%, about a factor 3 larger than the case of the 
starburst attenuation curve with constant star formation.  
The conclusion is still that a low filling factor is needed for the emitting 
regions, between 10\% and 30\%, for constant star formation, and around 15\%--50\% for exponentially 
decreasing star formation, in order to account for the observed trend between the 
70~$\mu$m emission and the SFSD for $\Sigma_{SFR}\lesssim$0.01~M$_{\odot}$~yr$^{-1}$~kpc$^{-2}$.

The overall trend of the high--metallicity datapoints is consistent with both increasing dust opacity 
and a smooth transition from a low 
filling factor to a high filling factor as the SFSD increases, with the transition point located around  
$\Sigma_{SFR}\sim$0.01~M$_{\odot}$~yr$^{-1}$~kpc$^{-2}$.
 
\begin{figure}
\figurenum{12}
\plotone{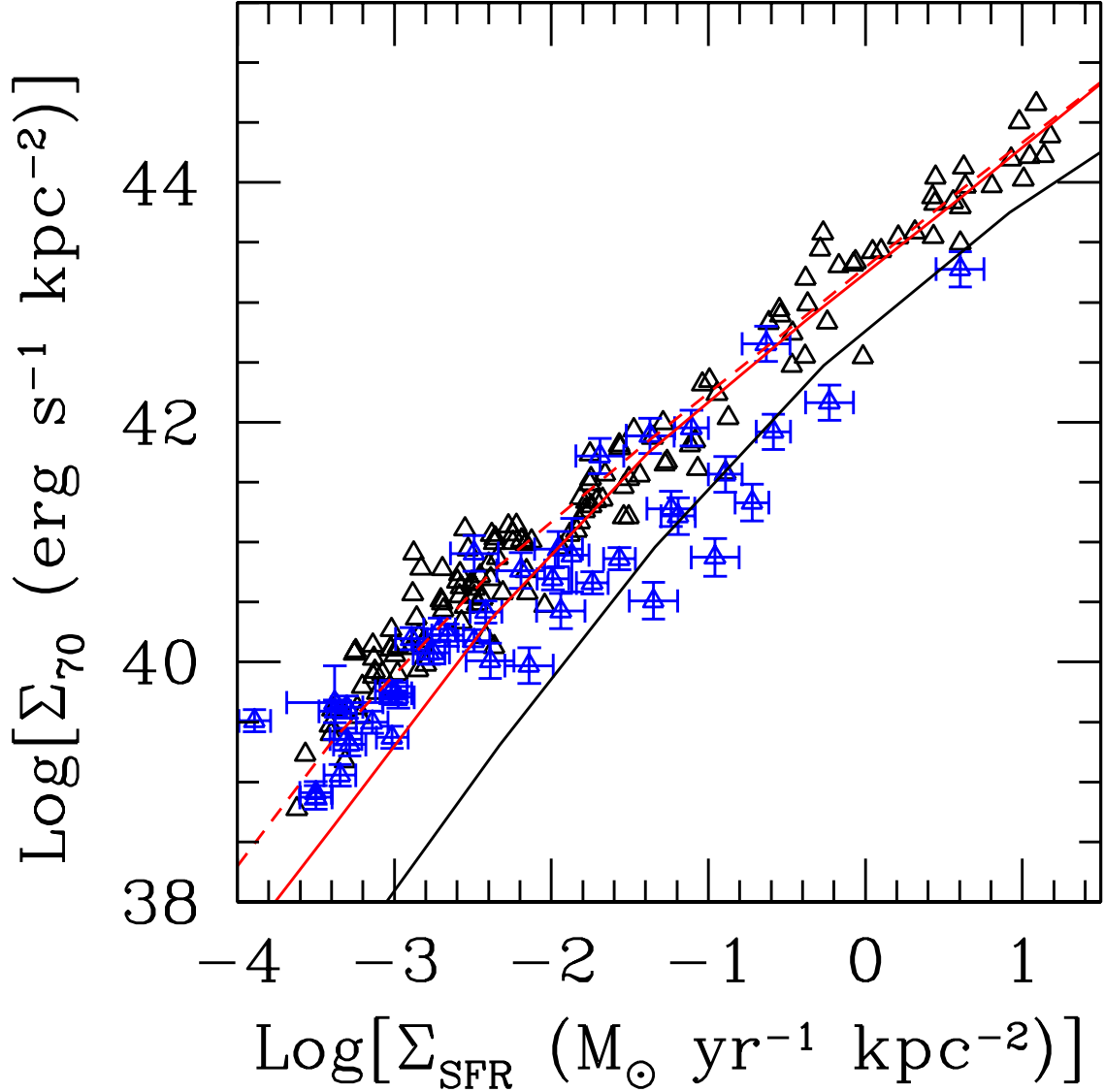}
%\plotone{figure11.eps}
\caption{The data on the 70~$\mu$m LSD, $\Sigma_{70}$, versus
 the SFSD, $\Sigma_{SFR}$ now include both metal--rich (black symbols) and metal--poor (blue 
 symbols) galaxies, as defined in the text. 1--$\sigma$ error bars are reported for the low metallicity datapoints, while the error bars for the metal--rich galaxies are omitted for clarity. The model of a 
 100~Myr continuous star formation population is used for reference, from Figure~\ref{fig11}. 
 The default solar metallicity model (red continuous line) and the 10\% filling--factor one 
 (red dashed line) are compared with a 0.1 solar metallicity model \citep[continuous black line, from][]{Calzetti2007}.  
\label{fig12}}
\end{figure}

 Including the low--metallicity data in the $\Sigma_{70}$--$\Sigma_{SFR}$ plot provides further insights 
 into the effect of dust opacity, as low metallicity galaxies tend to be more transparent than higher 
 metallicity ones \citep{Cannon2005,Walter2007}. These data are located typically below the trend 
 marked by the high metallicity data (Figure~\ref{fig12}), as expected for decreasing dust amounts in the ISM. The inclusion of the model line for a 1/10 solar metallicity ISM from \citet{Calzetti2007} shows 
 that metallicity has a major effect on the location of galaxies on the plot. The model line for low--metallicity systems is located between 4 times (at high luminosity) and an order of magnitude (at low luminosity) below the analogous model for solar metallicity systems. There are a few galaxies that are located even below our low--metallicity model line (Figure~\ref{fig12}). Two galaxies, Tol~65 and 
 SBS~0335$-$052, are more than 3~$\sigma$ below the low--metallicity curve, while IZw18 is 
 about 2~$\sigma$ below the curve. These three galaxies, with metallicities between 7.20 and 7.45 
 \citep{Nagao2006}, are among the most metal--poor in our sample, and well below 1/10th solar.  
 Furthermore, Tol~65 and SBS~0335$-$052 have high effective dust temperature and their  peak infrared emission is located shortward of 70~$\mu$m \citep{Engelbracht2008}. Thus, it is not surprising that the three galaxies lie below our 1/10th solar model line. 

\begin{figure}
\figurenum{13}
\plotone{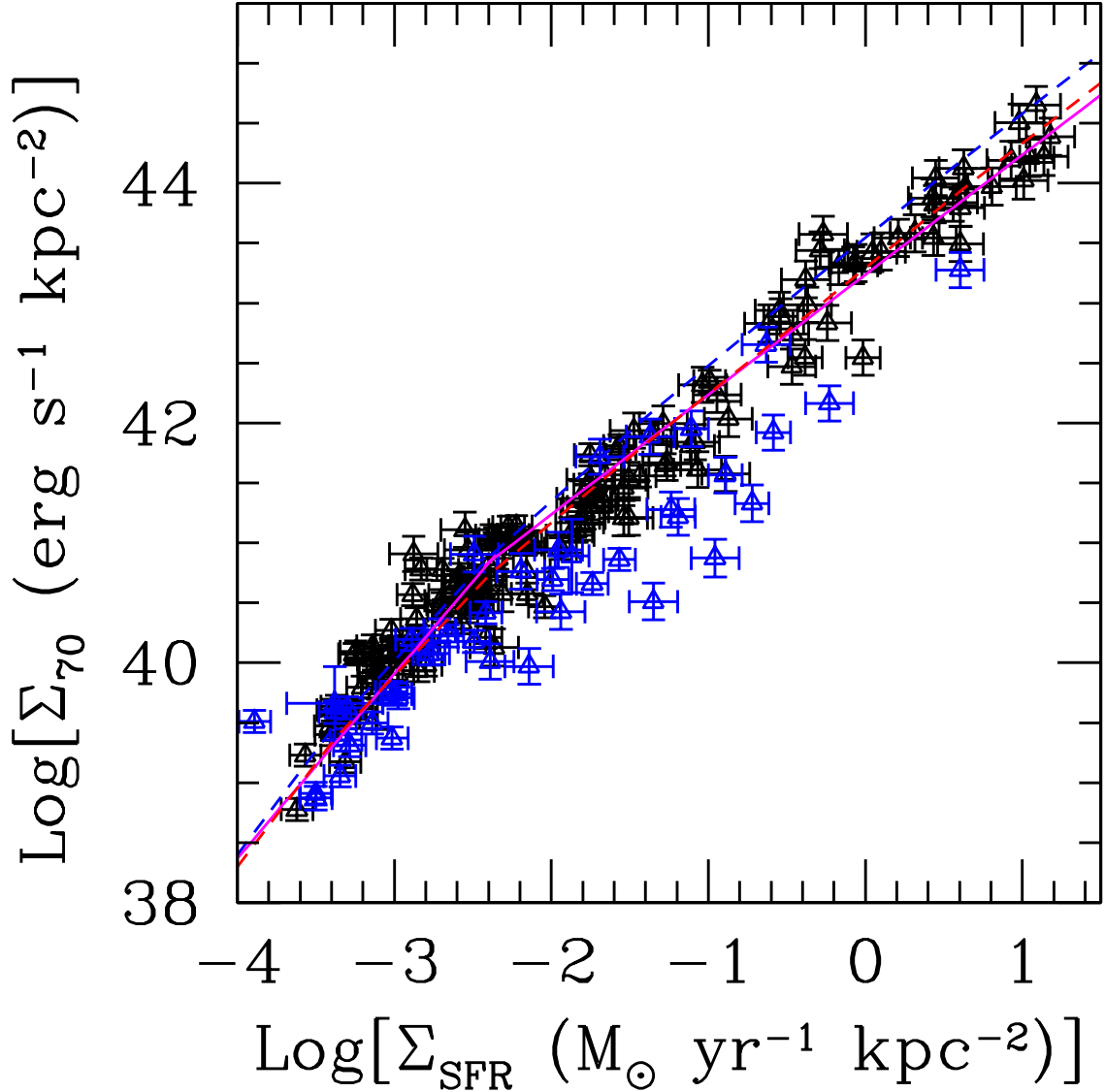}
%\plotone{figure12.eps}
\caption{The data on the 70~$\mu$m LSD, $\Sigma_{70}$, versus
 the SFSD, $\Sigma_{SFR}$ for both metal--rich (black symbols) and metal--poor (blue 
 symbols) galaxies. The data are the same as in Figure~\ref{fig12}, except that now errorbars are included for all datapoints. Models of stellar populations with constant star formation over 100~Myr (red dashed  
 line) and 10~Gyr (blue dashed line)  surrounded by a solar metallicity ISM with a 10\% area filling factor 
 at the low luminosity end are also reported, and compared with the best power--law fits to the 
 asymptotic trends at the low and high luminosity ends (broken magenta line, equations~20 and 21). 
\label{fig13}}
\end{figure}

\begin{figure}
\figurenum{14}
\plotone{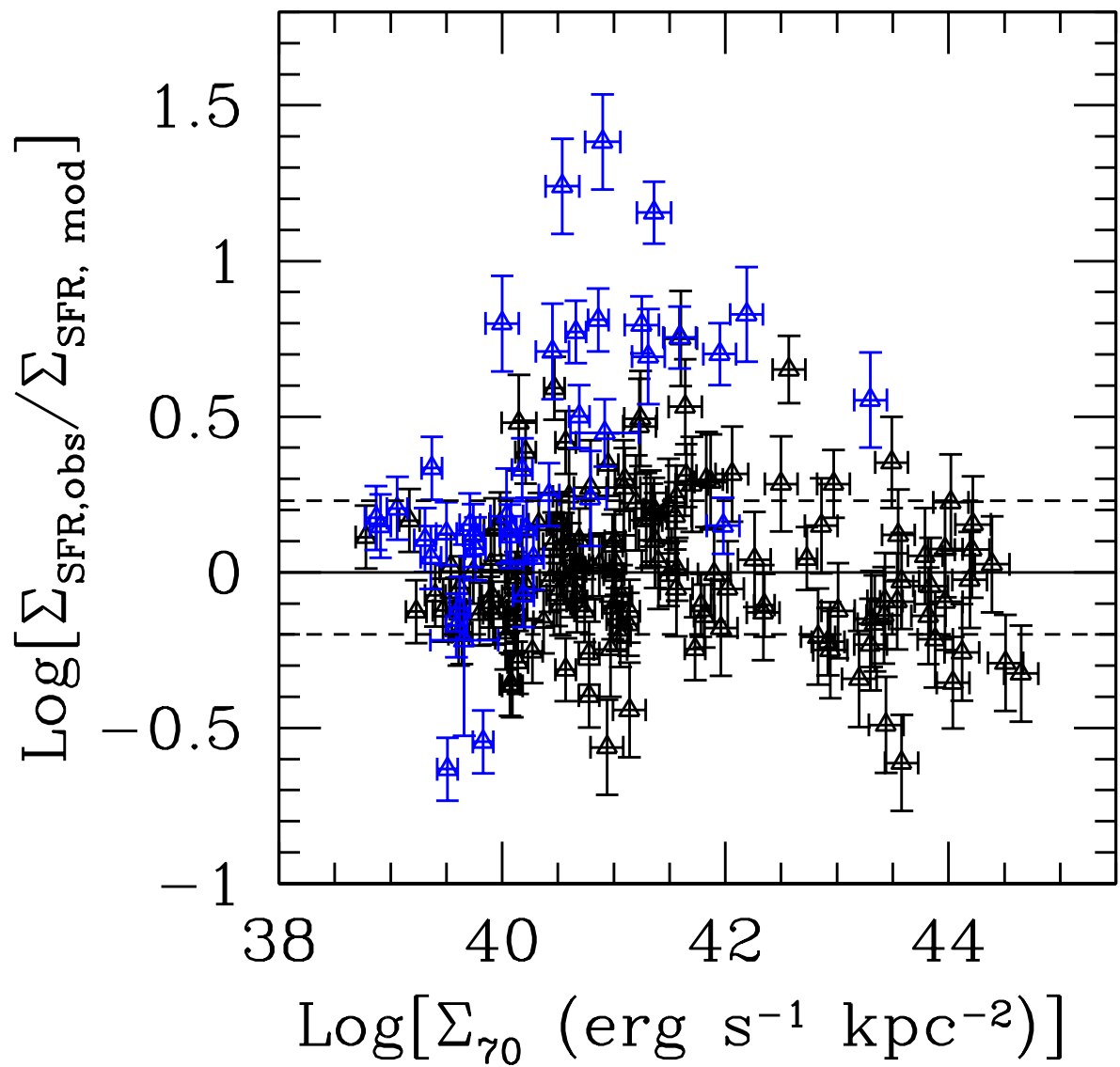}
%\plotone{figure13b.eps}
\caption{The difference between the data and the fitting lines described in equations~20 and 21 
is reported as a function of the 70~$\mu$m LSD, together with horizontal lines 
showing identity between models and data (continuous line) and the 1~$\sigma$ dispersion 
value ($-$0.20, $+$0.23, dashed lines).
\label{fig14}}
\end{figure}

Although metallicity has an important effect on the applicability of the 70~$\mu$m emission as a 
SFR indicator, luminous infrared galaxies tend to be, on average, metal and dust rich. Indeed, 
most of the low metallicity galaxies in our sample cluster at the low--luminosity and low $\Sigma_{SFR}$ 
end of the range (Figure~\ref{fig12}), an effect of typically `hotter' IR SEDs relative to the high metallicity 
galaxies \citep{Dale2005}.  From the discussion above, we approximate the trend of the 
high metallicity data (Figure~\ref{fig13}) with: 
\begin{equation}
\Sigma_{SFR}\ (M_{\odot}~yr^{-1}~kpc^{-2}) = {\Sigma_{70}\ (erg~s^{-1}~kpc^{-2}) \over 1.7\times 10^{43}}, 
\end{equation}
for 7$\times$10$^{40}$~erg~s$^{-1}$~kpc$^{-2}\lesssim\Sigma_{70}\lesssim$5$\times$10$^{44}$~erg~s$^{-1}$~kpc$^{-2}$, and:
\begin{equation}
\Sigma_{SFR}\ (M_{\odot}~yr^{-1}~kpc^{-2})= \Bigl({\Sigma_{70}\ (erg~s^{-1}~kpc^{-2})\over 3.3\times 10^{44}}\Bigr)^{0.65}, 
\end{equation}
for 6$\times$10$^{38}$~erg~s$^{-1}$~kpc$^{-2}\lesssim\Sigma_{70}\lesssim$7$\times$10$^{40}$~erg~s$^{-1}$~kpc$^{-2}$. The non--linear relation between the  70~$\mu$m LSD and 
the SFSD (equation~21) attempts to take into account the increasing transparency of the medium 
for decreasing SFSDs, and is a rough approximation to a more complicated trend (Figure~\ref{fig13}). 
The important caveat about equation~21 is that the calibration is highly dependent on the type of 
objects from which it is derived (whole galaxies in this case), and should not be applied to 
different physical systems (e.g., HII regions or complexes within galaxies) or metal--poor systems, 
 without proper checks. Nevertheless, the two 
power laws, equations~20 and 21, provide a reasonable description of the datapoints
in our sample with 12$+$Log(O/H)$\gtrsim$8.1, 
which show a relatively symmetric dispersion around those mean trends, with 1~$\sigma\sim$0.2~dex 
(Figure~\ref{fig14}), or about 60\% in linear scale.  The linear relation at the high luminosity end 
offers the opportunity to derive directly a star formation rate:
\begin{equation}
SFR (70)\ (M_{\odot}~yr^{-1})= {L(70)\ (erg~s^{-1})\over 1.7\times 10^{43}}, 
\end{equation}
 applicable to luminosities 
L(70)$\gtrsim$1.4$\times$10$^{42}$~erg~s$^{-1}\sim$3.7$\times$10$^8$~L$_{\odot}$. Again,  the relation between luminosity and luminosity/area has at least one order of magnitude dispersion in our sample.
Adopting the SFR--L(H$\alpha$) calibration of \citet{Kennicutt1998} would change the denominator in equation~22 to 1.2$\times$10$^{43}$. The change is due to a combination of adopted 
stellar IMF \citep[][adopts a Salpeter IMF in the range 0.1--100~M$_{\odot}$]{Kennicutt1998} and small 
($\sim$10\%) variations in the stellar population models. 

\section{Can we also derive SFR(160)?}

\begin{figure}
\figurenum{15}
\plottwo{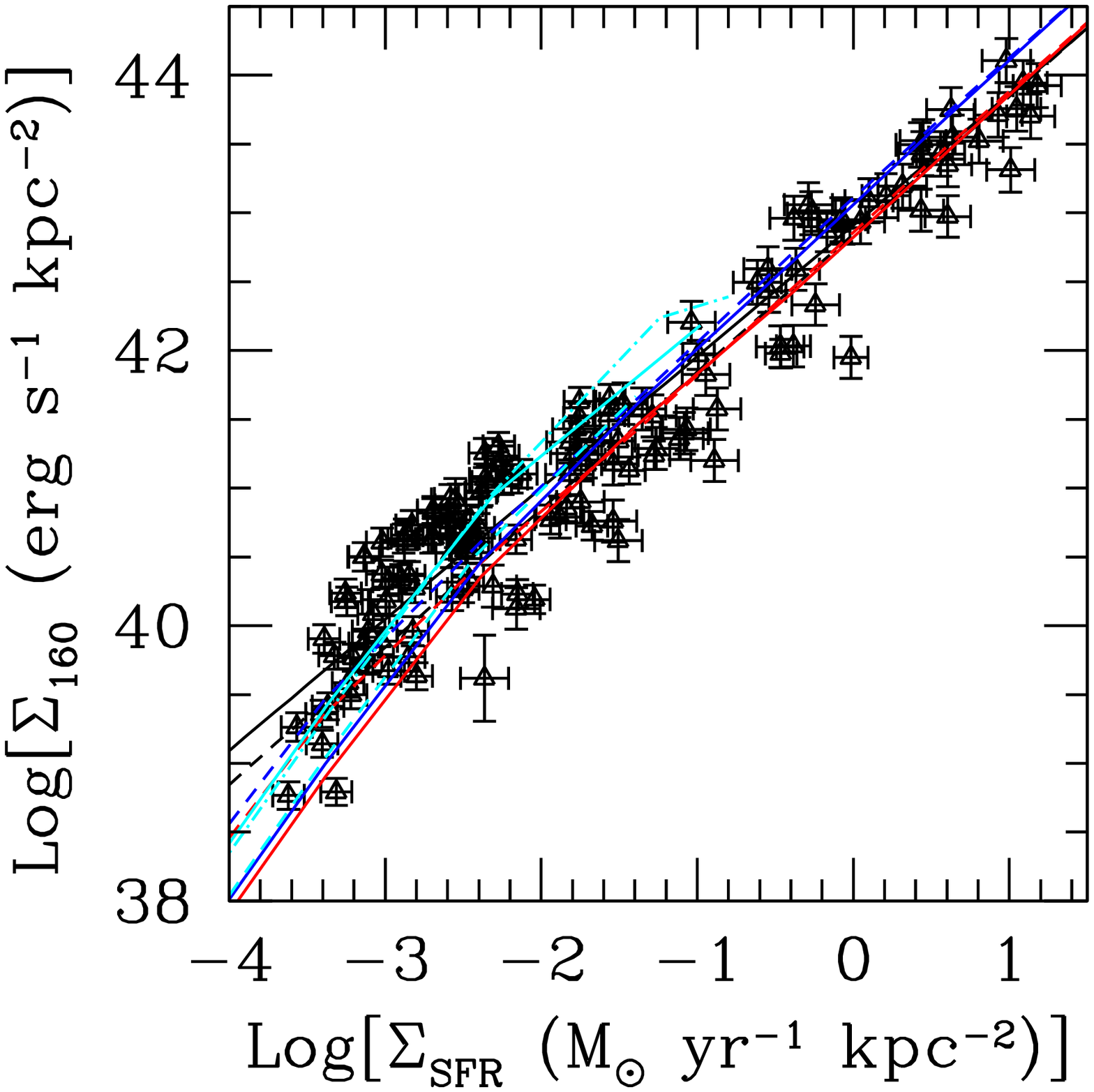}{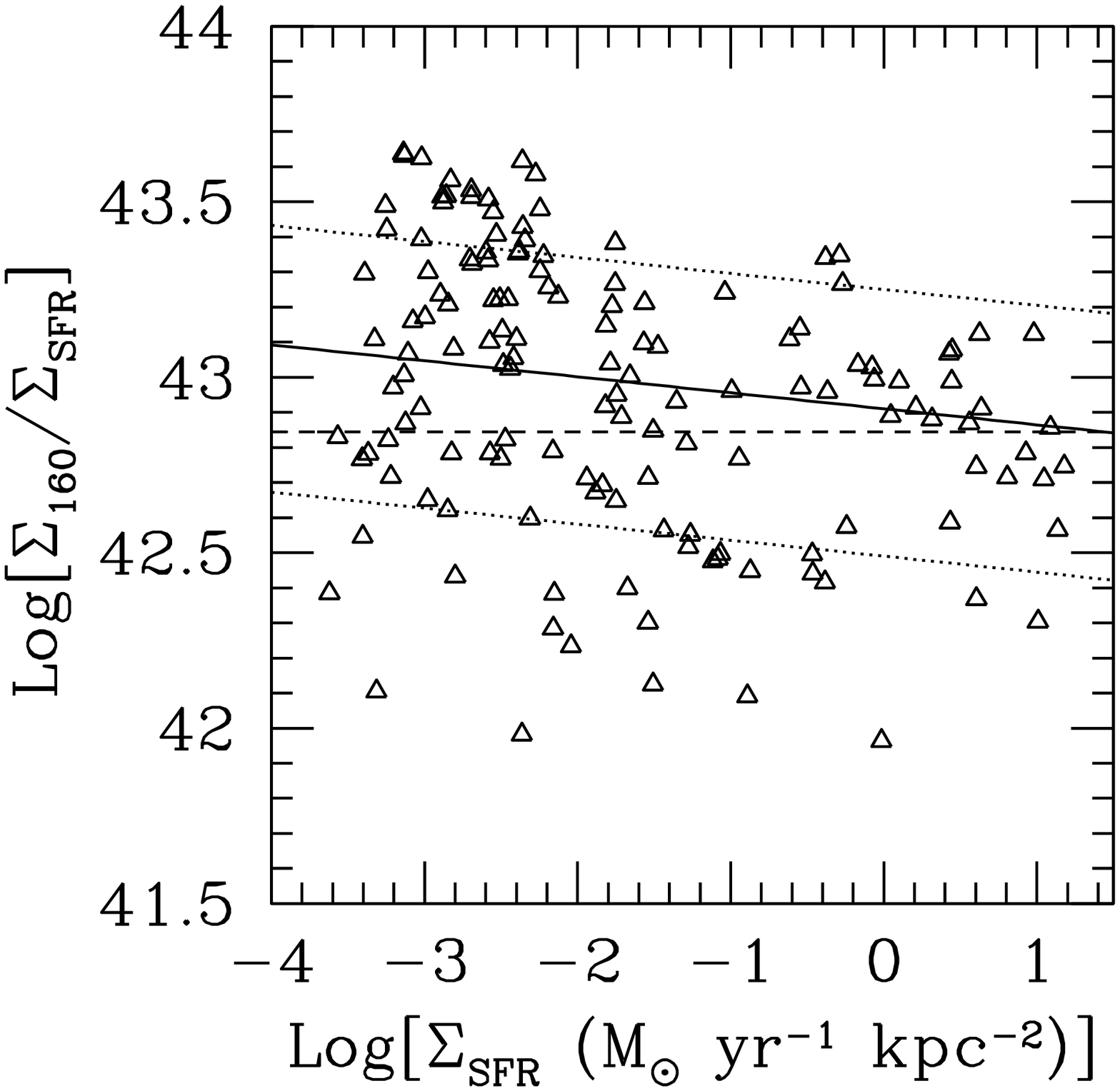}
%\plottwo{figure14.eps}{figure14b.eps}
\caption{{\bf Left.} The 160~$\mu$m LSD, $\Sigma_{160}$=L(160)/area, versus
 the SFSD, $\Sigma_{SFR}$, for the metal--rich galaxies in our sample. Both best linear fits and model 
 lines are overplotted on the data. The best fit has a slope slightly less than 
 unity (continuous black line), and a line through the data with unity slope is shown as a dashed black  
 line. The models are the same of Figure~\ref{fig11}. {\bf Right.} Similarly to the right--hand panel of Figure~\ref{fig9},  the same data and best linear fit (continuous line) are shown, together with the 1~$\sigma$ envelope to the data (dotted lines), for the ratio L(160)/SFR as a function of the SFSD.  The best fit with unity slope is shown as a dashed line.  This plot better shows the deviation of the datapoints from a one--to--one relation. The error bars on the data are not shown, for clarity.
\label{fig15}}
\end{figure}

The 160~$\mu$m luminosity surface density correlates, like $\Sigma_{70}$, with SFSD, with linear best--fit 
(Figure~\ref{fig15}):
\begin{equation}
Log [\Sigma_{160}\ (erg~s^{-1}~kpc^{-2})] = (0.954\pm 0.012) Log [\Sigma_{SFR}\ (M_{\odot}~yr^{-1}~kpc^{-2})] + (42.910\pm0.028).
\end{equation}
The shallower--than--unity slope is consistent with our interpretation that galaxies of decreasing luminosity have `cooler' infrared SEDs (Figure~\ref{fig5}). This result is partially counteracted by a 
sharp decline in 
160~$\mu$m luminosity for $Log(\Sigma_{SFR})<-$3.2, indicating the regime where the galaxies 
become measurably  transparent. Like in the case of the 70~$\mu$m LSD, the best 
fit line tends asymptotically to a slope of unity for galaxies of increasing SFSD; in the case of the 
160~$\mu$m LSD, the best fit has slope of unity within 1~$\sigma$ already for 
$Log(\Sigma_{SFR})>-$2. 

\begin{figure}
\figurenum{16}
\plottwo{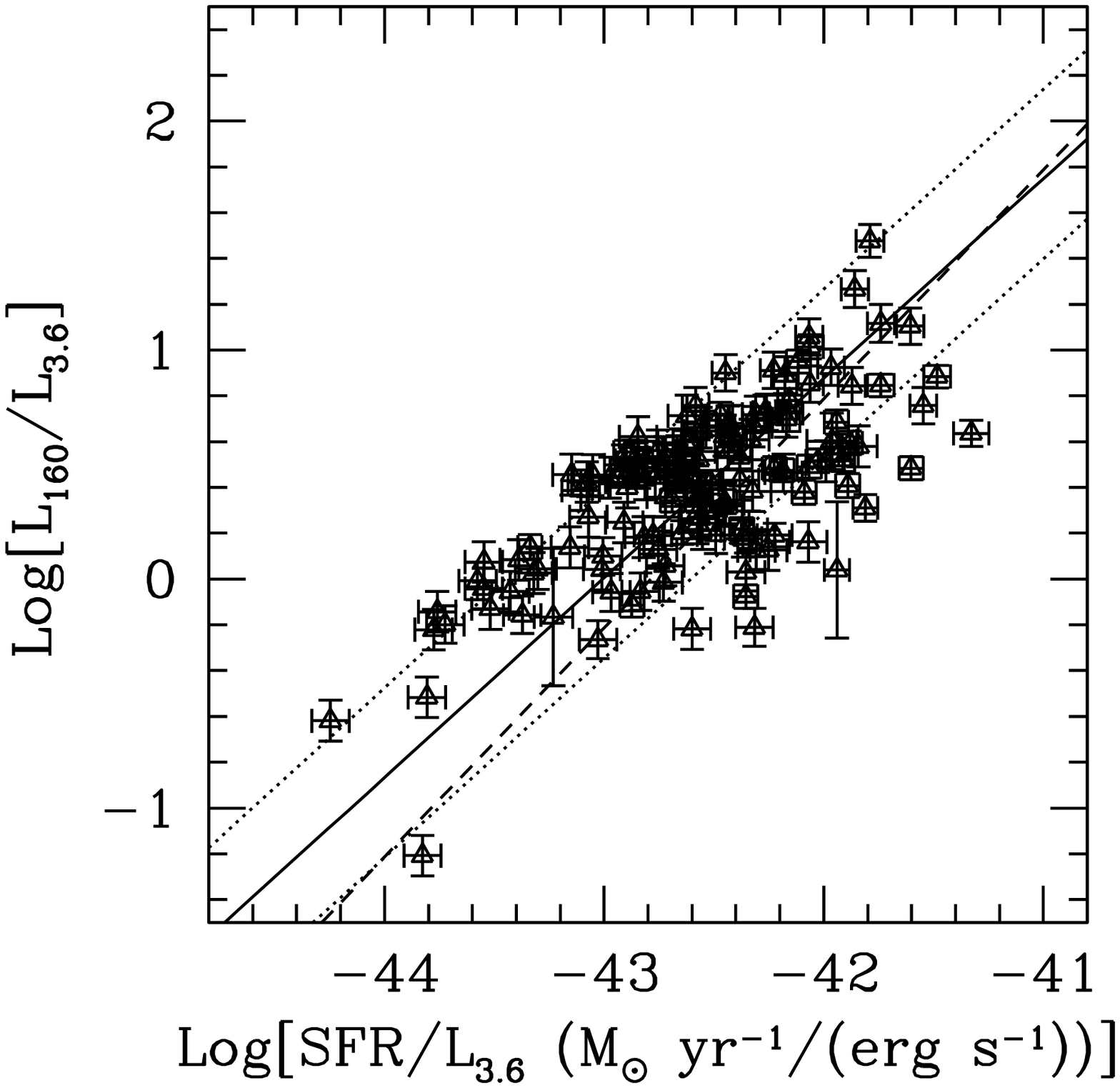}{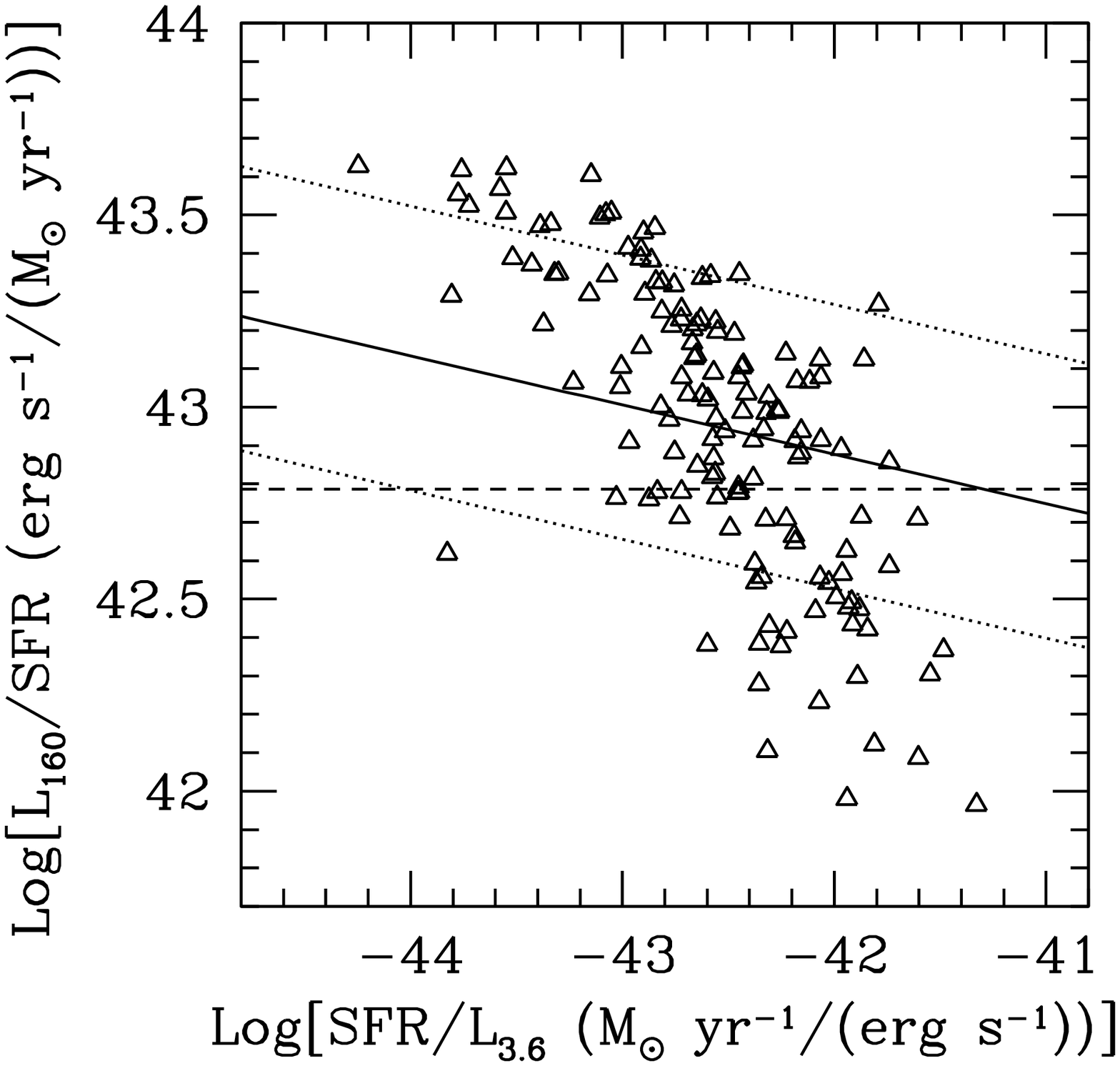}
%\plottwo{figure15a.eps}{figure15b.eps}
\caption{{\bf (left)} and  {\bf (right).} The same data as Figure~\ref{fig15}, with the data normalized by the 
3.6~$\mu$m luminosity, instead of the emitting area. The best fit (continuous line) and the 1~$\sigma$ envelope to the data (dotted lines, which correspond to the range $-$0.35,$+$0.39~dex) are also shown. A line through the data with unity slope is shown in both panels (dashed line).
\label{fig16}}
\end{figure}

The dispersion of the data about the best fit line, ($-$0.42,$+$0.34)~dex,  is larger for $\Sigma_{160}$ 
than for  $\Sigma_{70}$ by a factor $\sim$1.3--1.5 (Figures~\ref{fig9} and \ref{fig14}, left). A similar ratio, 
$\sim$1.2--1.4, between the 1~$\sigma$ dispersions holds when quantities normalized to the 
3.6~$\mu$m luminosity are considered (Figure~\ref{fig16}). Furthermore, the dispersion on 
the positive side, i.e., for large 160~$\mu$m emission relative to the SFSD, is dominated by galaxies 
at the low SFSD end, $Log(\Sigma_{SFR})<-$2 (Figure~\ref{fig15}, right). These two facts, when taken together, suggest that 
the 160~$\mu$m luminosity is affected by emission from dust heated by stellar populations unrelated to 
those powering the current star formation. This effect becomes more and more evident as total luminosities decrease. Indeed, the same models with 10\% area filling factor of the emitting regions 
that satisfactorily account for the trend of the data in the $\Sigma_{70}$--$\Sigma_{SFR}$ plane provide an 
unsatisfactory description of the data in the $\Sigma_{160}$--$\Sigma_{SFR}$ plane   for 
$\Sigma_{SFR}<$0.01, for both the 100~Myr and the 10~Gyr constant star formation population: the models 
underpredict, by roughly  0.5~dex or more, the median locus of the data for most of the data in this SFSD 
range (Figure~\ref{fig15}, left). 
A similar result is obtained when normalizing the 160~$\mu$m luminosity and the SFR to the 
3.6~$\mu$m luminosity, instead of the area (Figure~\ref{fig16}). In this case, the best fit to the data 
has a shallower--than--unity slope:
 \begin{equation}
Log [L(160)/L(3.6)] = (0.87\pm 0.01) Log [SFR/L(3.6)] + (37.48\pm0.02),
\end{equation}
(SFR is from equation~17, and the units of SFR/L(3.6) are M$_{\odot}$~yr$^{-1}$~(erg~s$^{-1}$)$^{-1}$)
and the slope approaches unity only for increasingly luminous galaxies.  Equation~24 supports, again,  
the presence of two effects for low luminosity galaxies: `cooler' infrared SEDs for lower luminosity galaxies and 
 a proportionally increasing contribution, for decreasing luminosity,  to the 160~$\mu$m emission
from dust heated by stellar populations that are unassociated with the current star formation. 

All of the above strengthens  the consideration made in section~5 that a fraction of the dust emission detected in the 70~$\mu$m band towards low SFSDs  is due to stellar populations unrelated to the current star formation, 
and this effect is, expectedly, larger in the 160~$\mu$m showing up as both a larger dispersion of the data around the mean trends and an excess deviation of the data from the model predictions at the low luminosity end. 

This conclusion is not new: many authors have noted  over the past $\sim$2 decades the presence of 
a contribution to the infrared emission of galaxies from  stellar populations not associated to the current star formation \citep[e.g.][]{Lonsdale1987, Sauvage1992, Buat1996, Walterbos1996}. The presence of such contribution limits the usefulness, especially at low galaxy luminosity, of SFR calibrations based on monochromatic IR emission long-ward of the SED peak. 

In summary, although defining a SFR calibration based on the 160~$\mu$m emission may be tempting: 
\begin{equation}
SFR (160)\ (M_{\odot}~yr^{-1}) = {L(160)\ (erg~s^{-1}) \over 7.0\times10^{42}},
\end{equation}
 with applicability for L(160)$>$1$\times$10$^{42}$~erg~s$^{-1}\sim$2.6$\times$10$^8$~L$_{\odot}$, the presence of a large dispersion, 1~$\sigma_{160}\sim$2$\times$(1~$\sigma_{70}$), and the likelihood that a portion of the 160~$\mu$m emission (and a large fraction of it at the low luminosity end) is due to dust heating by non--star--forming populations prevents such definition from having the same confidence level as the calibration of SFR(70).  As will be seen in the next section, a safer range of applicability of equation~25 is for L(160)$>$2$\times$10$^{42}$~erg~s$^{-1}\sim$5.2$\times$10$^8$~L$_{\odot}$. With the \citet{Kennicutt1998} calibration for SFR, the numerical constant in equation~25 changes 
 to 4.8$\times$10$^{42}$.

\section{Discussion and Conclusions}

Monochromatic SFR indicators have the undeniable advantage of immediate application, 
especially when observational data cover only a limited range of wavelengths, as is often the case for distant galaxies.  Within this framework, Spitzer observations of nearby galaxies are offering a unique opportunity to test the applicability and limitations of a number of SFR indicators based on the emission from the dust heated by stars.  The main goal of this paper has been to investigate whether long wavelength infrared 
fluxes at 70 and 160~$\mu$m can be reasonably
calibrated as SFR indicators. 

Previous works  \citep{Roussel2001,Boselli2004,Forster2004,Calzetti2005,Dale2005,Wu2005,AlonsoHerrero2006,PerezGonzalez2006,Calzetti2007,Dale2007,Rellano2007,Zhu2008,Rieke2009} have concentrated on the short--wavelength infrared, in the range $\sim$5--40~$\mu$m, commonly referred to as the mid--IR. 
Although this wavelength range is below the peak infrared emission and only accounts for a few to a few tens of percent of 
the total IR emission, the dust heated by the hot, massive stars in a recent star formation episode can have high effective temperatures and may preferentially emit at the shorter infrared wavelengths. 
This general picture translates into a  correlation between SFRs and luminosities in the mid--IR, such 
as L(8) and L(24), centered on the Spitzer 8~$\mu$m and 24~$\mu$m bands. However, 
the correlation is also non--linear, and careful analysis shows that other contributors, in addition to the 
SFR, determine the short infrared wavelength luminosity, including transiently heated dust by single 
UV or optical photons in the general radiation field of a galaxy \citep{Haas2002,Boselli2004,Bendo2008}, variations in the dust effective temperature 
\citep[for L(24), see][]{Calzetti2007}, and, especially for L(8), gas metallicity \citep{Engelbracht2005,Rosenberg2006,Wu2006,Madden2006,Jackson2006,Draine2007}.  

\subsection{Establishing a reference SFR indicator}

In the present paper, we first establish a ``reference'' SFR calibration against which
we can compare the 70 and 160~$\mu$m luminosities
and LSDs.  The adopted reference SFR is given in equation~17.  We choose this particular three--part
formulation after comparing a variety of published 24~$\mu$m  
and composite H$\alpha$ plus 24~$\mu$m calibrations. 

Above a luminosity L(24)$\sim$5$\times$10$^{43}~$erg~s$^{-1}$, 
the contribution of direct (unabsorbed by dust) massive stars emission to SFR measurements 
is less than $\sim$15\% (Figure~\ref{fig8}).
% in this regime, many calibrations using L(24) to 
%measure SFRs converge to common or similar mean values (Figure~\ref{fig6}). Thus, 
For L(24)$\gtrsim$5$\times$10$^{43}~$erg~s$^{-1}$,  the 24~$\mu$m emission is a reliable 
SFR indicator  {\em for whole galaxies}. However, in this luminosity regime, the 24~$\mu$m dust emission is also self--absorbed \citep{Rieke2009}, and a non--linear correction to the calibration 
needs to be applied \citep[third term of equation~17, re--calibrated from the original calibration of][]{Rieke2009}.  

\begin{figure}
\figurenum{17}
\plottwo{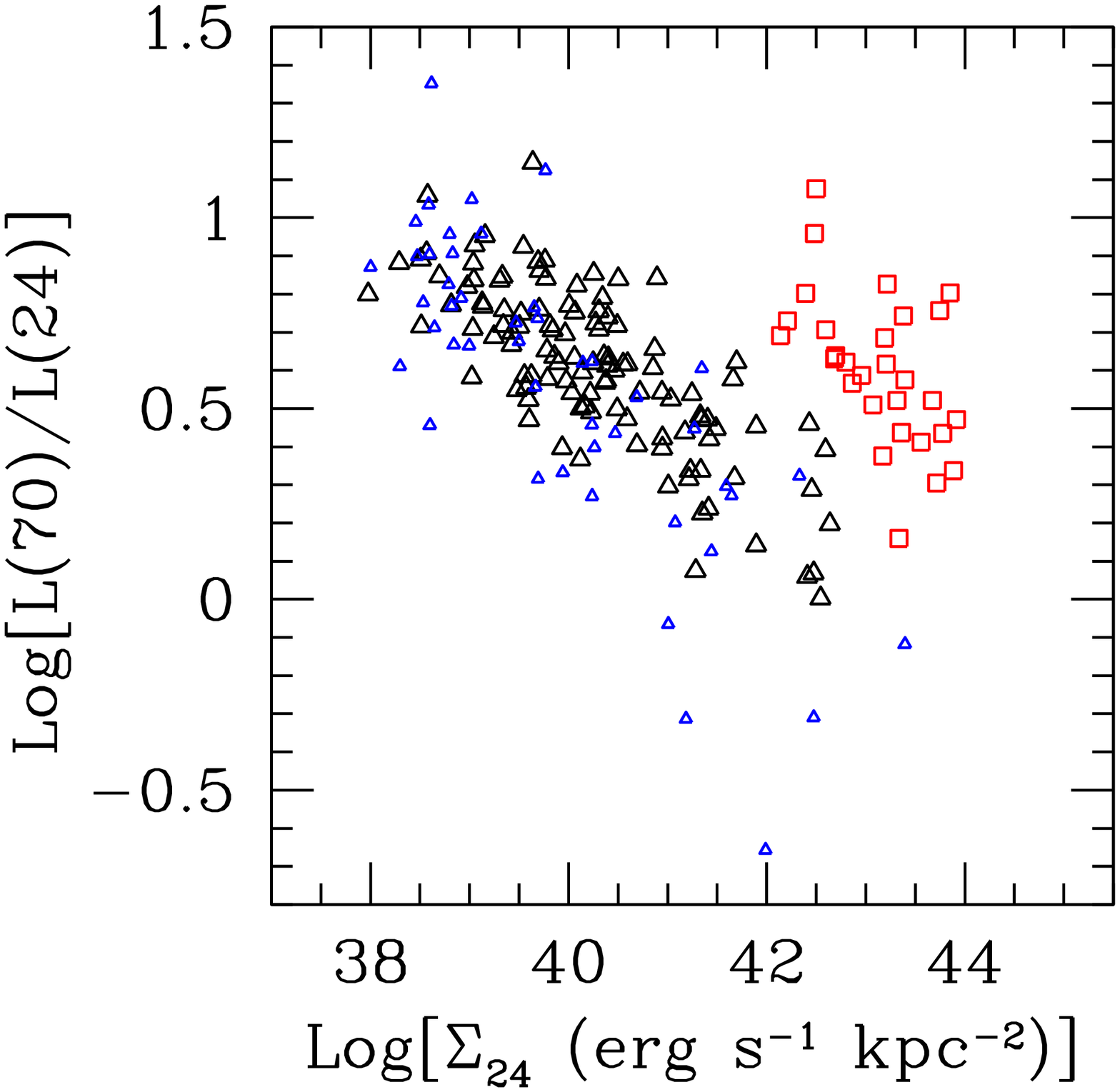}{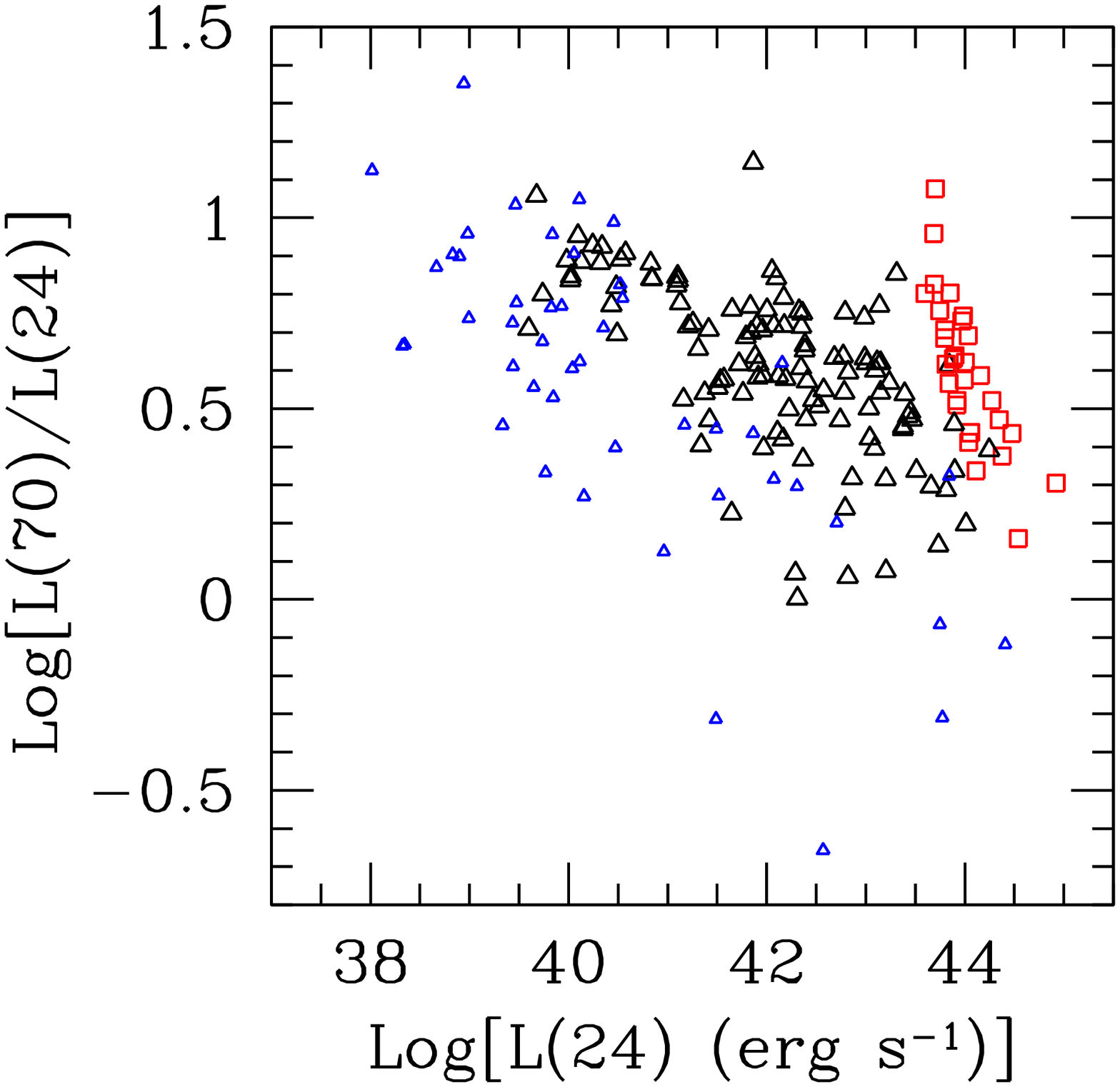}
%\plottwo{figure15a.eps}{figure15b.eps}
\caption{The L(70)/L(24) ratio as a function of the 24~$\mu$m LSD, $\Sigma_{24}$, (left) and 
luminosity, L(24), (right), respectively. As in Figure~\ref{fig1}, colors identify the star--forming 
and starburst high--metallicity (black symbols) and low--metallicity  (blue symbols) galaxies,  and the LIRGs (red). Errorbars are omitted for clarity. The decreasing trend of L(70)/L(24) as a function of increasing LSD for normal star--forming and starburst galaxies (black and blue symbols) observed in the left--hand--side panel shows the variation in the dust effective temperature for these systems, similarly to Figure~\ref{fig3}. The trend shows a large scatter, when the ratio L(70)/L(24) is plotted as a function of luminosity (right--hand--side panel).
\label{fig17}}
\end{figure}

Below a luminosity L(24)$\sim$5$\times$10$^{43}~$erg~s$^{-1}$, the galaxies become  transparent in the mid--infrared, and, as the luminosity further decreases, they become transparent at UV and optical wavelengths \citep[e.g.][]{Bell2003,Buat2007}. At L(24)$\sim$4$\times$10$^{42}~$erg~s$^{-1}$, the contribution to the SFR from emission at optical (H$\alpha$) and IR (24~$\mu$m) wavelengths is roughly equal (Figure~\ref{fig7}).  
Thus, below L(24)$\sim$5$\times$10$^{43}~$erg~s$^{-1}$, the 24~$\mu$m luminosity 
becomes increasingly insufficient, by itself, to fully characterize the SFR of a whole galaxy, as  the portion of the star formation unabsorbed by dust  becomes an increasingly significant contribution to the total SFR. Additionally, a correlation between the effective temperature and the luminosity/area of the thermal dust 
emission is present in HII~regions for decreasing total SFRs \citep{Calzetti2007}, and in whole galaxies  as well \citep[][and Figure~\ref{fig4}]{Chanial2007}. This correlation is further seen for the galaxies in our sample by plotting the ratio L(70)/L(24) as a function of $\Sigma_{24}$ (Figure~\ref{fig17}, left); the star--forming and starburst galaxies show a decreasing IR ratio for increasing 
LSD, as is expected for increasing effective dust temperature. This trend is less tight,  albeit still present to some extent,  when the L(70)/L(24) ratio is plotted as a function of luminosity (Figure~\ref{fig17}, right), rather than luminosity/area. The use of total luminosity introduces a scatter because galaxies are weighted not only by their intrinsic level of activity but also by their size (a low--activity, low IR/area, but large,  galaxy can have the same total luminosity as a high--activity, but small, galaxy). 

\begin{figure}
\figurenum{18}
\plotone{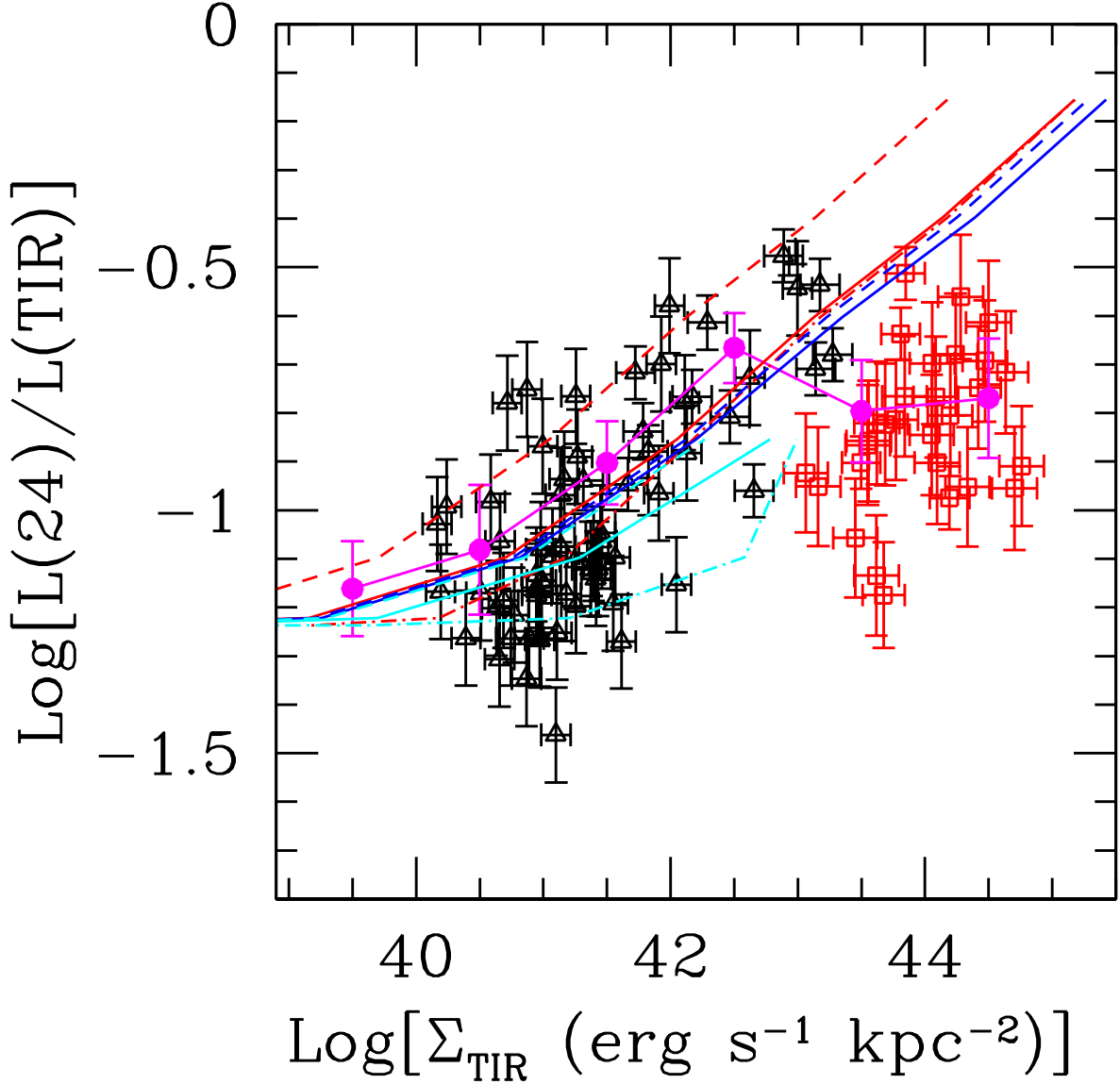}
%\plottwo{figure15a.eps}{figure15b.eps}
\caption{The L(24)/L(TIR) ratio as a function of the IRSD for the galaxies in our sample with 
oxygen abundance 12$+$Log(O/H)$>$8.5 (black and red points). The magenta points and 
line, and the models lines (blue, red, and cyan) are as in Figure~\ref{fig4}, left. The trend for the galaxies with 12$+$Log(O/H)$>$8.5  is similar to that of the whole sample with 12$+$Log(O/H)$>$8.1.
\label{fig18}}
\end{figure}

The relation between the effective temperature and the luminosity/area is present in 
galaxies up to  an IRSD $\Sigma_{TIR}\sim$0.3--1$\times$10$^{43}$~erg~s$^{-1}$~kpc$^{-2}$ (L(TIR)$\approx$1--3$\times$10$^{44}$~erg~s$^{-1}$$\sim$3--8$\times$10$^{10}$~L$_{\odot}$ or L(24)$\approx$2--6$\times$10$^{43}$~erg~s$^{-1}$$\sim$0.6--2$\times$10$^{10}$~L$_{\odot}$),  independent of the metal content of the galaxy  (Figures~\ref{fig3}, left, and \ref{fig17}).  This can be further seen in Figure~\ref{fig18}, 
where we plot L(24)/L(TIR) as a function of IRSD only for those galaxies in our sample that have  oxygen abundance 12$+$Log(O/H)$>$8.5. There are 66 normal star--forming and starburst galaxies above this oxygen abundance  value (or about 1/2 
of the normal star--forming and starburst galaxies in our high--metallicity sample), and the 29 
LIRGS.  A similar trend to that of Figure~\ref{fig4} of L(24)/L(TIR) as a function of IRSD is observed for this restricted, higher--metallicity, sample, implying that the effect is not primarily driven by the abundance of dust in the galaxy. 

The combination of the two effects discussed above, increasing transparency of the galaxy and decreasing effective dust temperature for decreasing luminosity/area,  may account for why different published  calibrations for SFR(24) diverge when extrapolated towards the low luminosity end of our sample  (Figure~\ref{fig6}).  

A more robust approach appears to involve measuring SFRs by combining the 24~$\mu$m emission and the observed 
H$\alpha$ emission, the first probing the dust--absorbed star formation and the second the unabsorbed one \citep{Kennicutt2007,Calzetti2007,Kennicutt2009}. Even in this case, though, a transition in the proportionality factor between 
L(24) and L(H$\alpha$)$_{obs}$ needs to be implemented for estimating accurate SFRs (equation~17, first and second terms). 
The transition point is marked by the condition L(H$\alpha$)$_{obs}$/(0.020 L(24))$\approx$1, which corresponds, for 
decreasing luminosity, to the transition from galaxies dominated by dust--obscured star formation to those dominated by 
unobscured star formation \citep[e.g.,][]{Buat2007}.  The transition corresponds to a change of roughly 50\% in the fraction of L(24) to be added to L(H$\alpha$)$_{obs}$, from 0.031 to 0.020 for L(24)$\lesssim$4$\times$10$^{42}~$erg~s$^{-1}$ (first and second relation in equation~17), or 
SFR$\sim$1~M$_{\odot}$~yr$^{-1}$. Our results thus indicate that around this SFR, 
the stellar population 
that dominates the heating of the dust transitions from being a relatively young one (constant star formation over the past 
$\sim$100~Myr), typical of starburst events, to a much more evolved one (constant star formation over the past 
$\sim$10~Gyr, or exponentially decreasing star formation), typical of widespread, but isolated, star formation amid an otherwise quiescent galaxy. In the latter case, 
the stellar populations not associated with the current star formation events contribute, in a proportionally larger fraction than 
the starburst case, to the heating of the 24~$\mu$m--emitting dust, whose emission contribution to the SFR calibration needs 
then to be accordingly removed, on average, by reducing L(24) via a smaller multiplicative factor \citep{Kennicutt2009} 
than the starburst or HII~region case \citep{Calzetti2007}. 

\subsection{The Emission at 70 and 160~$\mu$m as SFR Indicators}

Our reference SFR calibration (equation~17) enables the testing of other IR monochromatic luminosities as potential 
SFR indicators, over a wide range of luminosity surface densities.  
In this work, we have investigated the applicability of L(70) and L(160), i.e., the long wavelength  Spitzer bands, as such indicators. These two wavelength regions are attractive because each contains a larger fraction of the total 
infrared emission than the 24~$\mu$m emission. L(70) is 50\% of L(TIR) for  infrared LSD $\Sigma_{TIR}\gtrsim$10$^{41}$~erg~s$^{-1}$~kpc$^{-2}$, and  L(160) is about 20\% or larger fraction  of L(TIR) for the entire luminosity range investigated in this work (Figure~\ref{fig5}), while  L(24) only contains $\sim$10--20\%  
of the total infrared emission (Figure~\ref{fig4}). At the high luminosity LIRGs regime, both L(70) and L(160) 
represent a larger  fraction of L(TIR) than expected from models \citep[e.g.,][]{Draine2007}. This indicates 
a `cooler' (larger effective temperature) IR SED for those systems than predicted from models, an effect already previously observed \citep{Dunne2001}, and in line with the re--emission of the energy absorbed in the mid--IR wavelength range. Galaxies in the 
LIRG luminosity regime (L(TIR)$\gtrsim$1--3$\times$10$^{44}$~erg~s$^{-1}$ or $\Sigma_{TIR}\gtrsim$0.3--1$\times$10$^{43}$~erg~s$^{-1}$~kpc$^{-2}$) have their energy output dominated by a star formation event that is evolving in a dusty 
region of sufficient optical depth that even light emerging in the mid--IR gets absorbed, and that light is re--emitted at longer infrared wavelengths, thus producing an overall cool IR SED, specifically cooler 
than expected from models where the emission is emerging from a transparent (at IR wavelengths) 
medium. This requires that the dust attenuation at 24~$\mu$m be substantial, $\approx$1~mag, 
which translates to A$_V\sim$15--50~mag, depending on the extinction curve adopted \citep{DraineLi2007, Flaherty2007}. Such large extinctions are not uncommon in bright infrared galaxies.  
\citet{Genzel1998} determines that Ultraluminous InfraRed Galaxies (ULIRGs) have extinction values 
in the range A$_{V,screen}$=5--50~mag or A$_{V,mixed}$=50--1000~mag. Although LIRGs are 
typically an order of magnitude less luminous than ULIRGs, dust extinctions are still expected to be 
large in these systems. 

The 70~$\mu$m emission from galaxies correlates linearly with the SFR for luminosities L(70))$\gtrsim$1.4$\times$10$^{42}$~erg~s$^{-1}$, and a calibration is given in equation~22. Similarly, L(160) is linearly correlated with the SFR for 
luminosities L(160)$>$10$^{42}$~erg~s$^{-1}$ (equation~25). 

\subsection{Scatter and the Impact of Metallicity}

\begin{figure}
\figurenum{19}
\plottwo{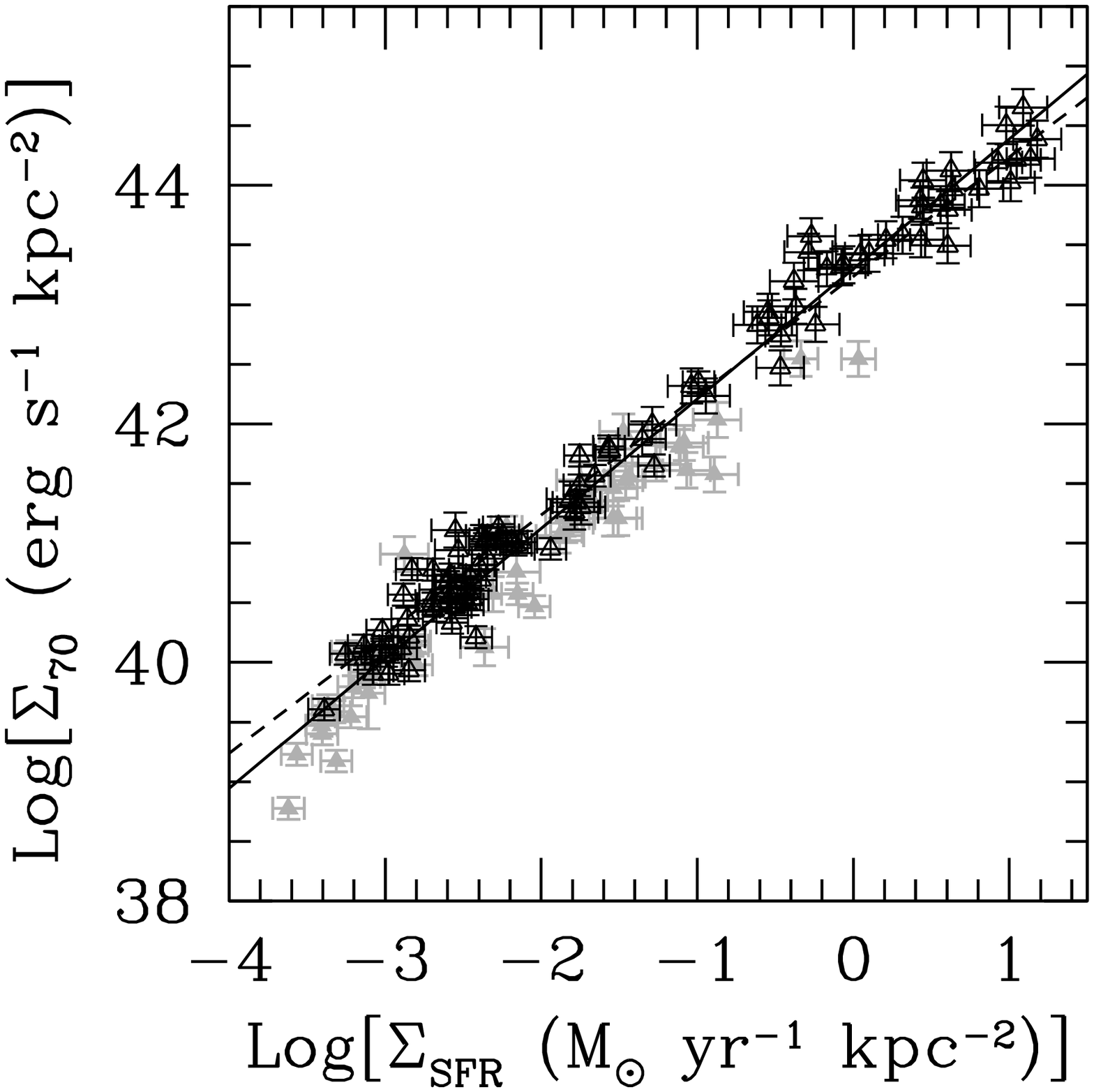}{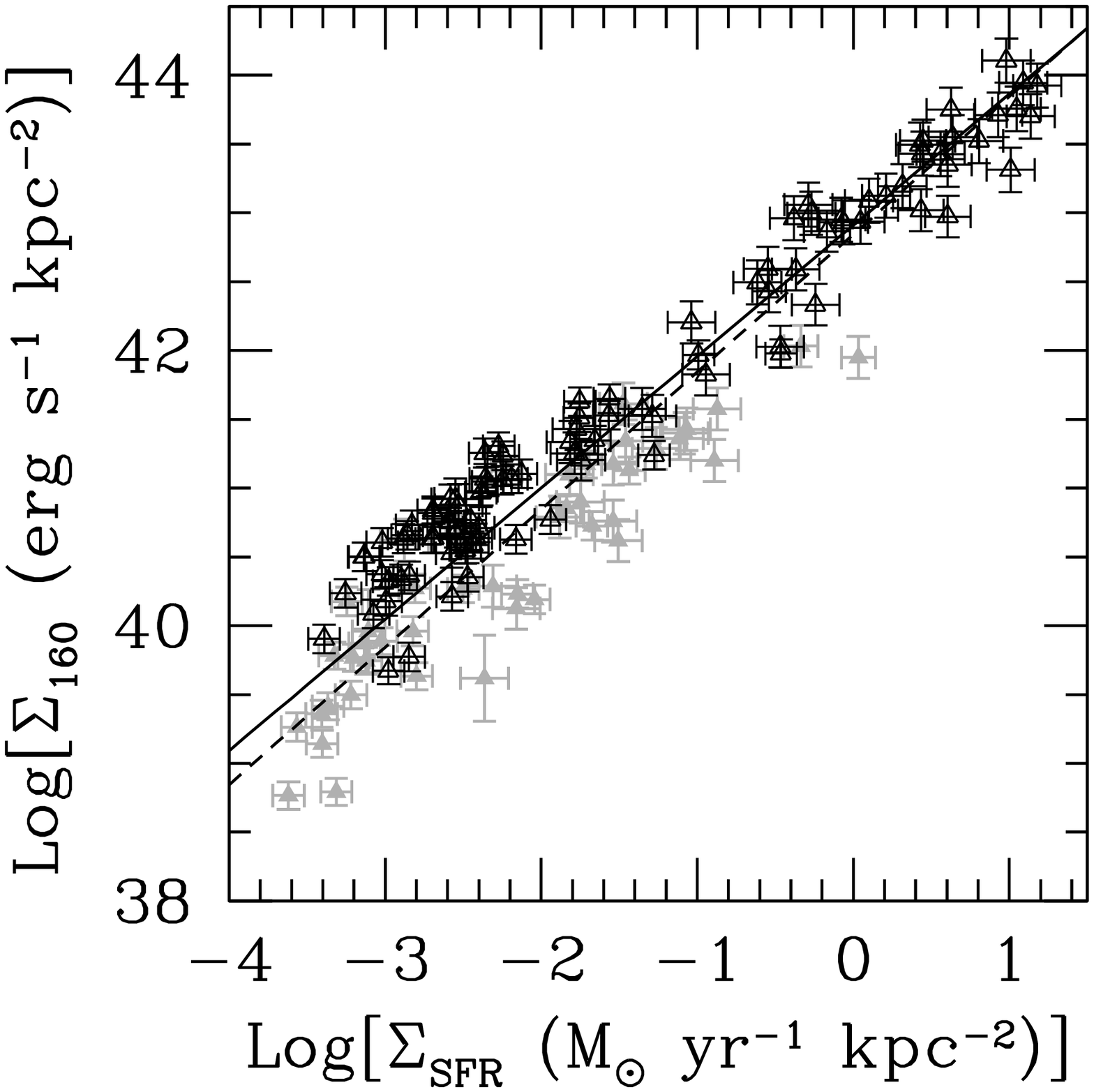}
%\plottwo{figure15a.eps}{figure15b.eps}
\caption{The 70~$\mu$m LSD {\bf (left)}  and 160~$\mu$m LSD {\bf (right)} as a function of 
$\Sigma_{SFR}$. The high--metallicity data have been divided into two subsamples according to 
the galaxies oxygen abundance: 8.1$<$12$+$Log(O/H)$\le$8.5 (grey triangles) and 12$+$Log(O/H)$>$8.5 (black triangles). See text for an explanation of the selected metallicity boundary. 
The scatter in the data decreases when restricting the sample to the 
highest metallicity values. The continuous and dashed lines in both panels are the best fit (equations~18 and 23) and the unity slope lines, respectively. 
\label{fig19}}
\end{figure}

Both calibrations, however, have significant scatter, 
$\sigma_{70}\sim$0.2~dex and $\sigma_{160}\sim$0.4~dex, around the mean trend, and larger than the scatter, 
$\sim$0.12--0.16~dex, 
of calibrations based on L(24) or a mix of L(24) and L(H$\alpha$)$_{obs}$ \citep{Rieke2009,Kennicutt2009}. Furthermore, 
the scatter increases for increasing wavelength, and appears to decrease for increasing luminosity (Figures~\ref{fig9} and 
\ref{fig15}). To further investigate these trends, we subdivide the high--metallicity sample into 
two sub--samples of oxygen abundance higher and lower than 12$+$Log(O/H)$=$8.5. This 
separating value is chosen after Figure~5 of \citet{Engelbracht2008}, which shows that galaxies 
above a metallicity of 12$+$Log(O/H)$=$8.5 tend to have a roughly constant f$_{\nu}$(70)/f$_{\nu}$(160) flux ratio (albeit with a large scatter), while galaxies below that metallicity value 
tend to have an increasing ratio for decreasing oxygen abundance. Figure~\ref{fig19} shows the 
high--metallicity data from our sample divided into the two sub--samples. The scatter of the data for both  $\Sigma_{70}$ and $\Sigma_{160}$ at constant $\Sigma_{SFR}$ decreases when 
the sample is restricted to the highest metallicity data (12$+$Log(O/H)$>$8.5); specifically, 
faint  $\Sigma_{70}$ and $\Sigma_{160}$  galaxies at fixed $\Sigma_{SFR}$ drop from 
the sub--sample. This is in the expected direction, as seen earlier in this paper when analyzing the 
impact of low--metallicity galaxies (Figure~\ref{fig12}). 

The results of Figure~\ref{fig19} should be considered  preliminary at this stage, since the oxygen abundances for a number of high metallicity galaxies are uncertain (sometimes by about 0.2~dex). Nevertheless, they confirm and support the results of section~5 for SFR(70), in the sense that as the galaxy metallicity increases and its dispersion in the sample decreases,  $\Sigma_{70}$ tends to approach a 1--to--1 relation with $\Sigma_{SFR}$ (Figure~\ref{fig19}, left).  Conversely, in the same metallicity range, $\Sigma_{160}$ tends to show a flattening in its relation with  $\Sigma_{SFR}$ and the dispersion of the datapoints around the mean trend remains large (Figure~\ref{fig19}, right). As already discussed in section~6, this is 
 in agreement with the expectation that the emission at long infrared 
wavelengths receives an increasing contribution (and an increasing scatter) from dust heated by evolved stellar populations 
unassociated with the current star formation event \citep{Lonsdale1987}.  Figure~\ref{fig19} (right) 
further stresses that the calibration of SFR(160), equation~25, should only be used for L(160)$\gtrsim$2$\times$10$^{42}$~erg~s$^{-1}$, equivalent to $\Sigma$(160)$\gtrsim$10$^{41}$~erg~s$^{-1}$~kpc$^{-2}$.

The calibrations just quoted apply to galaxies with mean gas--phase metallicity 12$+Log$(O/H)$\gtrsim$8.1. The scatter becomes even larger, and strongly asymmetric, when galaxies 
of lower oxygen abundance value are included (Figure~\ref{fig12}). Indeed, the general expectation is that lower metallicity 
galaxies will in general contain less dust and be more transparent than high metallicity ones.  In this case, the infrared will 
be a poor tracer of the current SFR, as most of the light produced by recently formed stars will emerge unabsorbed by dust. 
Thus, calibrations of SFR(70) and SFR(160) need to be applied with an awareness of these limitations.

A hybrid approach, which combines, e.g., L(70) with L(H$\alpha$)$_{obs}$, could likely provide a `remedy' for the dependence 
of far--infrared SFR indicators on metallicity, similarly to what has been already verified for L(24) and L(8), where the H$\alpha$ observed luminosity accounts for the unabsorbed portion of the SFR \citep{Calzetti2007, Kennicutt2007, Zhu2008, Kennicutt2009}. However, our current data do not enable us to test such hybrid indicator. We already use, in our analysis, the combination of 
L(24) and L(H$\alpha$)$_{obs}$ to trace SFR in galaxies in an unbiased fashion; this indicator is dominated by the 
H$\alpha$ luminosity at low SFSDs, as would be any hybrid combination of L(70) and L(H$\alpha$)$_{obs}$, thus resulting in  
a degeneracy for any test we may attempt. 

\subsection{Limits of Validity for the Calibrations}

For the high-metallicity galaxies, the presence of a contribution to the long wavelength infrared emission from dust heated 
by populations unassociated with the current star formation becomes more evident for decreasing galaxy luminosity  
(Figures~\ref{fig11} and \ref{fig15}), but it also competes with at least three more effects: (1) the effective dust temperature 
as measured from the IR SED tends to decrease (Figures~\ref{fig4}, \ref{fig5}, and \ref{fig17}); (2) the overall interstellar medium becomes 
progressively more transparent, thus a lower fraction of the SFR is traced in the infrared; (3) the area filling factor of star forming 
regions within galaxies decreases down to $\sim$10\%--50\% (depending on the dust extinction 
and stellar population models adopted), contributing 
to a change in the proportionality between $\Sigma_{70}$ or $\Sigma_{160}$ with $\Sigma_{SFR}$ relative to that of more 
luminous galaxies. These four effects all contribute, at various levels, to change the relation between $\Sigma_{70}$ (or 
$\Sigma_{160}$) with $\Sigma_{SFR}$ from linear to non--linear (e.g., equation~21), thus complicating any SFR calibration 
for luminosities below L(70)$\sim$1.4$\times$10$^{42}$~erg~s$^{-1}$ (L(160)$\sim$2$\times$10$^{42}$~erg~s$^{-1}$). 
These luminosity values correspond to SFR$\sim$0.1--0.3~M$_{\odot}$~yr$^{-1}$, thus they are not extremely restrictive especially in the context of current studies of distant galaxy populations. 

\subsection{Additional Uncertainties}

So far, we have not considered another potential source of heating for the dust associated with the diffuse interstellar medium of galaxies: photons leaking out of HII regions. About 20\%--50\% of the 
integrated H$\alpha$ luminosity in galaxies is associated with the diffuse interstellar medium 
\citep{Reynolds1990,Ferguson1996,Wang1997,Martin1997}, and those photons will produce dust heating. Since photons leaking out of HII regions are associated with the current star formation, 
in our analysis we have assumed that the diffuse ionized gas  follows the same scaling relations 
as the  star--forming regions, albeit possibly with different absolute values (e.g., 
the diffuse photons may heat the dust to typically lower temperatures than those spatially associated 
with HII regions). This assumption is justified by the findings of \citet{Wang1998}, according to which 
the fraction of diffuse-to-total ionized gas remains relatively constant in galaxies of increasing SFR,
 from normal star--forming to starbursts,  once the  diffuse gas surface brightness is normalized by the 
 mean star formation rate per unit area of the galaxy. 

The infrared emission from the galaxies in our sample is dominated by dust heated by current star formation, with, in some cases, a non--negligible contribution from  evolved 
stellar populations. Central AGNs do not constitute a dominant heating source in the present sample, 
and their impact on our calibrations could not be investigated. Thus, the application of 
equations~22 (or 25) to galaxy populations first requires the evaluation of any contamination from 
dust heating by AGNs. 

Finally, all our derivations are predicated on the assumption that a universal stellar IMF can be applied to all galaxies.  While 
this is generally considered a reasonable assumption, recent investigations using  the Sloan 
Digital Sky Survey  galaxy sample \citep{Hoversten2008}, a complete sample of nearby galaxies 
\citep{Lee2009b}, and HI--selected galaxies \citep{Meurer2009} suggest that the galaxy--integrated  IMF may become increasingly bottom--heavy in low luminosity and/or low surface brightness galaxies. Changes in the IMF will clearly impact the calibration of SFR indicators across the full electromagnetic wavelength range, and establishing whether such variations are present and what physical parameters may 
be causing them is an avenue of future investigation.

\acknowledgments

This work is based in part on observations made with the Spitzer Space Telescope, which is operated by the Jet Propulsion Laboratory, California Institute of Technology under a contract with NASA. This work is part of LVL and SINGS, two of the Spitzer Space Telescope Legacy Science Programs; it has 
been partially supported by the JPL, Caltech, Contract Number 1316765. 

AGdP is partially financed by the Spanish Ram\'on y Cajal program and the Programa Nacional de Astronom'a y Astrof'sica under grant AYA 2006-02358. 

C.-N. Hao acknowledges the support from NSFC key project 10833006. 

We would like to thank an anonymous referee for many comments that have helped improve the manuscript.

This research has made use of the NASA/IPAC Extragalactic Database (NED) which is operated by the Jet Propulsion Laboratory, California Institute of Technology, under contract with the National Aeronautics and Space Administration.

%\appendix

%\section{Placeholder}

\clearpage

%% Use the figure environment and \plotone or \plottwo to include 
%% figures and captions in your electronic submission.

\clearpage 

\clearpage 

\clearpage 

\clearpage 

\clearpage 

\clearpage 

\clearpage 

\clearpage 

\clearpage 

\clearpage 

\clearpage 

\clearpage 

\clearpage 

\clearpage 

\clearpage 

\clearpage 

\clearpage 

\clearpage

\begin{deluxetable}{lrrrrrrrrrr}
\tablecolumns{11}
\rotate
\tabletypesize{\scriptsize}
\tablecaption{Properties of Normal Star--Forming and Starburst Galaxies.\label{tbl-1}}
\tablewidth{0pt}
\tablehead{
\colhead{Name\tablenotemark{a}} & \colhead{D\tablenotemark{b}} 
& \colhead{12$+$Log(O/H)\tablenotemark{c}} 
& \colhead{R$_{25}$\tablenotemark{d}}  & \colhead{R$_{H\alpha}$\tablenotemark{e}} 
& \colhead{Ref.\tablenotemark{e}}
& \colhead{Log($\Sigma_{SFR}$)\tablenotemark{f}} & \colhead{Log($\Sigma_{70}$)\tablenotemark{g}}
& \colhead{Log($\Sigma_{160}$)\tablenotemark{g}} 
& \colhead{Log($\Sigma_{TIR}$)\tablenotemark{g}} 
& \colhead{Ref.\tablenotemark{h}}
\\
\colhead{} & \colhead{(Mpc)} & \colhead{} & \colhead{($^{\prime\prime}$)} 
& \colhead{($^{\prime\prime}$)} & \colhead{} 
& \colhead{(M$_{\odot}$~yr$^{-1}$~kpc$^{-2}$)} 
& \colhead{(erg~s$^{-1}$~kpc$^{-2}$)} & \colhead{(erg~s$^{-1}$~kpc$^{-2}$)} 
& \colhead{(erg~s$^{-1}$~kpc$^{-2}$)} & \colhead{} 
\\
%\cline{1-11}\\
%\multicolumn{11}{c}{High Metallicity Galaxies}
}
\startdata
\cutinhead{High Metallicity Galaxies}
NGC0625     &  4.07&  8.1&  172.4&  60.2&  L  &$-$1.838$\pm$0.101& 41.10$\pm$0.09& 40.85$\pm$0.10& 41.43$\pm$0.11&K08,D09 \\
NGC1522     &  9.99&  8.2&  36.05&  13.5&  L  &$-$1.435$\pm$0.101& 41.56$\pm$0.09& 41.13$\pm$0.10& 41.80$\pm$0.11&K08,D09 \\
NGC1800     &  8.83& 8.36&  59.85&  19.6&  L  &$-$1.713$\pm$0.101& 41.34$\pm$0.09& 41.17$\pm$0.10& 41.62$\pm$0.11&K08,D09 \\
UGC4278     &  8.13& 8.08&  140.3&  144.8&  L  &$-$3.407$\pm$0.101& 39.47$\pm$0.09& 39.36$\pm$0.10& 39.77$\pm$0.11&K08,D09 \\
UGC5829     &  8.44&  8.3&  140.3&  140.8&  L  &$-$3.404$\pm$0.101& 39.40$\pm$0.09& 39.14$\pm$0.10& 39.63$\pm$0.11&K08,D09 \\
NGC3368     & 11.27& 9.04&  227.6&   84.8&  L  &$-$2.274$\pm$0.101& 41.13$\pm$0.09& 41.30$\pm$0.10& 41.61$\pm$0.11&K08,D09 \\
UGC5923     &  7.67& 8.26&     28&  19.6&  L  &$-$2.501$\pm$0.101& 40.65$\pm$0.09& 40.27$\pm$0.10& 40.83$\pm$0.12&K08,D09 \\
NGC3510     &  9.18& 8.08&  119.5&  99.6&  L  &$-$3.123$\pm$0.101& 39.92$\pm$0.09& 39.75$\pm$0.10& 40.19$\pm$0.11&K08,D09 \\
NGC3623     &  9.59& 9.06&  293.1&  125.2&  L  &$-$2.693$\pm$0.101& 40.49$\pm$0.09& 40.84$\pm$0.10& 41.11$\pm$0.11&K08,D09 \\
UGC6900     &     8&  8.1&   62.7&  27.8&  L  &$-$3.107$\pm$0.103& 39.74$\pm$0.30& 39.96$\pm$0.30& 40.26$\pm$0.38&K08,D09 \\
NGC4248     &  7.76& 8.15&   90.6&  27.4&  L  &$-$2.508$\pm$0.101& 40.66$\pm$0.09& 40.71$\pm$0.10& 41.08$\pm$0.11&K08,D09 \\
NGC4288     &  8.22& 8.52&  64.15&  48.2&  L  &$-$2.486$\pm$0.101& 40.61$\pm$0.09& 40.55$\pm$0.10& 40.95$\pm$0.11&K08,D09 \\
UGC7490     &     9& 8.46&  99.35&  83.5&  L  &$-$3.325$\pm$0.101& 39.59$\pm$0.09& 39.78$\pm$0.10& 40.08$\pm$0.11&K08,D09 \\
UGC7699     &  7.34& 8.15&  114.1&  89.6&  L  &$-$3.204$\pm$0.101& 39.80$\pm$0.09& 39.77$\pm$0.10& 40.15$\pm$0.11&K08,D09 \\
UGC7698     &   6.1& 8.04&  193.7&  145.2&  L  &$-$3.620$\pm$0.101& 38.78$\pm$0.09& 38.77$\pm$0.10& 39.14$\pm$0.12&K08,D09 \\
NGC4656     &   9.2& 8.78&    454&  282.6&  L  &$-$2.979$\pm$0.101& 39.90$\pm$0.09& 39.67$\pm$0.10& 40.17$\pm$0.11&K08,D09 \\
UGC7950     &  8.48& 8.37&  38.65&  40.4&  L  &$-$3.024$\pm$0.101& 40.11$\pm$0.09& 39.89$\pm$0.10& 40.35$\pm$0.12&K08,D09 \\
NGC4707     &  7.97& 8.43&  67.15&   67.2&  L  &$-$3.219$\pm$0.101& 39.54$\pm$0.09& 39.50$\pm$0.10& 39.89$\pm$0.12&K08,D09 \\
UGCA320     &  7.76& 8.08&  168.7&  155.2&  L  &$-$3.313$\pm$0.101& 39.17$\pm$0.09& 38.79$\pm$0.10& 39.36$\pm$0.12&K08,D09 \\
UGC8320     &  4.33& 8.29&  108.9&   95.2&  L  &$-$3.367$\pm$0.101& 39.64$\pm$0.09& 39.42$\pm$0.10& 39.87$\pm$0.12&K08,D09 \\
NGC5068     &  6.24& 8.82&  217.4&  206.7&  L  &$-$2.438$\pm$0.101& 40.53$\pm$0.09& 40.59$\pm$0.10& 40.94$\pm$0.11&K08,D09 \\
NGC5477     &  8.25& 8.14&   49.8&   55.6&  L  &$-$2.798$\pm$0.101& 39.98$\pm$0.09& 39.63$\pm$0.10& 40.17$\pm$0.12&K08,D09 \\
NGC0024     &  8.13& 8.62&  172.7&    143.&L,S  &$-$3.075$\pm$0.101& 39.91$\pm$0.09& 40.09$\pm$0.10& 40.39$\pm$0.11&K09,D07 \\
NGC0337     & 24.69& 8.56&   86.5&    109.&  S  &$-$2.399$\pm$0.101& 40.81$\pm$0.09& 40.71$\pm$0.10& 41.14$\pm$0.11&K09,D07 \\
NGC0628     &  7.32& 8.67&  314.1&    261.&L,S  &$-$2.583$\pm$0.101& 40.55$\pm$0.09& 40.75$\pm$0.10& 41.07$\pm$0.11&K09,D07 \\
NGC0925     &  9.12& 8.51&  314.1&    296.&  S  &$-$2.995$\pm$0.101& 40.06$\pm$0.09& 40.18$\pm$0.10& 40.51$\pm$0.11&K09,D07 \\
NGC1097     & 16.88& 8.78&   280.&    282.&  S  &$-$2.453$\pm$0.101& 40.72$\pm$0.09& 40.77$\pm$0.10& 41.16$\pm$0.11&K09,D07 \\
NGC1291     &  9.83& 8.72&  293.1&    329.&L,S  &$-$3.392$\pm$0.101& 39.61$\pm$0.09& 39.90$\pm$0.10& 40.19$\pm$0.11&K09,D07 \\
NGC1377     & 24.17& 8.37&  53.35&    44.3&  S  &$-$1.674$\pm$0.101& 41.36$\pm$0.09& 40.73$\pm$0.10& 41.74$\pm$0.11&K09,D07 \\
NGC1482     & 22.03& 8.53&  73.65&    59.1&  S  &$-$1.567$\pm$0.101& 41.81$\pm$0.09& 41.53$\pm$0.10& 42.11$\pm$0.11&K09,D07 \\
NGC1512     & 10.14& 8.71&  267.4&    185.&L,S  &$-$2.977$\pm$0.101& 40.13$\pm$0.09& 40.32$\pm$0.10& 40.63$\pm$0.11&K09,D07 \\
NGC1566     & 18.27& 8.82&  249.6&    269.&  S  &$-$2.704$\pm$0.101& 40.52$\pm$0.09& 40.63$\pm$0.10& 40.98$\pm$0.11&K09,D07 \\
NGC1705     &   5.1& 8.32&  57.15&    57.2&L,S  &$-$2.043$\pm$0.101& 40.47$\pm$0.09& 40.19$\pm$0.10& 40.69$\pm$0.11&K09,D07 \\
NGC2403     &  3.22& 8.56&  656.3&    413.&L,S  &$-$2.492$\pm$0.101& 40.55$\pm$0.09& 40.64$\pm$0.10& 40.99$\pm$0.11&K09,D07 \\
NGC2798     & 24.68& 8.69&   77.1&    26.0&  S  &$-$0.992$\pm$0.101& 42.35$\pm$0.09& 41.97$\pm$0.10& 42.62$\pm$0.11&K09,D07 \\
NGC2841     &  9.81& 8.86&  243.9&    209.&  S  &$-$2.418$\pm$0.101& 40.21$\pm$0.09& 40.64$\pm$0.10& 40.89$\pm$0.11&K09,D07 \\
NGC2915     &  3.78& 8.16&  57.15&    51.4& G03 &$-$2.152$\pm$0.101& 40.57$\pm$0.09& 40.23$\pm$0.10& 40.77$\pm$0.11&G03,D07 \\
NGC2976     &  3.56& 8.64&  176.7&    117.&L,S  &$-$2.189$\pm$0.101& 41.02$\pm$0.09& 41.07$\pm$0.10& 41.43$\pm$0.11&K09,D07 \\
NGC3049     &  19.1& 8.75&  65.65&    81.3&  S  &$-$2.471$\pm$0.101& 40.49$\pm$0.09& 40.35$\pm$0.10& 40.87$\pm$0.11&K09,D07 \\
NGC3031     &  3.63& 8.69&  692.7&    680.&L,S  &$-$3.020$\pm$0.101& 40.11$\pm$0.09& 40.37$\pm$0.10& 40.65$\pm$0.11&P06,D07 \\
NGC3034     &  3.89& 8.72&  336.6&    143.&L,S  &$-$0.463$\pm$0.101& 42.74$\pm$0.09& 41.98$\pm$0.10& 43.00$\pm$0.11&K09,D07 \\
NGC3190     & 16.81&  8.7&  130.9&    103.&  S  &$-$2.882$\pm$0.101& 40.57$\pm$0.09& 40.63$\pm$0.10& 40.97$\pm$0.11&K09,D07 \\
NGC3184     &  8.53& 8.81&  222.4&    217.&  S  &$-$2.859$\pm$0.101& 40.37$\pm$0.09& 40.66$\pm$0.10& 40.95$\pm$0.11&K09,D07 \\
NGC3198     & 13.68&  8.6&  255.4&    233.&  S  &$-$2.897$\pm$0.101& 40.12$\pm$0.09& 40.34$\pm$0.10& 40.66$\pm$0.11&K09,D07 \\
NGC3265     & 19.49& 8.65&  38.65&    44.3&  S  &$-$2.162$\pm$0.101& 40.99$\pm$0.09& 40.63$\pm$0.10& 41.26$\pm$0.11&K09,D07 \\
Mrk33           & 21.92& 8.56&    30.&    26.1&  S  &$-$1.276$\pm$0.101& 41.65$\pm$0.09& 41.24$\pm$0.10& 41.99$\pm$0.11&K09,D07 \\
NGC3351     &  9.33& 8.91&  222.4&    189.&L,S  &$-$2.597$\pm$0.101& 40.67$\pm$0.09& 40.76$\pm$0.10& 41.13$\pm$0.11&K08,D07 \\
NGC3521     &  8.99& 8.68&  328.9&    284.&L,S  &$-$2.583$\pm$0.101& 40.74$\pm$0.09& 40.92$\pm$0.10& 41.25$\pm$0.11&K08,D07 \\
NGC3621     &  6.64&  8.5&  369.1&    531.&  S  &$-$3.246$\pm$0.101& 40.09$\pm$0.09& 40.18$\pm$0.10& 40.53$\pm$0.11&K09,D07 \\
NGC3627     &  9.38&  8.8&  273.6&    244.&L,S  &$-$2.358$\pm$0.101& 41.03$\pm$0.09& 41.07$\pm$0.10& 41.45$\pm$0.11&K09,D07 \\
NGC3773     & 12.53& 8.64&  35.25&    35.1&  S  &$-$1.941$\pm$0.101& 40.95$\pm$0.09& 40.77$\pm$0.10& 41.27$\pm$0.11&K09,D07 \\
NGC3938     & 12.22& 8.71&  161.1&    191.&  S  &$-$2.687$\pm$0.101& 40.43$\pm$0.09& 40.64$\pm$0.10& 40.95$\pm$0.11&K09,D07 \\
NGC4125     & 22.91& 8.89&  172.7&     95.&  S  &$-$2.848$\pm$0.101& 39.93$\pm$0.09& 39.77$\pm$0.10& 40.24$\pm$0.11&K09,D07 \\
NGC4236     &  4.45& 8.31&  656.3&    582.&L,S  &$-$3.567$\pm$0.101& 39.23$\pm$0.09& 39.26$\pm$0.10& 39.63$\pm$0.11&K08,D07 \\
NGC4254     & 33.29&  8.8&  161.1&    185.&  S  &$-$2.126$\pm$0.101& 41.01$\pm$0.09& 41.10$\pm$0.10& 41.46$\pm$0.11&K09,D07 \\
NGC4321     & 14.32& 8.84&  222.4&    195.&  S  &$-$2.343$\pm$0.101& 40.87$\pm$0.09& 41.05$\pm$0.10& 41.37$\pm$0.11&K09,D07 \\
NGC4450     & 27.07& 8.85&  157.4&    113.&  S  &$-$3.019$\pm$0.101& 40.27$\pm$0.09& 40.61$\pm$0.10& 40.87$\pm$0.11&K09,D07 \\
NGC4536     & 14.45&  8.6&  227.6&    229.&  S  &$-$2.574$\pm$0.101& 40.63$\pm$0.09& 40.53$\pm$0.10& 40.99$\pm$0.11&K09,D07 \\
NGC4559     & 11.57& 8.51&  321.4&    267.&  S  &$-$2.845$\pm$0.101& 40.22$\pm$0.09& 40.36$\pm$0.10& 40.69$\pm$0.11&K09,D07 \\
NGC4569     &    16& 8.88&  286.5&    270.& K01 &$-$3.254$\pm$0.101& 40.07$\pm$0.09& 40.24$\pm$0.10& 40.58$\pm$0.11&B02,D07 \\
NGC4579     & 20.62& 8.93&  176.7&    116.&  S  &$-$2.387$\pm$0.101& 40.69$\pm$0.09& 40.97$\pm$0.10& 41.26$\pm$0.11&K09,D07 \\
NGC4594     &  9.33& 8.99&  261.3&    201.&L,S  &$-$3.138$\pm$0.101& 40.14$\pm$0.09& 40.50$\pm$0.10& 40.77$\pm$0.11&K09,D07 \\
NGC4625     &  9.51& 8.65&  65.65&    38.0&L,S  &$-$2.245$\pm$0.101& 41.00$\pm$0.09& 41.06$\pm$0.10& 41.41$\pm$0.11&K09,D07 \\
NGC4631     &  8.92& 8.44&  464.6&    257.&L,S  &$-$2.222$\pm$0.101& 41.14$\pm$0.09& 41.12$\pm$0.10& 41.51$\pm$0.11&K09,D07 \\
NGC4725     & 17.12& 8.73&  321.4&    240.&  S  &$-$3.132$\pm$0.101& 40.03$\pm$0.09& 40.50$\pm$0.10& 40.74$\pm$0.11&K09,D07 \\
NGC4736     &  5.01& 8.66&  336.6&    238.&L,S  &$-$2.380$\pm$0.101& 41.06$\pm$0.09& 40.98$\pm$0.10& 41.39$\pm$0.11&K08,D07 \\
NGC4826     &  5.54& 8.92&    300&     84.&L,S  &$-$1.753$\pm$0.101& 41.73$\pm$0.09& 41.63$\pm$0.10& 42.05$\pm$0.11&K09,D07 \\
NGC5055     &  8.32& 8.78&  377.7&    226.&L,S  &$-$2.361$\pm$0.101& 40.99$\pm$0.09& 41.25$\pm$0.10& 41.55$\pm$0.11&K09,D07 \\
NGC5194     &  8.13& 8.86&  336.6&    296.&L,S  &$-$2.244$\pm$0.101& 41.07$\pm$0.09& 41.23$\pm$0.10& 41.56$\pm$0.11&K09,D07 \\
Tol89            &  14.9& 8.54&  84.55&     81.&  S  &$-$2.572$\pm$0.101& 40.33$\pm$0.09& 40.21$\pm$0.10& 40.72$\pm$0.11&K09,D07 \\
NGC5474     &   6.8& 8.32&  143.6&    131.&L,S  &$-$2.810$\pm$0.101& 40.18$\pm$0.09& 40.27$\pm$0.10& 40.60$\pm$0.11&K09,D07 \\
NGC5713     & 26.49& 8.64&  82.65&    52.2&  S  &$-$1.563$\pm$0.101& 41.78$\pm$0.09& 41.65$\pm$0.10& 42.13$\pm$0.11&K09,D07 \\
NGC5866     & 15.14& 8.72&  140.3&    100.&  S  &$-$2.830$\pm$0.101& 40.78$\pm$0.09& 40.73$\pm$0.10& 41.10$\pm$0.11&K09,D07 \\
IC4710          &  8.48& 8.37&  108.9&    109.&  S  &$-$2.823$\pm$0.101& 40.14$\pm$0.09& 39.96$\pm$0.10& 40.42$\pm$0.11&K09,D07 \\
NGC6822     &  0.47& 8.35&  464.6&    765.&  S  &$-$3.133$\pm$0.101& 39.88$\pm$0.09& 39.87$\pm$0.10& 40.24$\pm$0.11&K09,D07 \\
NGC6946     &   6.8& 8.72&  344.4&    217.&  S  &$-$1.755$\pm$0.101& 41.49$\pm$0.09& 41.51$\pm$0.10& 41.91$\pm$0.11&K09,D07 \\
NGC7331     & 14.52&  8.7&  314.1&    295.&  S  &$-$2.693$\pm$0.101& 40.78$\pm$0.09& 40.82$\pm$0.10& 41.18$\pm$0.11&K09,D07 \\
NGC7552     & 22.27& 8.74&  101.7&    113.&  S  &$-$1.657$\pm$0.101& 41.57$\pm$0.09& 41.35$\pm$0.10& 41.93$\pm$0.11&K09,D07 \\
NGC7793     &  3.82& 8.53&    255&    286.&L,S  &$-$2.549$\pm$0.101& 40.46$\pm$0.09& 40.67$\pm$0.10& 40.97$\pm$0.11&K09,D07 \\
Mrk170      &  20.4& 8.09&   31.4&  18.1&  A  &$>-$2.877$\pm$0.153& 40.91$\pm$0.15& 40.62$\pm$0.15& 41.14$\pm$0.15&...,E08 \\
Mrk930      &    77& 8.11&   25.8&   14.9&  A  &$>-$1.748$\pm$0.153& 41.53$\pm$0.15& 40.90$\pm$0.15& 41.79$\pm$0.15&...,E08 \\
NGC1569     &   1.9& 8.13&  108.9&  85.9& H04 &$-$0.890$\pm$0.153& 41.58$\pm$0.15& 41.20$\pm$0.15& 41.93$\pm$0.15&H04,E08 \\
Mrk1094     &    41& 8.15&  20.75&  19.8& G03 &$-$1.540$\pm$0.153& 41.21$\pm$0.15& 40.76$\pm$0.15& 41.40$\pm$0.15&G03,E08 \\
NGC3310     &  21.3& 8.18&   92.7&  67.2& J04 &$-$1.263$\pm$0.153& 41.67$\pm$0.15& 41.29$\pm$0.15& 41.98$\pm$0.15&J04,E08 \\
Mrk162      &    98& 8.19&   18.1&  10.5&  A  &$>-$1.474$\pm$0.153& 41.94$\pm$0.15& 41.61$\pm$0.15& 42.23$\pm$0.15&...,E08 \\
NGC1156     &   7.8& 8.19&  99.35&  88.0& H04 &$-$2.159$\pm$0.153& 40.76$\pm$0.15& 40.13$\pm$0.15& 40.94$\pm$0.15&H04,E08 \\
Tol2        &    22& 8.22&     25&  19.4& G03 &$-$1.506$\pm$0.153& 41.21$\pm$0.15& 40.62$\pm$0.15& 41.38$\pm$0.15&G03,E08 \\
MinkObj     &    78& 8.22&     15&  13.9& C06 &$-$2.364$\pm$0.153& 40.13$\pm$0.15& 39.62$\pm$0.31& 40.34$\pm$0.32&C06,E08 \\
NGC4449     &   4.2& 8.23&    185&  133.5&  L  &$-$1.819$\pm$0.153& 41.17$\pm$0.15& 41.10$\pm$0.15& 41.53$\pm$0.15&K08,E08 \\
NGC7714     &    40& 8.26&  57.15&  24.2& J04 &$-$0.873$\pm$0.153& 42.04$\pm$0.15& 41.58$\pm$0.15& 42.41$\pm$0.15&J04,E08 \\
UGC4703     &    57& 8.31&   13.7&    6.0& M99 &$-$1.068$\pm$0.153& 41.61$\pm$0.15& 41.43$\pm$0.15& 41.96$\pm$0.16&M99,E08 \\
NGC1140     &  21.2& 8.32&   49.8&   44.0& H94 &$-$1.881$\pm$0.153& 41.06$\pm$0.15& 40.79$\pm$0.15& 41.37$\pm$0.15&H94,E08 \\
NGC1510     &  11.8& 8.33&  39.55&  10.9&  S  &$-$1.114$\pm$0.153& 41.81$\pm$0.15& 41.36$\pm$0.15& 42.06$\pm$0.15&K09,E08 \\
NGC3125     &    12& 8.34&   32.9&  22.9& G03 &$-$1.083$\pm$0.153& 41.84$\pm$0.15& 41.40$\pm$0.15& 42.11$\pm$0.15&G03,E08 \\
NGC4214     &   2.9& 8.36&  255.4&  188.6& H04 &$-$2.310$\pm$0.153& 40.57$\pm$0.15& 40.29$\pm$0.15& 40.86$\pm$0.15&K08,E08 \\
NGC4670     &  23.2& 8.38&   42.4&  28.1& G03 &$-$1.539$\pm$0.153& 41.46$\pm$0.15& 41.18$\pm$0.15& 41.73$\pm$0.15&G03,E08 \\
He2$-$10      &     9& 8.55&  52.15&   20.7& J00 &$-$0.468$\pm$0.153& 42.47$\pm$0.15& 42.03$\pm$0.15& 42.89$\pm$0.15&J00,E08 \\
NGC3628     &  13.1& 8.57&  443.8&    240.& F90 &$-$2.530$\pm$0.153& 40.94$\pm$0.15& 40.88$\pm$0.15& 41.31$\pm$0.15&K08,E08 \\
NGC3079     &  21.8& 8.57&  238.3&    180.& R07 &$-$2.551$\pm$0.153& 41.11$\pm$0.15& 40.92$\pm$0.15& 41.38$\pm$0.15&L96,E08 \\
NGC2782     &    42& 8.59&    104&     30.& E96 &$-$1.354$\pm$0.153& 41.87$\pm$0.15& 41.58$\pm$0.15& 42.17$\pm$0.15&M06,E08 \\
NGC3077     &   3.8&  8.6&  161.1&  27.6& J04 &$-$0.944$\pm$0.153& 42.24$\pm$0.15& 41.83$\pm$0.15& 42.47$\pm$0.15&K08,E08 \\
NGC3367     &    49& 8.62&  75.35&  50.3& G01 &$-$1.744$\pm$0.153& 41.30$\pm$0.15& 41.21$\pm$0.15& 41.72$\pm$0.15&M06,E08 \\
NGC5236     &   4.5& 8.62&  386.4&    305.& L99 &$-$1.774$\pm$0.153& 41.34$\pm$0.15& 41.43$\pm$0.15& 41.82$\pm$0.15&K08,E08 \\
NGC5953     &    35& 8.67&  48.65&     17.& H03 &$-$1.037$\pm$0.153& 42.32$\pm$0.15& 42.20$\pm$0.15& 42.66$\pm$0.15&K87,E08 \\
NGC4194     &    42& 8.67&   54.6&   13.5&HT04 &$-$0.242$\pm$0.153& 42.84$\pm$0.15& 42.33$\pm$0.15& 43.17$\pm$0.15&HT04,E08\\
NGC2903     &   8.9& 8.68&  377.7&    146.& B02 &$-$1.814$\pm$0.153& 41.37$\pm$0.15& 41.33$\pm$0.15& 41.78$\pm$0.15&K08,E08 \\
Mrk25       &    48& 8.68&   16.5&   9.5& A   &$>-$1.287$\pm$0.153& 41.99$\pm$0.15& 41.52$\pm$0.15& 42.29$\pm$0.15&...,E08 \\
Mrk331      &    78& 8.76&   20.3&     10.&HT04 &$-$0.368$\pm$0.153& 42.99$\pm$0.15& 42.59$\pm$0.15& 43.27$\pm$0.15&HT04,E08\\
IC342       &   3.3& 8.85&  641.4&    380.& H05 &$-$1.787$\pm$0.153& 41.26$\pm$0.15& 41.25$\pm$0.15& 41.67$\pm$0.15&W99,E08 \\
IIZw40      &   9.2& 8.11&   12.5&  10.4&G03  &$-$0.015$\pm$0.108& 42.54$\pm$0.15& 41.95$\pm$0.15& 42.94$\pm$0.15&C07,E08 \\
NGC5253     &     4& 8.19&   150.&  22.2&C99  & 0.017$\pm$0.108& 42.95$\pm$0.15& 42.43$\pm$0.15& 43.39$\pm$0.15&C07,E08 \\
NGC2537     &   6.9& 8.44&  52.15&  28.4&J04  &$-$1.506$\pm$0.108& 41.53$\pm$0.15& 41.34$\pm$0.15& 41.83$\pm$0.15&C07,E08 \\
NGC2146     &  17.9& 8.68&  180.8&  35.5&J04  &$-$0.540$\pm$0.108& 42.89$\pm$0.15& 42.43$\pm$0.15& 43.15$\pm$0.15&C07,E08 \\
\cutinhead{Low Metallicity Galaxies}
WLM         &  0.92& 7.77&  344.4&  107.9&  L  &$-$2.751$\pm$0.101& 40.08$\pm$0.09& 40.00$\pm$0.10& 40.40$\pm$0.11&K08,D09 \\
UGC5272     &  7.61& 7.83&   62.7&   62.6&  L  &$-$2.991$\pm$0.101& 39.74$\pm$0.09& 39.38$\pm$0.10& 39.93$\pm$0.12&K08,D09 \\
NGC3109     &  1.34& 7.77&  571.6&  286.7&  L  &$-$3.020$\pm$0.101& 39.76$\pm$0.09& 39.65$\pm$0.10& 40.06$\pm$0.11&K08,D09 \\
SextansA    &  1.32& 7.54&  176.7&  144.3&  L  &$-$3.015$\pm$0.101& 39.37$\pm$0.09& 39.15$\pm$0.10& 39.62$\pm$0.12&K08,D09 \\
UGC5764     &  7.59& 7.95&  59.85&  45.21&  L  &$-$3.285$\pm$0.102& 39.31$\pm$0.10& 39.11$\pm$0.11& 39.59$\pm$0.12&K08,D09 \\
UGCA281     &   5.7&  7.8&  24.95&  26.08&  L  &$-$1.739$\pm$0.101& 40.66$\pm$0.09& 39.89$\pm$0.10& 40.87$\pm$0.12&K08,D09 \\
UGC8508     &  2.69& 7.89&  50.95&   56.8&  L  &$-$3.141$\pm$0.101& 39.50$\pm$0.09& 39.26$\pm$0.10& 39.74$\pm$0.12&K08,D09 \\
UGC8651     &  3.02& 7.85&  70.35&  83.03&  L  &$-$3.500$\pm$0.102& 38.87$\pm$0.10& 38.76$\pm$0.11& 39.18$\pm$0.13&K08,D09 \\
UGC8837     &  8.89&  7.7&    128&  93.46&  L  &$-$3.308$\pm$0.101& 39.36$\pm$0.09& 39.43$\pm$0.10& 39.78$\pm$0.12&K08,D09 \\
UGC9128     &  2.24& 7.75&   49.8&  23.91&  L  &$-$3.379$\pm$0.307& 39.66$\pm$0.31& 39.88$\pm$0.31& 40.18$\pm$0.44&K08,D09 \\
UGC9240     &   2.8& 7.95&   54.6&  39.99&  L  &$-$2.891$\pm$0.101& 40.19$\pm$0.09& 39.97$\pm$0.10& 40.47$\pm$0.12&K08,D09 \\
Mrk475      &  9.66& 7.97&  11.69&  10.43&  L  &$-$1.566$\pm$0.101& 40.86$\pm$0.09& 40.06$\pm$0.11& 40.99$\pm$0.13&K08,D09 \\
%\cutinhead{Low Metallicity Galaxies}
UGC9992     &  9.17& 7.88&  48.65&  45.64&  L  &$-$3.378$\pm$0.101& 39.59$\pm$0.09& 39.30$\pm$0.10& 39.83$\pm$0.12&K08,D09 \\
ESO347$-$G17  & 10.04&  7.9&   41.4&  40.39&  L  &$-$2.779$\pm$0.101& 40.07$\pm$0.09& 39.73$\pm$0.10& 40.26$\pm$0.12&K08,D09 \\
UGC12613    &  0.76& 7.93&  150.4&  86.51&  L  &$-$3.891$\pm$0.101& 39.51$\pm$0.09& 39.69$\pm$0.10& 40.01$\pm$0.12&K08,D09 \\
UGCA442     &  4.27& 7.72&  106.5&  101.4&  L  &$-$3.500$\pm$0.101& 38.91$\pm$0.09& 38.89$\pm$0.10& 39.30$\pm$0.12&K08,D09 \\
HolmII      &  3.39& 7.88&  238.3&   224.& L,S &$-$2.976$\pm$0.101& 39.71$\pm$0.09& 39.43$\pm$0.10& 39.95$\pm$0.11&K09,D07 \\
M81DwA      &  3.55& 7.49&   0.01&   39.3& L,S &$-$3.595$\pm$0.101& 39.83$\pm$0.09& 39.46$\pm$0.10& 40.10$\pm$0.11&K09,D07 \\
DDO053      &  3.56& 7.82&  46.45&   43.2& L,S &$-$2.493$\pm$0.101& 40.18$\pm$0.09& 39.92$\pm$0.10& 40.44$\pm$0.11&K09,D07 \\
HolmI       &  3.84& 8.00&  108.9&    83.& L,S &$-$3.302$\pm$0.101& 39.62$\pm$0.09& 39.60$\pm$0.10& 39.96$\pm$0.11&K09,D07 \\
HolmIX          &  3.28&  7.8&  75.35&   29.4& L,S &$-$2.712$\pm$0.101& 40.28$\pm$0.09& 40.22$\pm$0.10& 40.70$\pm$0.11&K09,D07 \\
M81DwB      &  7.08& 8.02&  26.15&   20.1& L,S &$-$2.418$\pm$0.101& 40.42$\pm$0.09& 40.47$\pm$0.10& 40.83$\pm$0.11&K09,D07 \\
DDO154      &   4.3& 7.78&   90.6&   63.5& L,S &$-$3.346$\pm$0.101& 39.06$\pm$0.09& 39.43$\pm$0.10& 39.71$\pm$0.11&K09,D07 \\
DDO165      &  4.57& 7.80&    104&   25.3& L,S &$-$2.659$\pm$0.101& 40.22$\pm$0.09& 40.20$\pm$0.10& 40.62$\pm$0.11&K09,D07 \\
IC2574         &   2.8&  7.93&  395.5&    305.&L,S  &$-$3.237$\pm$0.101& 39.62$\pm$0.09& 39.58$\pm$0.10& 39.97$\pm$0.11&K08,D07 \\
NGC5408     &  4.81& 8.02&  48.65&   71.3&  S  &$-$1.991$\pm$0.101& 40.69$\pm$0.09& 40.19$\pm$0.10& 40.93$\pm$0.11&K09,D07 \\
IZw18       &  12.6& 7.19&   8.85&   8.39& G03 &$-$1.349$\pm$0.153& 40.51$\pm$0.15& 40.51$\pm$0.32& 40.98$\pm$0.32&G03,E08 \\
Tol65       &    34& 7.45&  10.45&   4.85& G03 &$-$0.957$\pm$0.153& 40.87$\pm$0.16& 40.99$\pm$0.32& 41.63$\pm$0.32&G03,E08 \\
UGC4483     &   3.2& 7.55&  33.65&  16.35& G03 &$-$1.939$\pm$0.154& 40.43$\pm$0.15& 39.95$\pm$0.19& 40.60$\pm$0.19&K08,E08 \\
ESO146$-$G14  &  23.8& 7.66&   84.6&  48.84&  A  &$>-$4.506$\pm$0.156& 39.45$\pm$0.15& 39.30$\pm$0.17& 39.72$\pm$0.17&...,E08 \\
Mrk178      &   4.7& 7.82&   36.9&     36& G03 &$-$2.141$\pm$0.154& 39.97$\pm$0.15& 39.46$\pm$0.16& 40.07$\pm$0.16&K08,E08 \\
Mrk153      &    41& 7.83&   24.4&  14.09&  A  &$>-$2.491$\pm$0.153& 40.91$\pm$0.15& 40.34$\pm$0.16& 41.14$\pm$0.16&...,E08 \\
UM462       &  13.4& 7.91&  19.35&  11.17&  A  &$>-$1.691$\pm$0.153& 41.72$\pm$0.15& 41.15$\pm$0.15& 41.95$\pm$0.15&...,E08 \\
Haro11      &    87& 7.92&   14.7&    4.3&  H  & 0.604$\pm$0.153& 43.28$\pm$0.15& 42.54$\pm$0.15& 43.76$\pm$0.15&S06,E08 \\
UGC4393     &    35& 7.95&  67.15&  33.72& J04 &$-$2.195$\pm$0.153& 40.76$\pm$0.15& 40.98$\pm$0.15& 41.29$\pm$0.15&J04,E08 \\
POX4        &    52& 7.96&  16.71&   14.6& G03 &$-$1.238$\pm$0.153& 41.28$\pm$0.15& 40.64$\pm$0.15& 41.59$\pm$0.16&G03,E08 \\
Tol2138$-$405 &   246& 8.01&   15.5&  11.14&  A  &$-$1.956$\pm$0.153& 40.94$\pm$0.15& 40.55$\pm$0.16& 41.44$\pm$0.16&...,E08 \\
NGC4861     &  15.2& 8.01&  119.5&  118.5& G03 &$-$2.739$\pm$0.153& 40.01$\pm$0.15& 39.53$\pm$0.15& 40.30$\pm$0.15&G03,E08 \\
Mrk206      &  25.4& 8.04&   18.1&  10.45&  A  &$>-$1.371$\pm$0.153& 41.89$\pm$0.15& 41.42$\pm$0.15& 42.19$\pm$0.15&...,E08 \\
UM448       &    87& 8.06&  13.25&   7.65&  A  &$>-$0.632$\pm$0.153& 42.65$\pm$0.15& 42.16$\pm$0.15& 42.94$\pm$0.15&...,E08 \\
SHOC391     &   106& 8.06&   8.49&    4.9&  A  &$-$0.228$\pm$0.153& 42.16$\pm$0.15& 41.43$\pm$0.16& 42.79$\pm$0.16&K04,E08 \\
Mrk1450     &    20& 7.99&     13&    5.1& C07 &$-$0.586$\pm$0.108& 41.92$\pm$0.15& 41.22$\pm$0.16& 42.19$\pm$0.16&C07,E08 \\
SBS0335$-$052 &    57& 7.25&     15&    4.1& C07 &$-$0.721$\pm$0.108& 41.33$\pm$0.15& 41.05$\pm$0.32& 42.26$\pm$0.32&C07,E08 \\
VIIZw403    &   4.3& 7.71&  43.35&    6.0& C07 &$-$1.107$\pm$0.108& 41.95$\pm$0.15& 41.33$\pm$0.40& 42.12$\pm$0.40&C07,E08 \\
HS0822$+$3542 &    11&  7.4&      8&    4.1& C07 &$-$1.192$\pm$0.108& 41.22$\pm$0.15& 41.00$\pm$0.32& 41.53$\pm$0.32&C07,E08 \\
UGCA292     &   3.1& 7.27&     30&    5.0& C07 &$-$1.869$\pm$0.108& 40.89$\pm$0.31& 40.65$\pm$0.32& 41.11$\pm$0.39&C07,E08 \\
UM461       &  13.4&  7.8&   9.08&    5.0& C07 &$-$0.892$\pm$0.108& 41.57$\pm$0.15& 40.69$\pm$0.21& 41.89$\pm$0.21&C07,E08 \\
 \enddata

%% Text for table notes should follow after the \enddata but before
%% the \end{deluxetable}. Make sure there is at least one \tablenotemark
%% in the table for each \tablenotetext.

\tablenotetext{a}{Galaxy name, as listed in \citet{Dale2007}, \citet{Dale2009}, or \citet{Engelbracht2008}.}
\tablenotetext{b}{Distance in Mpc, rescaled, where necessary, to H$_o$=70~km~s$^{-1}$~Mpc$^{-1}$,  as reported in \citet{Dale2007}, \citet{Kennicutt2008}, and \citet{Engelbracht2008}.}
\tablenotetext{c}{Oxygen abundances. For the SINGS galaxies (marked as `S' in column~6), 
the reported number is an average of the `high' and `low' oxygen abundance values 
derived in \citet{Moustakas2009} \citep[see discussion in][]{Calzetti2007}. For the LVL galaxies 
(marked as `L' in column~6), the reported oxygen abundances are from \citet{Marble2010}. For the 
other galaxies, oxygen abundances are from a variety of literature sources, as reported in 
\citet{Engelbracht2008}.}
\tablenotetext{d}{The optical radius R$_{25}$ in arcseconds, defined as D$_{25}$/2. 
When available, the preferred source 
is the RC3 catalog \citep{DeVaucouleurs1991}, as reported in  NED.}
\tablenotetext{e}{The radius of the ionized gas emission, R$_{H\alpha}$, in arcseconds (column~5), 
and source of the measurement (column~6). 
Measurements performed on images: 
S=SINGS \citep[available at: http://data.spitzer.caltech.edu/popular/sings/20070410\_enhanced\_v1/; ][]{Kennicutt2009}; L=LVL \citep{Kennicutt2008}; H=archival HST/ACS image; C99=\citet{Calzetti1999}; 
C07=\citet[][using P$\alpha$~$\lambda$1.8756~$\mu$m images]{Calzetti2007}. Measurements performed on images 
retrieved from NED:  E96=\citet{Evans1996}; G01=\citet{Garcia2001}; G03=\citet{GildePaz2003}; 
H04=\citet{Hunter2004}; J00=\citet{Johnson2000}; J04=\citet{James2004}; K01=\citet{Koopman2001}; 
L99=\citet{Larsen1999}. Measurements performed on published hard--copy images only: 
B02=\citet{Boselli2002}; C06=\citet{Croft2006}; F90=\citet{Fabbiano1990}; 
H94=\citet{Hunter1994}; H03=\citet{Hernandez2003}; HT04=\citet{Hattori2004}; 
H05=\citet{Hernandez2005}; M99=\citet{Mendez1999};  R07=\citet{Robitaille2007}. When no image 
is available, values adopted from Figure~\ref{fig1} are indicated with `A'.}
\tablenotetext{f}{Star formation rate per unit area, in units of M$_{\odot}$~yr$^{-1}$~kpc$^{-2}$, derived 
from equation~17, and dividing by the area calculated from the ionized gas emission 
radius of column~5. The $\Sigma_{SFR}$ of galaxies for which the H$\alpha$ flux information 
is missing (see last column of Table) is marked as a lower limit.}
\tablenotetext{g}{\ Luminosity at 70~$\mu$m (column~8), 160~$\mu$m (column~9), 
and integrated over the full 3--1,100~$\mu$m wavelength range (TIR, column~10), normalized to the ionized gas emission area, in units of erg~s$^{-1}$~kpc$^{-2}$.} 
\tablenotetext{h}{\ Literature source of the H$\alpha$ (first reference) and infrared (second reference) 
flux data.  B02=\citet{Boselli2002}; C06=\citet{Croft2006};  C07=\citet[][$\Sigma_{SFR}$ derived from extinction--corrected P$\alpha$ fluxes]{Calzetti2007}; D07=\citet{Dale2007};  D09=\citet{Dale2009}; E08=\citet{Engelbracht2008};  H94=\citet{Hunter1994}; K87=\citet{Kennicutt1987};  K04=\citet{Kniazev2004}; K08=\citet{Kennicutt2008}; K09=\citet{Kennicutt2009}; 
L96=\citet{LehnertHeckman1996}; M99=\citet{Mendez1999}; M06=\citet{Moustakas2006}; 
P06=\citet{PerezGonzalez2006};  S06=\citet{Schmitt2006}; W99=\citet{Wang1999}.}

\end{deluxetable}

\clearpage

\begin{deluxetable}{lrrrrrrrrr}
\tablecolumns{10}
\rotate
\tabletypesize{\scriptsize}
\tablecaption{Properties of Luminous Infrared Galaxies.\label{tbl-2}}
\tablewidth{0pt}
\tablehead{
\colhead{Name\tablenotemark{a}} & \colhead{D\tablenotemark{b}} 
& \colhead{R$_{25}$\tablenotemark{c}}  & \colhead{R$_{H\alpha}$\tablenotemark{d}} 
& \colhead{Log($\Sigma_{SFR}$)\tablenotemark{e}} & \colhead{Log($\Sigma_{24}$)\tablenotemark{f}}
& \colhead{Log($\Sigma_{70}$)\tablenotemark{f}}
& \colhead{Log($\Sigma_{160}$)\tablenotemark{f}} 
& \colhead{Log($\Sigma_{TIR}$)\tablenotemark{f}} 
& \colhead{Log($\Sigma_{3.6}$)\tablenotemark{g}} 
\\
\colhead{} & \colhead{(Mpc)} & \colhead{($^{\prime\prime}$)} 
& \colhead{($^{\prime\prime}$)} & \colhead{(M$_{\odot}$~yr$^{-1}$~kpc$^{-2}$)} 
& \colhead{(erg~s$^{-1}$~kpc$^{-2}$)} 
& \colhead{(erg~s$^{-1}$~kpc$^{-2}$)} & \colhead{(erg~s$^{-1}$~kpc$^{-2}$)} 
& \colhead{(erg~s$^{-1}$~kpc$^{-2}$)} & \colhead{(erg~s$^{-1}$~kpc$^{-2}$)} 
}
\startdata
NGC0023        & 63.8& 62.7&  10.&    0.602$\pm$0.154 &43.36$\pm$ 0.15 &43.93$\pm$ 0.16 &43.59$\pm$ 0.16 &44.19$\pm$ 0.18 &43.03$\pm$ 0.15\\
MCG+12-02-001  & 68.9&  6.6&  4.9&    1.179$\pm$0.154 &43.92$\pm$ 0.15 &44.39$\pm$ 0.16 &43.92$\pm$ 0.16 &44.63$\pm$ 0.18 &...$\pm$ ...\\
NGC0633        & 72.7& 38.6&  4.9&    0.807$\pm$0.154 &43.56$\pm$ 0.15 &43.97$\pm$ 0.16 &43.52$\pm$ 0.16 &44.24$\pm$ 0.18 &42.68$\pm$ 0.15\\
UGC1845        & 66.4& 36.1&  6.2&    0.428$\pm$0.154 &43.19$\pm$ 0.15 &43.88$\pm$ 0.16 &43.49$\pm$ 0.16 &44.09$\pm$ 0.18 &42.60$\pm$ 0.15\\
NGC1614        & 62.7& 39.5& 13.2&    0.606$\pm$0.146 &43.33$\pm$ 0.15 &43.49$\pm$ 0.15 &42.97$\pm$ 0.15 &43.85$\pm$ 0.15 &42.09$\pm$ 0.15\\
UGC3351        & 65.2& 47.6& 12.7& $-$0.287$\pm$0.154 &42.48$\pm$ 0.15 &43.44$\pm$ 0.15 &43.06$\pm$ 0.16 &43.62$\pm$ 0.17 &42.16$\pm$ 0.15\\
NGC2369        & 47.1&106.5& 17.5& $-$0.169$\pm$0.148 &42.60$\pm$ 0.15 &43.30$\pm$ 0.15 &42.87$\pm$ 0.15 &43.50$\pm$ 0.15 &42.24$\pm$ 0.15\\
NGC2388        & 61.9& 30.0& 13.3&    0.045$\pm$0.154 &42.80$\pm$ 0.15 &43.42$\pm$ 0.16 &42.94$\pm$ 0.16 &43.62$\pm$ 0.18 &42.01$\pm$ 0.15\\
MCG+02-20-003  & 72.4& 19.0&  2.8&    0.982$\pm$0.154 &43.75$\pm$ 0.15 &44.51$\pm$ 0.16 &44.11$\pm$ 0.16 &44.70$\pm$ 0.18 &42.84$\pm$ 0.15\\
NGC3110        & 78.7& 46.5& 20.&    0.012$\pm$0.154 &42.77$\pm$ 0.15 &43.50$\pm$ 0.15 &43.15$\pm$ 0.16 &43.72$\pm$ 0.17 &42.24$\pm$ 0.15\\
NGC3256        & 35.5&114.1& 23.3&    0.434$\pm$0.146 &43.17$\pm$ 0.15 &43.55$\pm$ 0.15 &43.02$\pm$ 0.15 &43.81$\pm$ 0.15 &42.17$\pm$ 0.15\\
NGC3690/IC694  & 51.1& 85.2& 16.2&    1.008$\pm$0.154 &43.72$\pm$ 0.15 &44.02$\pm$ 0.16 &43.31$\pm$ 0.16 &44.28$\pm$ 0.18 &42.56$\pm$ 0.15\\
ESO320-G030    & 40.4& 67.2&  8.8&    0.448$\pm$0.147 &43.22$\pm$ 0.15 &44.04$\pm$ 0.15 &43.53$\pm$ 0.15 &44.19$\pm$ 0.15 &42.51$\pm$ 0.15\\
MCG-02-33-098  & 77.7& 49.8&  3.5&    1.367$\pm$0.154 &44.12$\pm$ 0.15 &44.45$\pm$ 0.16 &43.93$\pm$ 0.16 &44.73$\pm$ 0.18 &43.33$\pm$ 0.17\\
IC860          & 63.3& 16.5&  3.3&    1.088$\pm$0.154 &43.85$\pm$ 0.15 &44.65$\pm$ 0.15 &43.94$\pm$ 0.16 &44.76$\pm$ 0.17 &42.83$\pm$ 0.15\\
NGC5135        & 55.9& 77.1&  7.4&    0.638$\pm$0.154 &43.39$\pm$ 0.15 &43.97$\pm$ 0.16 &43.55$\pm$ 0.16 &44.20$\pm$ 0.18 &42.82$\pm$ 0.15\\
NGC5653        & 65.1& 52.2& 12.7& $-$0.076$\pm$0.154 &42.68$\pm$ 0.15 &43.31$\pm$ 0.15 &42.95$\pm$ 0.16 &43.55$\pm$ 0.17 &42.23$\pm$ 0.15\\
NGC5734        & 63.5& 45.4& 13.0& $-$0.379$\pm$0.154 &42.39$\pm$ 0.15 &43.20$\pm$ 0.15 &42.96$\pm$ 0.16 &43.45$\pm$ 0.17 &42.20$\pm$ 0.15\\
IC4518         & 74.9&  8.4&  6.2&    0.606$\pm$0.154 &43.36$\pm$ 0.15 &43.79$\pm$ 0.16 &43.35$\pm$ 0.16 &44.06$\pm$ 0.18 &...$\pm$ ...\\
Zw049.057      & 63.3& 26.8& 13.0& $-$0.269$\pm$0.154 &42.50$\pm$ 0.15 &43.58$\pm$ 0.15 &43.00$\pm$ 0.16 &43.67$\pm$ 0.17 &41.52$\pm$ 0.15\\
NGC5936        & 65.1& 43.4&  8.4&    0.315$\pm$0.154 &43.07$\pm$ 0.15 &43.58$\pm$ 0.15 &43.20$\pm$ 0.16 &43.84$\pm$ 0.17 &42.47$\pm$ 0.15\\
IRAS17138-1017 & 81.2& 22.5&  5.1&    0.929$\pm$0.154 &43.67$\pm$ 0.15 &44.19$\pm$ 0.16 &43.71$\pm$ 0.16 &44.42$\pm$ 0.18 &...$\pm$ ...\\
IC4687/6       & 79.4& 38.7&  5.8&    1.046$\pm$0.154 &43.78$\pm$ 0.15 &44.21$\pm$ 0.16 &43.76$\pm$ 0.16 &44.47$\pm$ 0.18 &42.65$\pm$ 0.15\\
IC4734         & 73.5& 38.7&  5.6&    0.626$\pm$0.154 &43.38$\pm$ 0.15 &44.12$\pm$ 0.15 &43.75$\pm$ 0.16 &44.33$\pm$ 0.17 &42.69$\pm$ 0.15\\
NGC6701        & 60.6& 46.5&  6.8&    0.443$\pm$0.154 &43.21$\pm$ 0.15 &43.82$\pm$ 0.16 &43.43$\pm$ 0.16 &44.05$\pm$ 0.18 &42.70$\pm$ 0.15\\
NGC7130        & 70.7& 45.4& 11.7&    0.210$\pm$0.154 &42.96$\pm$ 0.15 &43.54$\pm$ 0.16 &43.12$\pm$ 0.16 &43.77$\pm$ 0.18 &42.27$\pm$ 0.15\\
IC5179         & 50.0& 70.4& 16.5& $-$0.062$\pm$0.154 &42.70$\pm$ 0.15 &43.34$\pm$ 0.16 &42.93$\pm$ 0.16 &43.56$\pm$ 0.18 &42.21$\pm$ 0.15\\
NGC7591        & 70.2& 58.5&  5.9&    0.559$\pm$0.154 &43.32$\pm$ 0.15 &43.84$\pm$ 0.16 &43.43$\pm$ 0.16 &44.08$\pm$ 0.18 &42.73$\pm$ 0.15\\
NGC7771        & 61.2& 75.4& 30.&    0.080$\pm$0.154 &42.83$\pm$ 0.15 &43.52$\pm$ 0.15 &43.19$\pm$ 0.16 &43.76$\pm$ 0.17 &42.51$\pm$ 0.15\\
 \enddata

%% Text for table notes should follow after the \enddata but before
%% the \end{deluxetable}. Make sure there is at least one \tablenotemark
%% in the table for each \tablenotetext.

\tablenotetext{a}{Galaxy name, as listed in \citet{AlonsoHerrero2006}.}
\tablenotetext{b}{Distance in Mpc from \citet{Sanders2003} and \citet{Surace2004}, rescaled to
 H$_o$=70~km/s/Mpc.}
\tablenotetext{c}{The optical radius R$_{25}$ in arcseconds. When available, the preferred source 
is the RC3 catalog \citep{DeVaucouleurs1991}, as reported in NED.}
\tablenotetext{d}{The radius of the ionized gas emission, R$_{H\alpha}$, in arcseconds.  When 
available, the H$\alpha$ sizes are from the images in \citet{Hattori2004}; otherwise 
they are from \citet{AlonsoHerrero2006}, using HST NICMOS P$\alpha$~$\lambda$1.8756~$\mu$m images.   
For NGC1614, NGC3256, NGC3690, NGC5653, and Zw049.057 the data are from 
\citet{AlonsoHerrero2000, AlonsoHerrero2001,AlonsoHerrero2002}.}
\tablenotetext{e}{Star formation rate per unit area, in units of M$_{\odot}$~yr$^{-1}$~kpc$^{-2}$, derived 
from equation~17, and dividing by the area calculated from the ionized gas emission 
radius of column~4.}
\tablenotetext{f}{\ Luminosity per unit area at 24~$\mu$m, 70~$\mu$m, 160~$\mu$m, 
and integrated over the full 8--1,000~$\mu$m wavelength range (TIR), in units of 
erg~s$^{-1}$~kpc$^{-2}$, using the ionized gas emission area. The luminosities in the MIPS bands 
are extrapolated from the IRAS measurements as described in section~3. Exceptions are NGC1614 and 
NGC3256, with MIPS data reported in \citet{Engelbracht2008}, and NGC2369, ESO320$-$G030, and 
Zw049.057, with IR SED published in \citet{Rieke2009}.} 
\tablenotetext{g}{\ Luminosity per unit area at 3.6~$\mu$m, in units of 
erg~s$^{-1}$~kpc$^{-2}$, extrapolated from the K$_S$--band luminosity, as described 
in section~3. K$_S$--band luminosities are from 2MASS, as reported in NED, and corrected for the 
effect of both Galactic foreground \citep{Schlegel1998} and internal extinction. }

\end{deluxetable}

\end{document}